\documentclass[twocolumn,noshowpacs,preprintnumbers,amsmath,amssymb]{revtex4}

\usepackage{graphicx,subfigure}
\usepackage{dcolumn}
\usepackage{bm}
\usepackage{gensymb}
\usepackage{wasysym}
\usepackage{textcomp}

\begin{document}

\title{The education of Walter Kohn \\ and  the creation of density functional theory}

\author{Andrew Zangwill}
 \email{andrew.zangwill@physics.gatech.edu}

\affiliation{%
School of Physics \\ Georgia Institute of Technology \\ Atlanta, GA 30332\\
}%


\begin{abstract}
  The theoretical solid-state physicist Walter Kohn was awarded one-half of the 1998 Nobel Prize in Chemistry for his mid-1960's  creation of an approach to the many-particle problem in quantum mechanics called density functional theory (DFT). In its exact form, DFT establishes that the total charge density of any system of electrons and nuclei provides all the information needed for a complete description of that system. This was a breakthrough for the study of atoms, molecules, gases, liquids, and solids. Before DFT, it was thought that only the vastly more complicated many-electron wave function was needed for a complete description of such systems. Today, fifty years after its introduction, DFT (in one  of its approximate forms) is the method of choice used by most scientists to calculate the physical properties of materials of all kinds. In this paper, I present a biographical essay of Kohn's educational experiences and professional career up to and including the creation of DFT. My account begins with Kohn's student years in Austria, England, and Canada during World War II and continues with his graduate and post-graduate training at Harvard University and Niels Bohr's Institute for Theoretical Physics in Copenhagen. I then study the research choices he made during the first ten years of his career (when he was a faculty member at the Carnegie Institute of Technology and a frequent visitor to the Bell Telephone Laboratories) in the context of the theoretical solid-state physics agenda of the late 1950's and early 1960's. Subsequent sections discuss  his move to the University of California, San Diego, identify the research issue which led directly to DFT, and analyze the two foundational papers of  the theory. The paper concludes with an explanation of how the chemists came to award ``their'' Nobel Prize to the physicist Kohn and a discussion of why he was unusually well-suited to create the theory in the first place.

\end{abstract}

\maketitle

\section{Introduction}

 The 1998 Nobel Prize in Chemistry recognized the field of quantum chemistry, a theoretical enterprise where quantum mechanics is used to study molecules and address chemical problems (Barden and Schaefer 2000, Gavroglu and Sim\~{o}es 2012). The prize was shared equally by ``Walter Kohn for his development of density functional theory and John Pople for his development of computational methods in quantum chemistry'' (Nobel 2013). The award to Pople surprised no one.\,\footnote{John Anthony Pople (1925-2004) earned his BA (1946) and PhD (1951) degrees in Mathematics from Cambridge University.  His PhD thesis, ``Lone pair orbitals'' was supervised by John Lennard-Jones, head of the Department of Theoretical Chemistry. A native Englishman, Pople spent more than a decade teaching and conducting research at Cambridge before moving to the United States in 1964 to take a position as Professor of  Chemical Physics at the Carnegie Institute of Technology in Pittsburgh, PA. During his thirty-year career at Carnegie Tech, Pople made a gradual transition for the development of semi-empirical methods in molecular orbital theory to the development of computer codes to solve the Schr\"{o}dinger equation for small molecules at the Hartree-Fock level and beyond. He moved to Northwestern University in 1993 and earned one-half of the 1998 Nobel Prize in Chemistry for his achievements in computational quantum chemistry (Buckingham 2006).} Ten years earlier, an  international conference, ``Forty Years of Quantum Chemistry'', had honored Pople's career-long  commitment to developing  semi-empirical and  first-principles methods to predict the structure and properties of molecular systems   (Handy and Schaefer 1990). A notable early contribution was the 1970 free release of his group's GAUSSIAN computer program which solved the Schr\"{o}dinger equation for molecules in the Hartree-Fock approximation. Today, after decades of improvements in accuracy and functionality, it is estimated that  90\% of all quantum  chemistry calculations are performed using the  (now commercial) GAUSSIAN suite of programs (Crawford {\it et al.} 2001).

By contrast, the fact that Walter Kohn earned a share of the Nobel Prize in {\it chemistry} surprised many because his international reputation was gained as a theoretical solid-state physicist.  Indeed, his body of work had previously earned him the   Oliver E. Buckley Prize (1961)  and the Davisson-Germer Prize (1977) of the American Physical Society for his contributions to, respectively, ``the foundations of the electron theory of solids'' and ``understanding the inhomogeneous electron gas and its application to electronic phenomena at surfaces'' (APS 2013a). A card-carrying chemist could reasonably ask: what do these things have to do with quantum chemistry? The answer lies in the research Kohn conducted in the period 1963-1965 which created the density functional theory (DFT) cited by the Nobel committee. This theory focuses on the density of electrons in an atom, molecule, or solid rather than on the many-electron wave function that is the focus of traditional quantum chemistry.

The connection between the work of the two 1998 Chemistry Nobel laureates can be understood from the `hyperbola of quantum chemistry'' which  Pople published in the proceedings of a 1965 symposium on atomic and molecular quantum theory (Pople 1965). The horizontal axis of this graph (see Fig. 1) indicates the number of electrons $N$ in the system of interest. The vertical axis lists a sequence of quantum mechanical methods (in order of increasing accuracy) used to determine the system's `electronic structure', {\it i.e.}, its  many-electron wave function,  electronic charge density, electron energy levels, and other properties calculable from these. Pople suggested that the activities of most quantum chemists tended to cluster around the extremities of the hyperbolic solid line. Those interested in the highest accuracy were forced by computational constraints to focus on small molecules (small $N$)  while those  interested in large molecules (large $N$) were forced by computational constraints to use methods that were capable of only low accuracy. He noted that progress would occur by moving off the hyperbola either horizontally from left to right or vertically from bottom to top.

\begin{center}
\includegraphics[scale=.45]{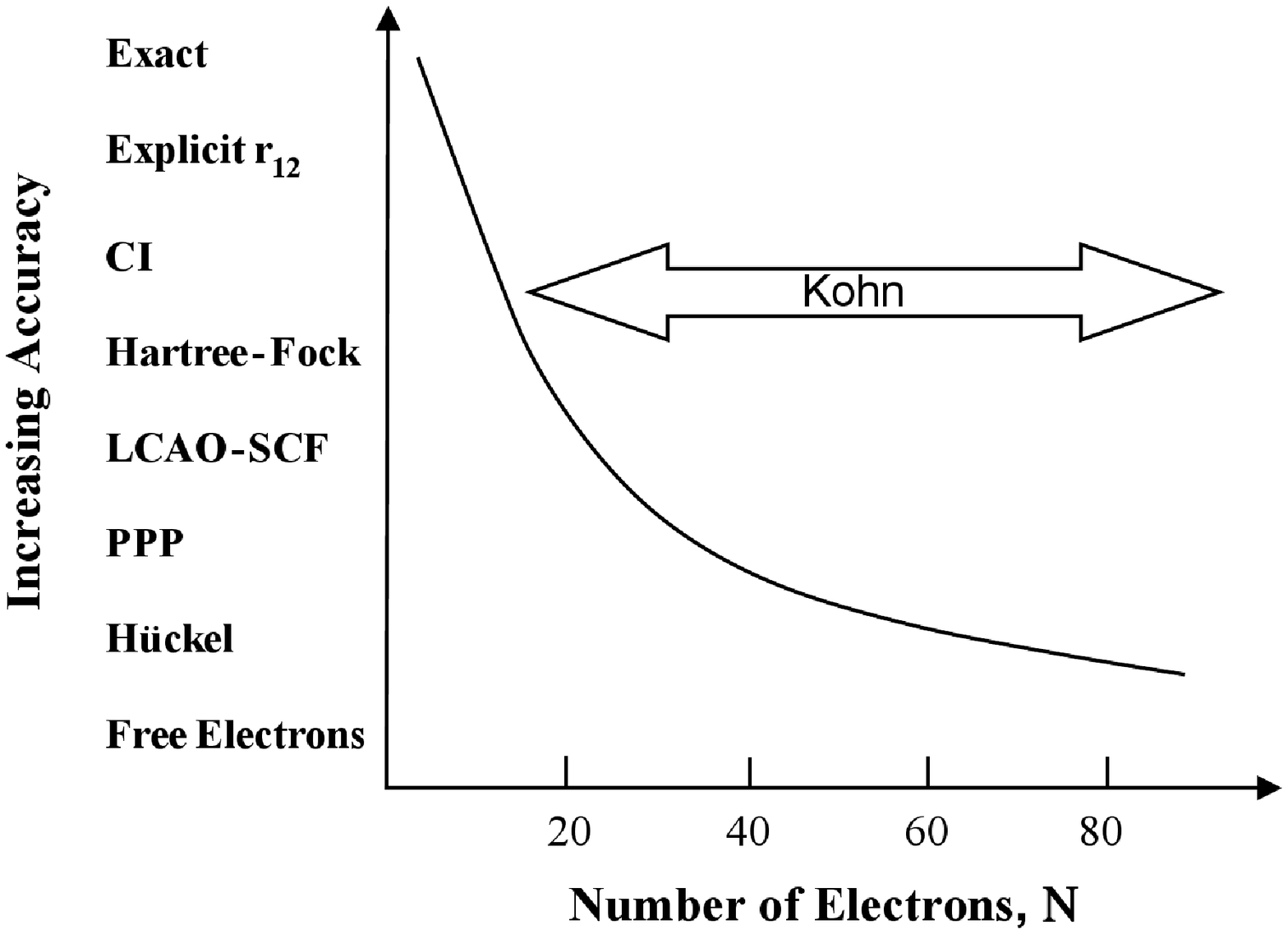}

\small Fig. 1. The applicability range of Kohn's density \\
      functional theory placed on Pople's hyperbola of \\
      quantum chemistry. Adapted from Pople (1965).
\end{center}

The density functional theory created by Walter Kohn represented a dramatic move away from Pople's hyperbola, particulary for systems with a large number of electrons (see the arrow in Fig. 1). Many chemists were wary (or dismissive) of this theory at first because it made no use of the $N$-electron wave function, a quantity thought to be indispensable for a proper description of any atom or molecule. Eventually, improvements to Kohn's theory  made  by chemists themselves led to quantitative successes that could not be denied. Today, it is the method of choice used by most scientists who wish to calculate the properties of real materials.

Unlike many other scientific achievements, the technical question which led Kohn to create DFT in the mid-1960's was not   ``in the air'' among physicists, chemists, or anyone else. It is entirely possible that the theory would be unknown today if Kohn's background, technical skills, and scientific experiences  differed  very much from what they were.  Accordingly, this article (i) recounts the unusual personal and intellectual journey which led to Walter Kohn's  success as a theoretical physicist; (ii) identifies the scientific issue which motivated the creation of DFT in the context of theoretical solid state physics research in the 1960's; (iii) summarizes the key elements of the founding papers of DFT; (iv) argues that Kohn's scientific background made him particularly well-suited to create density functional theory; and (v) explains in brief how Kohn came to win a share of the Nobel Prize in Chemistry for the creation of DFT.

\section{An Unsentimental Education}
In 1933, ten-year-old  Walther Kohn began the eight-year course of study at the Akademische
 Gymnasium, the oldest and one of the best secondary schools in his home city of Vienna.\,\footnote{Walther Kohn began to use the name Walter Kohn in 1940.}  In doing so, he was not unlike the children of many middle-class Jewish parents  who were actively engaged in the intellectual and artistic life of their city.\,\footnote{Beller (1989) paints a cultural portrait of the Jewish community in early twentieth century Vienna.} His father Salomon owned an art publishing company that  specialized in the manufacture and distribution of high-quality art postcards. Despite a worldwide clientele, it was a struggle to keep the business going in the face of a global economic depression which hit Austria particulary hard. Nevertheless, there was a tacit understanding that Walther would eventually run the family business. Kohn's mother Gittel was a highly educated woman who spoke  four languages and it was she who chose the humanistically oriented Akademische Gymnasium to educate her son. Walther excelled at Latin and ancient Roman history but showed no apparent aptitude for mathematics. The only grade of C he ever received was in that subject (Kohn 1998).

 Besides marking the beginning of Kohn's secondary school education, 1933 was also the year that Adolf Hitler and his Nazi party took power in neighboring Germany. The Nazi party was illegal in Austria, but its many sympathizers worked tirelessly to undermine the existing democratically elected government. Finally, in May 1934, the country succumbed to a form of authoritarian rule known to historians as Austrofascism (Berkley 1988). Four years later, cheering crowds welcomed Hitler when the German army crossed the border and marched into Vienna. The  {\it Anschluss} (political union) of Germany and Austria was a {\it fait accompli}.  These events were vivid in Kohn's memory many years later (Kohn 1998, Kohn 2013a):
\begin{quote}
 The [March 1938] Anschluss changed everything. The family business was confiscated but my father was required to continue its management without any compensation. . . . He wrote to a London art distributor and business client named Charles Hauff (whom he had never met) and asked if he  and his wife would temporarily accept me and my older sister Minna into their home. Hauff  replied affirmatively and Minna emigrated to England very soon thereafter. For reasons of their own, the Nazis made it much more difficult for young Jewish boys to leave. I remained in Vienna, but was expelled from my school.
\end{quote}

Many expelled students never went to school again. Kohn was lucky because he was permitted to finish the school year at a segregated high school for Jews. Then, in August 1938, he was one of a few hundred high-achieving Jewish students from the various Viennese secondary schools who were given the opportunity to continue their  education at the  Zwi Perez Chajes Gymnasium, a private all-Jewish high school.\,\footnote{This would be Kohn's sixth (and last) year of secondary school in Vienna. Quite unusually for Austria at that time, boys and girls were not separated for instruction in Walther's small class at the Chajes school (Neuhaus 2003).}  This experience was transformative for Kohn because his interest in science was ignited by the physics teacher Emil Nohel and the mathematics teacher Victor Sabbath.\,\footnote{Emil Eliezer Nohel grew up on a small farm in Bohemia.  He studied mathematics at the  Karl-Ferdinand (German) University in Prague and served as an assistant to Albert Einstein when Einstein was a professor there from 1911 to 1914.  Nohel's descriptions of the difficult conditions endured by Jews in Bohemia awakened Einstein's concern for the plight of his co-religionists. For most of his career, Nohel  taught mathematics at the Handelsakademie Wien, a business-oriented secondary school in Vienna. The  Anschluss precipitated his dismissal and he found work as the physics teacher (and then the principal) of the Zwi Perez Chajes Gymnasium. Nohel was arrested on December 12 1942 and spent two years at the Theresienstadt labor camp before he was transferred to the Auschwitz concentration camp and murdered by the Nazis (Frank, 1947, Pais 1982).} He later recalled that (Hollander 2000, Kohn 2004)
\begin{quote}
Nohel was  a tall, quiet, noble man who devoted himself to his students. It was really a combination of my admiration for this man as a person and his deep knowledge of physics that started me off . . . Though I was only fifteen going on sixteen years old, I already understood---due to Nohel's role model and by comparing myself to others---what it meant to really comprehend something in physics. This is one of the most important insights for a future theoretical physicist.

Sabbath was also a fantastic guy. The thing I remember about him is that while he was teaching us he told us about a new book he was reading by the great French physicist [Louis] de Broglie called {\it Matter and Light}. . . Sabbath was a teacher with great enthusiasm and it was very exciting what he told us.
\end{quote}
Kohn's inspirational teachers Emil Nohel and Victor Sabbath did not survive the Holocaust.

November 10 1938 was particulary memorable at the Chajes school because it was the day following the night of the  infamous, state-sanctioned orgy of destructive violence against Jewish homes, businesses and synagogues known as {\it Kristallnacht} (the night of broken glass). The principal dismissed the students early to avoid attracting attention, but Kohn and a classmate were arrested on their way home. They were released after several terrifying hours in the police station, but Kohn returned home to find ``our apartment absolutely vandalized by a group of hooligans, including the person who had taken over my father's business'' (Hanta 1999). The classes at Chajes got smaller and smaller in the months after Kristallnacht. Emigration was on everyone's mind, but it was difficult and expensive to make it happen (Ehrlich 2003). Kohn was again lucky. He escaped from Austria to England just three weeks  before Germany invaded Poland and World War II began. His parents were unable to leave. Both were eventually deported, first to the Terezin concentration camp in Czechoslovakia and then to Auschwitz, where they were murdered in 1944.

\begin{center}
\includegraphics[scale=.5]{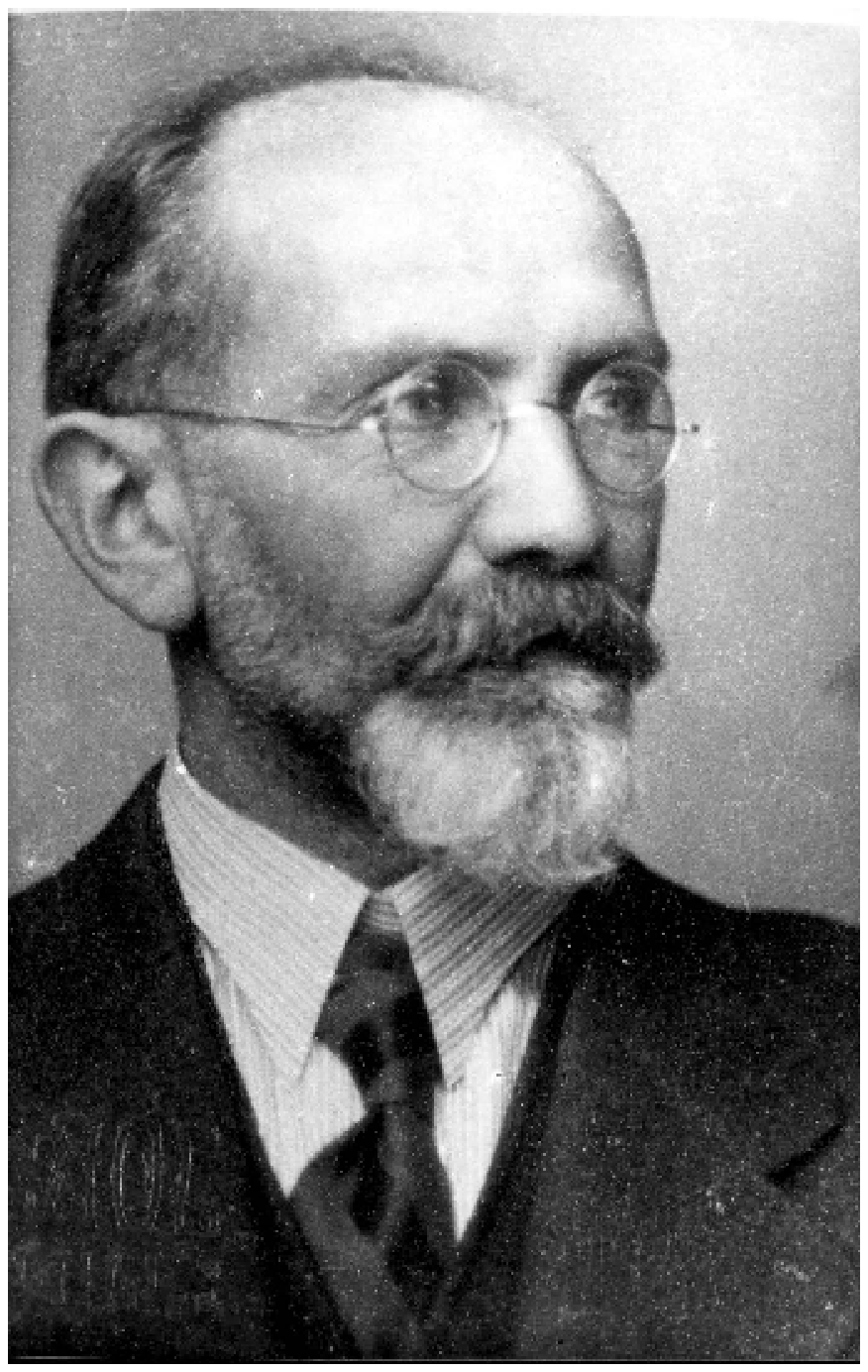}

\small Emil Eliezer Nohel (ca. 1938) was a high school physics \\
teacher who inspired Kohn to become a scientist. \\ Courtesy of the Yad Vashem Photo Archive.
\end{center}

Sixteen-year old Walther  was saved from probable extermination by the Nazis by the {\it Kindertransport}, an organized effort to evacuate Jewish children under the age of 18 from Germany, Austria, and Czechoslovakia. Between December 1 1938 and September 1 1939, this collaboration between British Quakers, World Jewish Relief, the Refugee Children's Movement (a British rescue organization), and the Jewish agencies in the affected countries made it possible for almost 10,000 unaccompanied children to board special trains bound for port cities in the Netherlands. Boats carried them from there to safety across the English Channel (Fast 2011). By agreement with the British government, the children were to be placed with sponsor families or volunteer foster parents for a period of up to two years, whereafter repatriation to their home countries was anticipated. The Jewish community of Britain bore full financial responsibility for the transport. The host families were expected to arrange educational opportunities for their guest children.

One quarter of  the Kindertransport children came from Austria and the Jewish Community of Vienna [Israelitsche Kultusgemeinde Wien (IKG)] was responsible for deciding which children would receive exit visas from amongst the great many applications they received (Curio 2004). Initially, the IKG focused on urgent cases, including stateless children under threat of expulsion and children in orphanages. Priority was given to children whose parents had been arrested or deported. Later, preference was given to children with guaranteed sponsors in Britain, but this alone did not ensure selection. The children were interviewed and the additional criteria of good health, a pleasant personality, likely success in school, and the ability to ``fit in'' were used to make the final choices. Kohn apparently met all these criteria and thereby earned a seat on one of the last Kindertransport trains. He arrived in England safely with the pre-arranged plan to live with Charles and Eva Hauff, the same couple who had previously welcomed his sister Minna. In fact, Minna met Walther at London's Victoria station and they traveled together by train to West Sussex where the Hauff's lived (Kohn 2013a).

Before leaving Vienna, Kohn and his parents had agreed that he should learn to be a farmer in England (Kohn 1998). They had seen too many unemployed intellectuals in pre-war Vienna and farming seemed like an occupation that would make him less subject to economic dislocations.\,\footnote{The unemployment rate in Austria stood at 20 percent at the time of the Anschluss (Senft 2003).} Moreover, Salomon Kohn was 66 in 1939 and it was understood that Walther would soon be responsible for his parents' financial security. All of this was communicated to the Hauffs ahead of time and they were able to arrange for Kohn's formal English education to begin at a training farm in Sittingbourne, Kent, about one hour away from the Hauff's home by car.  There, Kohn recalls (Hollander 2000),
 \begin{quote}
I pulled carrots and looked after  piglets. . . Unfortunately, on the farm, I contracted what turned out to be meningitis and so was very ill. Sulfa drugs had just been invented, so I pulled through, but it was touch and go. After that, I was very weak and going back to the farm was out of the question.
 \end{quote}
The Hauffs dealt swiftly with this setback and, in January 1940, Walther entered the nearby East Grinstead County Grammar School. His limited English skills led the  headmaster, Thomas W. Scott, to estimate that 2-3 years could be required for this new student to earn the school certificate needed to enter college. Scott's method to redress this situation was to enroll  Kohn in the lower $6^{\rm th}$ form (to avoid extra English requirements) and instruct his teachers to overlook their new student's deficiencies in English. He then created a  daily German class where Walther  was the only student. The instructor taught English to Kohn  for half the class and Kohn taught advanced German to the instructor for the other half (Kohn 2000, Ford 2013).

Kohn's preparation in math and science was equal to or exceeded that of his fellow 16-year-old English classmates (Kohn 2013a). The  level of physics he was exposed to at East Grinstead can be judged from two of his textbooks, both intended for students preparing for University Scholarship Examinations. {\it Heat} (1939) by R.G. Mitton is a thorough introduction to the thermal properties of matter, the kinetic theory of gases, heat engines, entropy, and the laws of thermodynamics. The prose is brisk, yet clear, and the author assumes familiarity with the laws of algebra and the geometrical meaning of the derivative. The latter occurs in a section devoted to the Clausius-Clapeyron equation, an advanced topic not often found in present-day American high school textbooks. The oddly-named {\it Properties of Matter} (1937) by D.N. Shorthose is a textbook of particle and continuum mechanics which includes chapters devoted to ballistic motion, circular motion, simple harmonic motion, rigid body motion, hydrostatics, friction, elasticity, and viscosity. The exposition moves freely between geometrical and algebraic reasoning and the reader is expected to understand first derivatives, second derivatives, and the law of integration which connects them. This is material found in present-day American textbooks intended for first-year college students.

Kohn's life at East Grinstead was happy and peaceful for five months. Then, on May 10 1940, Germany invaded Holland, Belgium, and Luxembourg.  Winston Churchill replaced Neville Chamberlain as Prime Minister of Great Britain and the British newspapers became filled with war hysteria and reports of `fifth columnists'.\,\footnote{The term `fifth column'  refers to the secret supporters of an enemy who live openly within the territory being defended.} The British War Office feared that an invasion was imminent and recommended to the Home Office that the government ``intern all enemy aliens in areas where German parachute troops are likely to land" (Gilman \& Gilman 1987). Arrests begin immediately and, when Italy entered the war on June 10, Churchill demanded that police officials `collar the lot'. This terse order expanded internment to all parts of the country and to all enemy aliens, {\it i.e.}, to all holders of passports from Italy or Nazi-occupied countries age 17 or older. It took Scotland Yard less than six weeks to intern 24,000 men and 4000 women (Cesarini and Kushner 1993). Those arrested included not just foreign nationals of quite short residency in England like  university students but also highly trained scientists, engineers, doctors, artists, and musicians who had lived in England for many years.\,\footnote{Vienna native Otto Frisch was working in the physics laboratory of Prof. Mark Oliphant at the University of Birmingham when the  roundup of aliens began. He avoided internment only because his employers made the case that he was engaged in important war work (Frisch 1979).}

Kohn had turned 17 on March 9 and thus was subject to the order of internment. He was arrested and shipped by train to a large camp that had been hastily constructed in the town of Huyton on the outskirts of Liverpool. It was at Huyton that  Walther met and began a lifelong friendship with Josef Eisinger, a boy who had been one year behind him at the Akademische Gymnasium in Vienna.\,\footnote{Eisinger's family smuggled him onto a Kindertransport train out of Vienna after failing to find an official  sponsor or a foster family for him in England. He fended for himself, working on a farm in Yorkshire and washing dishes in a hotel in Brighton, until the internment roundup of 1940 (Eisinger 2013).}
A week or so later, both boys were  transferred  to an internment camp on the Isle of Man. The Hauffs arranged for Walther's  East Grinstead teachers to send him his physics textbooks because they were told he would return home soon (Hanta 1999). Instead, he spent a month on the Isle of Man where there was no work and little food. He lost 30 pounds.

The burden of warehousing this great mass of civilian internees and the belief that many German prisoners of war would arrive soon led Churchill's government to seek help from the British Commonwealth nations. Accordingly, on July 4 1940, Kohn and his friend Eisinger found themselves bound for Canada aboard the  {\it Sobieski}, a Polish cruise ship that had been captured by the British and converted to a troop transport. The ship arrived in Quebec City two weeks later as part of a convoy that zigzagged across the Atlantic to avoid German submarines (Koch 1980, Auger 2005).\,\footnote{The {\it Sobieski} was the last of four ships  assigned the task to transport internees to Canada. The  {\it Ettrick} and the {\it Duchess of York} crossed the Atlantic safely.  The {\it Arandora Star} was torpedoed by a German submarine on July 2 and sunk with a loss of more than 800 lives.  Public outcry over this incident contributed to the cessation of the internments and the first releases of detainees from the Isle of Man (Gilman \& Gilman 1980).} By the time the British government discontinued its policy of internments and deportations in early August 1940, nearly 4400 civilian internees and 1950 prisoners of war had been relocated to Canada. What the Canadians did not know---because the British did not tell them---was that the majority of the civilian internees were not Nazi sympathizers but Jewish refugees from Nazi barbarism.

 Kohn spent the next eighteen months in four different Canadian internment camps, a burden  lessened only slightly
 by the Red Cross, which  made it possible for him to exchange letters with his parents every two weeks or so. He was also able to engage in learning much of the time, albeit not exactly of the sort he had experienced before. For example, on the evening of his arrival in Quebec City, he and 710 others internees were moved eighty miles down the St. Lawrence river to the town of Trois Rivi\`{e}res, where barbed wire had been strung around the perimeter of a local agricultural exhibition ground to create Camp T. Barracks were built  in an arena designed to house  livestock (Jones 1988). However, a baseball field  adjacent to the arena was the home of the Trois Rivi\`{e}res Renards of the Quebec Provincial Baseball League.  By standing on several tables stacked on top of one another to reach the windows, Walther and others could watch the games. They pieced together the rules of baseball during the month they were interned there  (Koch 1980, Kohn 2013a).

\begin{center}
\includegraphics[scale=.7]{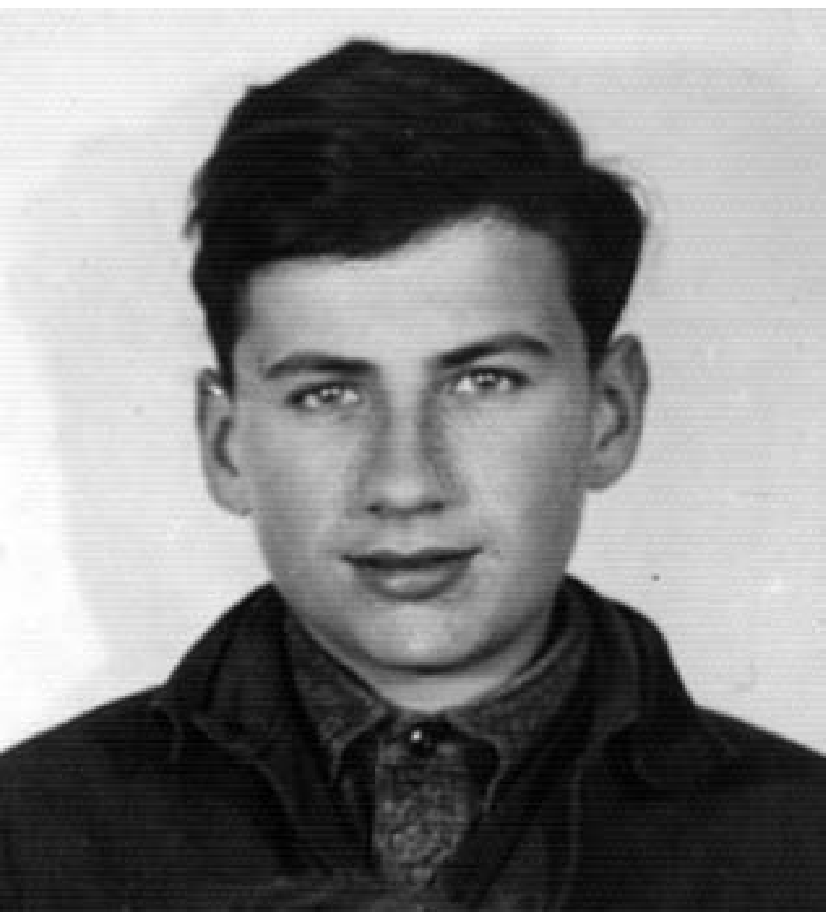}

\small Walther Kohn in Canada at age 18 (1941). \\ Courtesy of Josef Eisinger.
\end{center}

Camp T was closed on August 12, 1940 and its  population was transferred to Camp B, a 15 acre compound  built in the heavily wooded Acadia forest twenty miles east of Fredericton, the capital of the province of New Brunswick.\,\footnote{Camp B was built on the site of an abandoned, depression-era, unemployment relief facility.} Kohn spent nearly a full year at Camp B and a vivid portrait of the camp written by Ted Jones (1988) makes it possible to reconstruct life there in some detail. For our purposes, a salient fact is that the internees organized themselves into a highly-structured mini-society composed of sixteen distinct groups based on age, education, geography, past friendships, religious practices,  and political views. Kohn belonged to a group of academically-minded boys between 16 and 20. A different group was populated largely by academics from Cambridge University (graduate students, recent PhD's, post-doctoral fellows, and junior faculty), another group self-identified as communists, and another group was composed entirely of ultra-orthodox Jews.   Even before the construction of the camp was complete, each  group elected a leader and a deputy leader, as well as chairmen for the various departments of the internee camp Cabinet.

The camp's Canadian officials had a clear idea of how the able-bodied internees would spend most of their time: lumberjacking in the forest outside the camp for 20 cents a day. Kohn was not unhappy to do this because the  sheer physical labor of chopping trees warded off the Canadian winter cold. However, much more important to him were the many hours he spent at the  `camp school' where internees with special technical or academic knowledge offered classes  for the benefit of anyone who wanted to learn. Each  class met at a designated table in the camp's Recreation Hut and internees sought and received time off from their work details to attend.
Appeals to humanitarian organizations and letters written by the instructors to acquaintances around the world produced donations of textbooks, writing materials, exercise books, blackboards, painting supplies, musical instruments, etc. Two months after it opened in November 1940, the camp library boasted a collection of nearly 1000 volumes.

The Camp B school was organized by  the chair of the Education Department, Alfons Rosenberg, a 39-year-old former gymnasium teacher from Berlin who had taught briefly at the Cranbrook School in Kent, England before he was `collared' and deported to Canada. His motivation to create the camp school was simple: 60\% of the internees were under 20 years old and in desperate need of organized instruction.  The many Cambridge men in the camp offered their services and, over time, courses were given in accounting, acting, anthropology, architecture, art  history, astronomy, biology, composition, chemistry, economics, engineering, English, French, geography, German, history, Latin, law, literature, mathematics, music theory, philosophy, physics, physiology, political theory, psychology, sex education, Spanish, and typing.

In a brilliant stroke, Rosenberg arranged with the camp commandant for official examination booklets to be brought into the camp so internees could sit for  the nation-wide McGill University Junior Matriculation Exams---a necessary step to enter a Canadian college. Thomas Cassirer, a future professor of French at the University of Massachusetts, Amherst characterized Rosenberg as (Jones, 1988)
\begin{quote}
a schoolmaster in the good sense: he could listen to others, he had a wide outlook, and he gave suitable advice. . . . He was always at everybody's disposal. At discussions, and we had many with him, he would often solve everything with just a short sentence.
\end{quote}
Walther Kohn attended a daily physics class offered by Kurt Guggenheimer, a physical chemist who anticipated the shell model of the nucleus by pointing out similarities in the systematics of the binding energies of molecules and nuclei.\,\footnote{Kurt Martin Guggenheimer (1902-1975) studied chemistry at the University of Munich and physics at the University of Berlin. He earned his PhD in 1933 for work  on the ultraviolet absorption spectra of zinc, potassium, and cesium under the direction of Fritz Haber. Guggenheimer pursued post-doctoral work in Paris under the direction of Paul Langevin and  published his speculations about nuclei while there. He returned to Munich in 1935 but was arrested following Kristallnacht and spent several months at the Dachau detention center. He emigrated to England and was working at King's College (Cambridge) when he was interned as a enemy alien. After the war, Guggenheimer worked as a Lecturer at the University of Bristol and the University of Glasgow before retiring from academic life in 1967 (R\"{u}rup 2008, Fernandez 2013).} Kohn and future McGill University mathematics professor Joachim Lambek were the only students in a set theory course taught by Fritz Rothberger.\,\footnote{Fritz Rothberger (1902-2000) was a native of Vienna who graduated from the Akademisches Gymnasium and earned BS and PhD degrees in mathematics from the University of Vienna. He came under the influence of Waclaw Sierpinksi in Warsaw and began his lifelong work in combinatorial set theory. Rothberger emigrated to England just before the start of World War II  and he was working as a scholar at Trinity College (Cambridge) when he was interned. After the war, he served as a professor on the mathematics faculties of several Canadian universities: Acadia, Fredericton, Laval, and Windsor (Swaminathan 2000).} According to Lambek (1980),
   \begin{quote}
   Rothberger was an outstanding teacher with an inimitable style of lecturing. He gave unsparingly of his time and managed to bring the most abstract concepts down to earth. He instilled a love of mathematics in numerous young people.
   \end{quote}
A similar impression was left on Kohn (Kohn 1998, Kohn 2013a):
   \begin{quote}
   Rothberger normally taught us out-of-doors where he wore shorts and boots and nothing else. He used a stick and a sandy area as a blackboard to teach us about the different types of infinities . . . . He  was a most kind and unassuming man whose love for the intrinsic depth and beauty of mathematics was gradually absorbed by his students.
   \end{quote}

Walther  ``took the classes very seriously because I felt a huge responsibility to support my parents after the war'' (Kohn 2013a).  The level of his commitment is not difficult to demonstrate. First, he earned all passing marks when Rosenberg and his camp staff administered the McGill Junior Matriculation Exam in June of 1941 (Giannakis 2013).\,\footnote{The subjects tested were English literature, English composition, general history, elementary algebra, elementary geometry, physics, Latin authors, Latin composition, German grammar, German composition, intermediate algebra, intermediate geometry, and trigonometry. By his own account, Kohn performed better in Latin than in German and his worst subject was Canadian history (Kohn 2013a).} Second, Kohn took the 20 cents a day he earned lumberjacking and used the ``princely sum, carefully saved'' to order two books which were sent to him at the camp: {\it A Course in Pure Mathematics} (1938) by G.H. Hardy and {\it Introduction to  Chemical Physics} (1939) by J.C. Slater (Kohn 1998). The Cambridge-educated cohort at Camp B would have been very familiar with the Hardy book because its author was a Cambridge mathematician and the book was in its seventh edition at the time. By contrast, a  brand new book by the American physicist Slater would have been completely unknown to his fellow internees. Kohn purchased it entirely on the basis of a catalogue description (Kohn 2013b). In an uncanny way, the contents of both books foreshadow the mathematical rigor, taste for foundational issues, and deep  interest in the properties of matter that characterize much of this subsequent work.

 Hardy\,\footnote{Godfrey Harold Hardy (1877-1947) was a leading British mathematician of his time. He was a child prodigy who trained  at Trinity College (Cambridge) and served as a professor at both Cambridge and Oxford over the course of his career. He authored or co-authored eleven books and over 300 research papers,  mostly in the fields of analysis and number theory. He was particularly well-known for his  collaborations with Srinivasa Ramanujan and John Edensor Littlewood,  and for his disdain for any kind of ``applied'' work (Titchmarsh 1949).} and Slater\,\footnote{John Clarke Slater (1900-1976) was the chair of the physics department at the Massachusetts Institute of Technology (MIT) from 1930-1951. He wrote an experimental PhD thesis at Harvard, but then traveled to Europe where he made several important theoretical contributions to the early development of quantum mechanics. Dissatisfaction with `formal theory' led him to develop a large research group at MIT  devoted to solving the Schr\"{o}dinger equation numerically to calculate the physical properties of atoms, molecules, and solids. He authored or co-authored thirteen books and over 150 research papers. An important paper he published in 1951 turned out to bear directly on Kohn's density functional theory (Morse 1982).} were both first-rate researchers with a strong interest in the pedagogy of their fields. Hardy wrote the first edition of his book in 1908 for the express purpose of making the teaching of mathematics more rigorous at British universities.\,\footnote{{\it A Course in Pure Mathematics} is an elegant and rigorous introduction to mathematical analysis for serious first-year college students. The subject matter includes the notions of limit and convergence applied to series, sequences, functions, derivatives, and integrals. All the main theorems of the calculus of a real variable are discussed, as is the general theory of logarithmic, exponential and sinusoidal functions.}
 It was aimed specifically at first-year University students of `scholarship standard', {\it i.e.}, the top 10-20\% in ability. The preface to the first edition is explicit: this is  ``a book for mathematicians: I have nowhere made any attempt to meet the needs of students of engineering or indeed any class of students whose interest are not primarily mathematical."  Thirty years later, in the preface to Kohn's edition, Hardy remarks that ``the general plan of the book is unchanged" but  ``I have inserted a large number of new examples from the papers of the Mathematical Tripos".  Kohn would have learned from his camp-mates that these `new examples' were drawn from the demanding examinations used by Cambridge University to evaluate  its students who hoped to earn a BA degree in Mathematics.

The Slater book that Walther  bought and read (doubtless cover to cover) was very unusual for its time. The  preface states that the author worked hard ``to make it intelligible to a reader with a knowledge of calculus and differential equations, but unfamiliar with the more difficult branches of mathematical physics". For that reason ``the quantum theory  used is of a very elementary sort . . . and it has seemed desirable to omit wave mechanics.'' On the other hand, the content is far from elementary. Slater notes that ``it is customary to write books either on thermodynamics or on statistical mechanics; this one combines both." Moreover, ``atomic and molecular structure are introduced, together with a discussion of different types of substances, explaining their interatomic forces from quantum theory and their thermal and elastic behavior from our thermodynamic and statistical methods.'' The preface does not warn the reader about the author's frequent use of kinetic theory, which is a non-trivial subject of its own. All told, the first 100 pages cover  ``Thermodynamics, Statistical Mechanics, and Kinetic Theory'', the second 200 pages discuss  ``Gases, Liquids, and Solids'', and the final 200 pages concern themselves with ``Atoms, Molecules and the Structure of Matter.'' This material  would be challenging for a good American college student. It must have been even more so for a 17-18 year old student with a twice-interrupted high school career who was still learning English.\,\footnote{The undergraduate course Slater taught at MIT using {\it Introduction to Chemical Physics} proved to be too difficult for most of its intended audience. In later years, a separate ``modern physics'' course devoted to atoms, molecules, and the structure of matter  became a prerequisite for a senior-level ``thermal physics'' course which retained Slater's idea to combine thermodynamics and statistical mechanics in a single presentation. A textbook  written specifically for the latter by two of Slater's  MIT colleagues expanded his treatment of thermodynamics and contracted his  treatment of statistical mechanics (Allis and Herlin 1952).}

The Canadian military dissolved Camp B in the spring and summer of 1941. Some Jewish internees  were  returned to England and released when the British government finally acknowledged that they posed no threat to the war effort.\,\footnote{The future Nobel laureate Max Perutz (1914-2002) spent only six months as an internee in Canada. His  memoir {\it Enemy Alien} recounts how he was arrested by the British just four months after earning his PhD  under the direction of Sir Lawrence Bragg. He was deported to the Cove Fields internment camp in Quebec City (Camp L) where he organized a camp school with a faculty that included the physicist Klaus Fuchs and the future astrophysicists Hermann Bondi and Thomas Gold. Fuchs was recruited to the atomic bomb project after his release and gained notoriety in 1950 when it was discovered that he had betrayed the secrets of the Los Alamos laboratory to the Soviet Union (Perutz 1985).} Kohn was not so lucky. At the end of July,  he and Josef Eisinger were transferred to Camp A in the southern Quebec town of Farnham, 60 km southeast of Montreal (Eisinger 2013). Kohn was assigned to the knitting shop where he spent hundreds of hours making woolen socks and camouflage nets. On the other hand, Farnham  boasted a camp school every bit as good as the one at Fredericton and the many  hours of quiet repetitive work allowed him to think deeply about his schoolwork (Kohn 2013a, Auger 2005).

The Camp A school was organized by William Heckscher, a non-Jewish native of Hamburg who was working as an art historian in England when he was interned (Sears 1990). Kohn singles out Heckscher for special praise in his autobiography (Kohn 1998) and in the first history of the Canadian internment camps, Eric Koch (1980) reports that
\begin{quote}
William Heckscher was a remarkable figure. He was the ideal headmaster. The adjective with which several `old boys' of the Farnham camp school described him was `elegant'. He had grace, style, and patience.
\end{quote}
Heckscher told Koch that the Farnham camp commandant, Major Eric D.B. Kippen, once said to him, ``You know, Heckscher, I wish I could send my two sons to your school." At the end of September 1941, Heckscher escorted Kohn and a group of other internees  when they traveled  to Camp S on St. Helen's Island in Montreal to sit for McGill University's  Senior Matriculation Examination. The records show that Kohn did well in all the subjects tested: algebra, geometry, trigonometry, physics, chemistry, and coordinate geometry (Giannakis 2013).

Salvation for Kohn and Eisinger came in October 1941 when they received a letter from  the wife of a faculty member at the University of Toronto. She had heard about them from a former internee and offered to sponsor them to come live with her family after their release (Eisinger 2011).\,\footnote{The Canadian government announced in May 1941 that any internee under the age of 21 cleared by Scotland Yard would be released and given the opportunity to continue his education in Canada if he could find a sponsor willing to pay a fee of two thousand dollars (Jones 1988).}  Letters were exchanged with Scotland Yard and, by the end of January 1942, the boys found themselves sharing a comfortable attic space in the home of Hertha and Bruno Mendel.\,\footnote{Eisinger was released in early January. Kohn was released a few weeks later after a short stay at   Camp N, an internment camp outside the town of Sherbrooke, Quebec, about 130 km east of Montreal (Eisinger 2011).}  The  Mendels were refugees from the Nazis themselves who had earlier helped  young Jewish couples escape Germany and start new lives in Canada (Feldberg 1960).\,\footnote{Bruno Mendel (1897-1959) was the son of a research-active medical doctor who trained in Berlin and became a research-active physician himself. His medical practice slowly became less important as he increased the time he spent researching the metabolism of the cancer cell in his small private laboratory. Mendel read the political situation correctly and he took his wife and three children to Holland when Hitler came to power in 1933. In 1937, he emigrated to Canada and became (at first) an unpaid faculty member at the Banting Institute for Cancer Research of the University of Toronto. He returned to Europe in 1950 to accept a chair in Pharmacology at the University of Amsterdam (Feldberg 1960).} Their new mission was to arrange educational opportunities for recently released Jewish internees. Within days,  Eisinger was enrolled in a Toronto high school. Kohn, who was one year older, needed help from two complete strangers to get started at the   University of Toronto.\,\footnote{Eisinger followed Kohn to the University of Toronto the following year (Eisinger 2013).} He first received some valuable advice (Kohn 2013a),
\begin{quote}
Dr. Mendel worked at the university and his good friend, Leopold Infeld, came to their home very soon after I arrived. Infeld questioned me about my plans and I told him I wanted to be an engineer (another practical profession). He asked ``\,Is that your main interest?\,'' and I said no, it was mathematics and physics. He told me that engineering at the university was good but that the Math-Physics program was superb and that I should pursue a degree there. With the training I received, I could always do engineering.
\end{quote}
At that time, the University of Toronto had a Mathematics department, a Physics department, and a small (five-person) department of Applied Mathematics   (Robinson 1979). Leopold Infeld was a member of the Applied Mathematics faculty,  having come to Toronto  in 1938 after working  with Albert Einstein for two years at the Institute for Advanced Study in Princeton.\,\footnote{Leopold Infeld (1898-1968) earned the first PhD in theoretical physics awarded by a Polish university from the Jagellonian University in his native city of Kr\'{a}kow. He taught physics at Jewish high schools for nearly a decade before finding a senior assistantship in theoretical physics at  Lwow University. Infeld knew the importance of contacts with foreign physicists and successfully gained two-year visiting positions with Max Born in Cambridge and Albert Einstein in Princeton. With the latter, he co-authored {\it The Evolution of Physics}, a popular account of the history of ideas in physics. In 1938, Infeld accepted a position at the University of Toronto where he worked on a variety of theoretical problems in general relativity and cosmology. He returned to Poland in 1950 to found an Institute of Theoretical Physics at the University of Warsaw (Infeld 1980).} The Math-Physics program recommended to Kohn by Infeld was an honors curriculum  where all physics, mathematics, and applied mathematics students took the same courses for the first two years and then specialized in their final two years (Allin 1981).

Walther  attempted to  enroll in Toronto's Math-Physics program, but was rebuffed by the University registrar  because he lacked some of the mandatory prerequisites (Kohn 2013a). Mendel and Infeld arranged a meeting for Kohn (and five other camp boys with a similar problem) with the Dean of the College of Arts \& Science, Samuel Beatty, who was also the chair of the Mathematics department.\,\footnote{Samuel Beatty (1881-1970) earned the first PhD in mathematics awarded by a Canadian university at the University of Toronto. He remained at Toronto and from 1911-1959 rose from Lecturer in Mathematics to Professor and chair of Mathematics, to Dean of the College of Arts \& Science, and finally to Chancellor of the entire University. He  published 30 research papers, mostly in the field of algebraic functions, and devoted the bulk of his energy to teaching and to building the Mathematics department  (Robinson 1979).} Beatty was sympathetic, but he was unable to move the inflexible registrar. Therefore, he proposed to admit Walther and the others as `special students' who did not need the prerequisites. This artful maneuver required only the assent of the Department chairs whose departments were involved in the Math-Physics curriculum. This time, it was the chair of the Chemistry department, Frank Kenrick, who threw up a roadblock by refusing  to allow a foreign national from any Triple Axis country to enter his chemistry building where war research was being conducted.\,\footnote{Frank B. Kenrick (1874-1951) trained as a physical chemist with Wilhelm Ostwald in Leipzig. He served as chair of the Chemistry Department at the University of Toronto from 1937-1944. Like his mentor and predecessor as department chair, William Lash Miller (1866-1949), Kenrick favored an approach to chemistry that denied the reality of individual atoms and molecules. The war-related work carried out in Kenrick's department included experiments to develop the new explosive RDX as a replacement for TNT and the development of detectors for poison gas (Brook and McBryde 2007, Avery 1998).} Beatty arranged for Kohn to plead his case in person, first with Kenrick (who refused to acknowledge that Kohn was a refugee) and then with the University Chancellor,  Rev. Henry John Cody (who was unwilling to overrule one of his department chairs).

\begin{center}
\includegraphics[scale=.75]{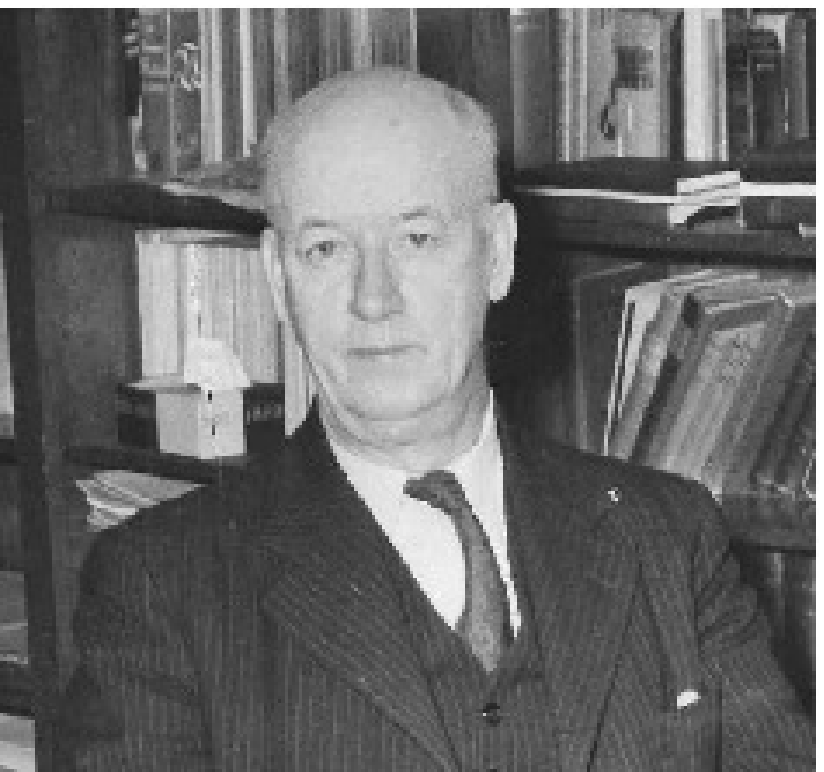}

\small Dean Samuel Beatty (ca. 1953) bent the admission rules \\
so Kohn could enter the University of Toronto.  \\ Courtesy of the University of Toronto Archives.
\end{center}

In the end, the creative Beatty simple redefined the meaning of `special student' to constitute the normal Math-Physics curriculum minus the usual chemistry requirement.  In this way, Kohn and his compatriots entered the Math-Physics program several weeks after the beginning of the spring 1942 term. Relatively little time was lost  because Beatty had permitted the group to audit his gateway mathematics course the entire time the admission negotiations were going on.  Even after the formalities were completed, Beatty tutored the group privately for a month to bring them up to speed with the rest of the class.

Kohn's undergraduate transcript shows that he completed the two-year Math-Physics ``common core'' in three semesters.\,\footnote{Dean Beatty apparently ``regularized'' Kohn's `special student' status at some point because his official transcript inaccurately states that he completed an introductory chemistry course and a chemistry laboratory course (Siochi 2013).} Besides algebra, analytic geometry, differential and integral calculus, differential equations, physics laboratory, mechanics, properties of matter, dynamics, electricity \& magnetism, and light \& acoustics, one finds two required courses in actuarial science, two required courses in French and German replaced by English courses, and two elected courses in `oriental literature' where texts in ancient Egyptian, Arabic, Hebrew, Persian, and Turkish  were read in translation. Kohn took twelve more advanced courses during the 1943-1944 academic year, including  algebraic geometry, differential geometry, partial differential equations, the theory of functions, group theory, thermodynamics, classical dynamics, quantum mechanics, variational principles in physics, English literature, and modern ethics. These two semesters turned out to be his last as an undergraduate because he was inducted into the Canadian army in September 1944.\,\footnote{Kohn and Josef Eisinger had volunteered for (and been rebuffed by) the Canadian Air Force immediately after their release from internment (Eisinger 2013).} He served until the war ended in August 1945 and was awarded his BA in Applied Mathematics at the summer 1945 convocation ceremony while still on active duty.\,\footnote{Kohn worked the summers of 1942-1943 for the Sutton-Horsley Company, a Toronto x-ray equipment manufacturer which began producing signalling lamps and instrument panels for fighter and bomber aircraft after the war began. His specific task  was the design and testing of compensation circuits to ensure that  cockpit instruments gave accurate readings when operated at unusually high and low temperatures (Kohn 1998, Horsley 2013, Kohn 2013b). Kohn worked the summers of 1945-1946   for the mineral surveying and exploration geophysics company Koulomzine, Geoffroy, Brossard \& Company of Val D'Or, Quebec. His job was to conduct magnetic field surveys in suspected gold-bearing regions of northern Ontario. A typical survey consisted of a grid of about 1000 magnetometer measurements with a grid-spacing of 100 meters (Geoffroy 1946, Kohn 1998, Kohn 2013b).}

The bare list of courses Walter took during his five undergraduate semesters  does not communicate the elite quality of the instructors who taught and mentored him.\,\footnote{See footnote 2.}  Leopold Infeld, who lectured to upperclassmen only, had been invited to join the Applied Mathematics  Department by its chair, the eminent Irish mathematician and theoretical physicist, J.L. Synge.\,\footnote{John Lighton Synge (1897-1995) is often regarded as the greatest mathematician of Irish descent since Sir William Rowan Hamilton. Synge studied mathematics at Trinity College Dublin and accepted a position as  Assistant Professor  at the University of Toronto in 1920. There he began  a lifelong interest in Einstein's theory of relativity and in geometrical methods to analyze dynamical systems. The peripatetic Synge subsequently held positions in Dublin, Toronto (again at the time Kohn was there), Ohio State, Carnegie Tech, and finally the Dublin Institute for Advanced Studies. He published 11 books and over  200 hundred papers (Florides 2008).}   Synge and his Applied Mathematics colleague Bernard Griffith were the authors of {\it Principles of Mechanics}, the introductory but quite sophisticated textbook used by all the students in the Math-Physics program. Synge was always  eager to add talent to his faculty and in 1941 he succeeded to recruit the Russian mathematical physicist Alexander Weinstein, a mature scientist with a strong reputation for his work on free boundary problems and variational principles.\,\footnote{Alexander Weinstein (1897-1979) was a PhD student of Hermann Weyl, who  considered him to be his most talented student. He worked  with Tullio Levi-Civit\`{a} in Rome and Jacques Hadamard in Paris before the German occupation of France in 1940 drove him from Europe permanently. Weinstein was a member of the Applied Mathematics faculty of the University of Toronto from 1941-1946, worked for some time at the US Naval Ordnance Laboratory, and spent 18 productive years  at the Institute for Fluid Dynamics and Applied Mathematics at the University of Maryland (Diaz 1978).}

It is significant to our story that variational methods were  something of a Toronto speciality at the time. Besides Weinstein, Synge, and Griffith, one should include Gilbert Robinson (Mathematics) and Arthur  Stevenson (Applied Mathematics) because Cornelius Lanczos thanks them in the preface to his now-classic 1949 text {\it The Variational Principles of Mechanics} because they together ``revised the entire manuscript''. In later years,  Kohn singled out the algebraist Richard Brauer and the non-Euclidean geometer H.S.M. (Donald) Coxeter as ``luminous faculty members whom I recall with special vividness'' (Kohn 1998). He also recalled the first-year electricity and magnetism lectures given by Lachlan Gilchrist, a 1913  PhD student of Robert Millikan, because Gilchrist told his Toronto students that it was he who had purchased the oil used by Millikan in his famous oil-drop experiment (Kohn 2003).

Weinstein's influence on the undergraduate Kohn is apparent from Kohn's first two published scientific papers. The first, submitted in July 1944, is  a two-page report on an exact solution for the oscillations of a spherical gyroscope which generalizes a method presented in the Synge and Griffith book but thanks Weinstein for ``his advice and criticism" (Kohn 1945). The second paper was completed and submitted in November 1944 at a time when Kohn was engaged in advanced basic training at Camp Borden, Ontario. This substantial piece of work (Kohn 1946) establishes bounds on the motion of a  heavy spherical top using a contour integration method used by Weinstein (1942) to study a spherical pendulum.
The text makes clear that Kohn had at least some familiarity with {\it \"{U}ber die Theorie des Kreisels} (1898), the great treatise on tops by Felix Klein and Arnold Sommerfeld.

In the 1945-1946 academic year (Kohn 1998),
\begin{quote}
 after my discharge from the army, I took an excellent crash masters program, including some senior courses I had missed, graduate courses, and a master's thesis consisting of my paper on tops  and a paper on scaling atomic wave functions.
\end{quote}
The atomic wave functions paper, ``Two Applications of the Variational Method to Quantum Mechanics'' (Kohn 1947) was the first of many papers to come (including the density functional papers) where Kohn exploits a variational principle. He  first learned about such principles from an advanced undergraduate course where Weinstein discussed  the Lagrangian and Hamiltonian formulations of classical mechanics. Weinstein regarded Kohn as a potential PhD student and thus shared with him his recent work on variational methods to study the vibrations of clamped plates and membranes (Aronszajn \& Weinstein 1941). A review paper by Weinstein (1941) summarized  the original contributions to this subject by Lord Rayleigh and Walter Ritz.

Notwithstanding the foregoing, it was Applied Mathematics Professor Arthur Stevenson  who broadened Kohn's perspective to include quantum problems and the use of variational methods to study them. He is thanked in Kohn (1947)  ``for his kind advice and interest".\,\footnote{Arthur Francis Chesterfield Stevenson (1899-1968) accepted a position in the Mathematics department of the University of Toronto immediately after earning his BA from Trinity College, Cambridge in 1922. He returned to Cambridge in 1928 where he  worked under the supervision of Ralph Fowler on a problem in theoretical spectroscopy which  eventually led to his PhD. He returned to Toronto where he published original research and lectured on atomic physics, quantum mechanics, electromagnetic theory, scattering theory, and the differential equations of mathematical physics. He spent the last dozen years of his academic career on the faculty of Wayne State University in Detroit, Michigan (Duff, 1969).} Stevenson's early research concerned quantum mechanical  methods to calculate the energy levels of electrons in atoms and he had performed variational calculations  for the helium atom in collaboration with a colleague in the Toronto physics department (Stevenson and Crawford 1938). Kohn surely read this paper because the helium  atom figured into his work also.

Kohn  assumed that the readers of his paper were familiar with the variational method to find approximate solutions to the  Schr\"{o}dinger equation. Indeed, most textbooks of quantum mechanics written between 1930 and 1945 devoted more than passing attention to this topic   because Egil Hylleraas (1929) had used it  with spectacular success to calculate  the ionization energy of the helium atom.
This provided the first convincing evidence that quantum mechanics could achieve quantitative success for a system of more than one electron.

For future reference, I sketch here a simple form of the Rayleigh-Ritz variational method appropriate to an $N$-electron system with ground state energy $E_0$ and Hamiltonian operator  $H$. If ${\bf r}=(x,y,z)$, the starting point is a trial wave function, $\psi({\bf r}_1, {\bf r}_2,\dots, {\bf r}_N)$,  which depends on the Cartesian coordinates of all the  electrons. One then computes a $3N$ dimensional integral  with the  configuration space volume element $d\tau = d{\bf r}_1 \cdots d{\bf r}_N$ and exploits the inequality,\,\footnote{It is necessary here that trial function satisfies $\int d\tau \psi^* \psi =1$ and that the integral $\int d\tau \psi^* H \psi$ converges.}
\begin{equation}
\label{one}
0\le E[\psi]=\int d\tau \psi^* (H-E_0) \psi = \int d\tau \psi^* H \psi - E_0.
\end{equation}
The exact ground state wave function $\psi_0$  satisfies the Schr\"{o}dinger equation,  $H\psi_0 = E_0 \psi_0$. Using the latter in Eq.~(\ref{one}) shows  that $E[\psi]=0$  when $\psi=\psi_0$ and suggests a strategy to find an upper bound to $E_0$: write the trial function  $\psi$ as a linear combination of a set of basis functions and minimize the integral on the right side of Eq.~(\ref{one}) with respect to the expansion coefficients. Increasing the number of  basis functions generally lowers the bound obtained for $E_0$. An important feature of this procedure emerges if we consider a trial function of the form $\psi = \psi_0 + \delta \psi$ where $\delta \psi$ is ``small'' by some measure. Inserting this trial function into Eq.~(\ref{one}) gives the variation $\delta E=E[\psi_0+\delta \psi]$ as

\begin{widetext}
\begin{equation}
\label{two}
\delta E = \int d \tau (\psi_0 + \delta \psi)^* (H-E_0) (\psi_0 + \delta \psi) = \int d \tau \delta \psi^* (H-E_0) \delta \psi =  O(\delta \psi^2).
\end{equation}
\end{widetext}

One says that the energy functional $E[\psi]$ is ``stationary'' in the sense that a trial function that differs from $\psi_0$ by a small amount (first order in $\delta \psi$) produces an energy which differs from $E_0$ by an amount that is {\it very} small (second order in $\delta \psi$). For that reason,  minimizing $E[\psi]$ with respect to a set of variational parameters produces a much better estimate for the ground-state energy than one might have supposed. The elegance and generality of this technique must have  appealed powerfully to the young Kohn, because  he ``read many of the  old papers on the subject and variational methods became the first tool in my theoretical physics toolbox'' (Kohn 2013b).

In the spring of 1946, Kohn completed his MS studies, taught calculus and analytic geometry as an Instructor for the Mathematics Department, and applied to a dozen or so PhD programs with the clear idea to study theoretical physics. Offers of admission with financial support came from  Rudolf Peierls at Birmingham and Eugene Wigner at Princeton, among others. Kohn accepted the Birmingham offer on the advice of Infeld, who knew Peierls personally (Kohn 2013b). One day later, an offer  arrived from Harvard which included a prestigious Arthur Lehman Fellowship. Kohn again consulted Infeld, who without hesitation told him to communicate his regrets to Peierls, accept the offer from Harvard,  and try to work for the young physics superstar Julian Schwinger.\,\footnote{Julian Seymour Schwinger (1918-1994) was one of the greatest theoretical physicists of the $20^{\rm th}$ century. By the age of 21, he had earned his PhD under the (nominal) supervision of Isador Rabi and published ten research papers in quantum mechanics and nuclear physics. He spent the war years working on waveguides for radar applications before turning his attention to quantum electrodynamics. This work earned him a one-third share of the 1965 Nobel Prize in physics. He supervised 73 PhD students over a long academic career  at Harvard (1945-1972) and UCLA (1972-1994) (Martin \& Glashow 1995, Mehra \& Milton 2000).}

Accordingly, a somewhat insecure twenty-three year old Walter Kohn arrived on the Harvard campus in the fall of 1946 as one of a group of about thirty first-year graduate students. The twenty-eight year old Schwinger  had joined the Harvard faculty the  previous spring and immediately began teaching a three-semester sequence of courses on special topics in theoretical physics (Schweber 1994). Kohn and his cohort stepped into the middle of  this sequence as a supplement to their required courses in classical mechanics, electrodynamics, quantum mechanics, and statistical mechanics. In principle, Schwinger's course was devoted to nuclear physics. In practice, he devoted part of the time to a  highly personal exposition of quantum mechanics in the style of Dirac (1935) and the rest of the time was given over to (Anderson 1999)
\begin{quote}
  essentially everything Schwinger knew about. All about Green functions, all about nuclear physics and so on. All the numerical tricks he had devised to solve quantum  mechanical problems and nuclear physics problems. . . there was a lot of physics and there were a lot of variational techniques, for example to solve the deuteron. . . . He was also starting to build the machinery that was going to solve the problems of quantum electrodynamics. We were treated to a lot of that machinery.
 \end{quote}
 Kohn himself has given one of the best descriptions of Schwinger's teaching style (Kohn 1996):
 \begin{quote}
 Attending one of his formal lectures was comparable to hearing a new major concert by a very great composer, flawlessly performed by the composer himself. . . Old and new material were treated from fresh points of view and organized in magnificent overall structures. The delivery was magisterial, even, carefully worded, irresistible like a mighty river. He commanded the attention of his audience entirely from the content and form of his material, and by his personal mastery of it, without a touch of dramatization.
 \end{quote}

Quite early on, the chairman of the Harvard Physics department, John Van Vleck, approached Kohn and asked him whether he would like to work with him on a solid state physics problem.\,\footnote{John Hasbrouk Van Vleck (1899-1980) was an eminent theoretical physicist who contributed widely to the fields of chemical physics, quantum electronics, solid state physics, and magnetism. He published 169 scientific articles and two books and served on the faculty at Harvard for  forty-six years. In 1977, he was awarded a one-third share of the Nobel Prize in  physics for his work in magnetism (Anderson 1987).} Kohn was not interested in solid state physics and instead presented himself to Schwinger as Infeld had urged him to do.\,\footnote{Kohn had learned from his fellow graduate students about Wolfgang Pauli's famously negative view that solid-state physics was insufficiently fundamental and too approximate to attract the attention of a serious young theoretical physicist (von Meyenn 1989).} He described to Schwinger his experiences in Toronto and (Kohn 1998, 2001b)
\begin{quote}
Luckily for me, we shared a common interest in the variational methods of theoretical physics. . . .  He accepted me within minutes as one of his 10 PhD students. He suggested  I should try to develop a Green function variational principle for three-body scattering problems, like low-energy neutron-deuteron scattering, while warning me ominously that  he himself had tried and failed.
\end{quote}
Schwinger was an acknowledged expert in the use of both variational principles and Green functions to solve a wide variety of problems. In Eq.~(\ref{one}), an energy functional $E[\psi]$ with the stationary property $\delta E=0$ made it possible to estimate the ground state energy of a bound-electron system like helium.  For a scattering problem, the energy is known and one is led to seek stationary variational functionals for other quantities. An example is the Green function, an energy-dependent operator defined in terms of the system Hamiltonian by
\begin{equation}
\label{three}
G(E)={1\over E-H}.
\end{equation}
For two-body scattering, one writes $H=H_f + V({\bf r})$ where $H_f \varphi = E_f \varphi$ is the Schr\"{o}dinger equation for a free particle and $V({\bf r})$ is the potential responsible for the scattering. Using a coordinate-space representation of the corresponding free-particle Green function $G_f$, the scattered wave function $\psi$ satisfies the integral equation,\,\footnote{It is necessary to replace  $E_f$ by the complex number $ E_f + i\epsilon$ in  Eq.~(\ref{three}) and let $\epsilon \to 0$ at the end to ensure that $\psi$ behaves like an outgoing spherical wave (Baym 1969).}
\begin{equation}
\label{four}
\psi({\bf r})= \varphi({\bf r}) + \int d^{\,3}r' G_f({\bf r},{\bf r'}|E_f) V({\bf r'}) \psi({\bf r'}).
\end{equation}

Kohn worked on the three-body Green function problem for half a year before abandoning it.\,\footnote{This difficult and subtle problem was solved in the early 1960's by the Russian mathematical physicist Ludvig Faddeev (Faddeev 1965). } Instead, he generalized Eq.~(\ref{one}) and developed a variational principle for the two-body scattering phase shift, a quantity  which characterizes the final state when two particles interact via a short-range potential.\,\footnote{At some point before he wrote up his thesis, Kohn learned that the Swedish physicist Lamek Hulth\'{e}n had independently derived a variational principle for the scattering phase shift very similar to his own. Hulth\'{e}n (1946) and Kohn (1948) begin with the same variational functional but propose slightly different variational procedures. The Kohn-Hulth\'{e}n variational principle later  found wide application in atomic, molecular, and nuclear scattering problems (Adhikari 1998, Nesbet 2003).} He also derived  a variational principle for the scattering amplitude for two-particle scattering with an arbitrary interaction potential. For both cases, Kohn borrowed from the Rayleigh-Ritz method and expanded the trial scattering wave function in a set of basis functions with the correct long-distance behavior. Finally, he derived  a variational principle for the elements of the scattering matrix for the special case of nuclear collisions where multiple disintegrations are energetically forbidden. It is interesting that Kohn made no use of Schwinger's ``beloved Green functions'' (Kohn 1998) in his thesis, ``Collisions of light nuclei'', or in the published version of his thesis (Kohn 1948).  He did, however, use his scattering amplitude variational principle to {\it rederive} an alternative  variational principle for the phase shift that  Schwinger had derived in his spring 1947 theoretical physics class using Eq.~(\ref{three}) and reported at a meeting of the American Physical Society (Schwinger 1947). The latter is commonly called ``Schwinger's variational principle for scattering" (Adhikari 1998, Nesbet 2003).

 \begin{center}
\includegraphics[scale=1.]{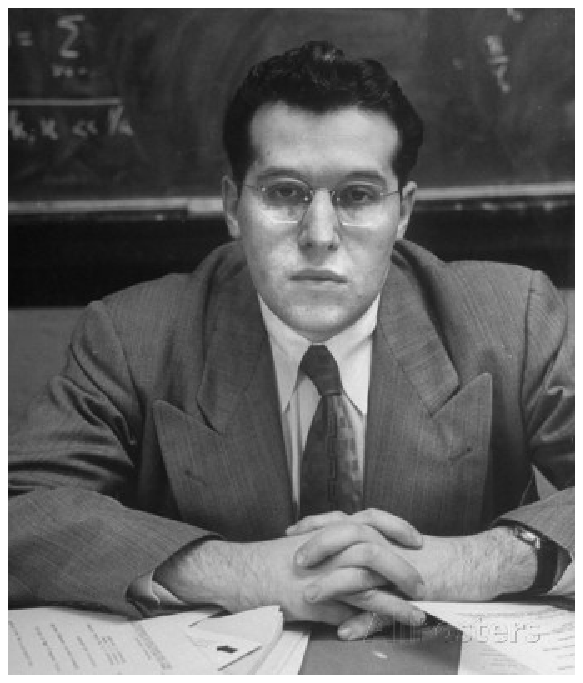}

\small Julian Schwinger (ca. late 1940's) was Kohn's PhD \\
supervisor at Harvard University. \\ Source unknown.
\end{center}

Kohn did not have a close personal relationship with his advisor.\,\footnote{This did not prevent Schwinger from later characterizing Kohn as ``my most illustrious student'' to one of Kohn's former graduate students (Rudnick 2003).} None of Schwinger's students did, in part because it was notoriously difficult to schedule a personal meeting with him. Kohn met with him only ``a few times a year'' (Kohn 1996) and according to John David Jackson, an MIT graduate student who listened to Schwinger's lectures at Harvard, there was one  occasion where ``Kohn was miffed by Julian's unavailability. He completed his thesis, wrote up a paper, and submitted it to {\it Physical Review} without ever consulting him'' (Mehra and
Milton 2000).  Such feelings must have passed quickly because, in a moving tribute at a memorial symposium after Schwinger's death, Kohn makes it clear that (Kohn 1996)
\begin{quote}
It was during these meetings, sometimes more than 2 hours long, that I learned the most from him. . . . to dig for the essential; to pay attention to the experimental facts; to try to say something precise and operationally meaningful, even if one cannot calculate everything {\it a priori}; not to be satisfied until one has embedded his ideas in a coherent, logical, and aesthetically satisfying structure. . . . I cannot even imagine my subsequent scientific  life without Julian's example and teaching.
\end{quote}

Besides Schwinger, Kohn benefitted from members of his graduate student cohort who either contributed materially to his education at Harvard or who played an important role subsequently. One group consisted of fellow Schwinger students:  Kenneth Case, Frederic de Hoffmann,  Roy Glauber, Julian Eisenstein, Ben Mottelson, and Fritz Rohrlich. Another group did their PhD work in other areas of physics: the theorists Thomas Kuhn, Rolf Landauer, and Philip Anderson, and the experimentalists Nicolaas Bloembergen, George Pake, and Charles Slichter. Joaquin Luttinger, another MIT graduate student who made the short trip to Harvard to audit Schwinger's classes, later became a close friend and a scientific collaborator.  All these members of Kohn's student network went on to have  successful scientific
 careers.\,\footnote{Josef Eisinger remained a close friend. He did his graduate work at MIT, just two miles down the Charles river from Harvard, and earned his PhD in physics in 1951 for an experimental determination of the magnetic moment of ${\rm K}^{40}$. He spent thirty years at Bell  Laboratories where he made a transition from solid state physics to biophysics.  From 1985 until his retirement in 1998, he taught and conducted research at the  Mount Sinai School of Medicine in New York City (Eisinger 2013).} Glauber, Mottelson, Anderson, and Bloembergen won Nobel Prizes themselves.

Walter's life changed profoundly in two important ways when he accepted an offer by Schwinger to remain at Harvard as a  post-doctoral fellow. First, the income from this job permitted him to bring to Boston and marry Lois Mary Adams, a former nursing student he had met at the University of Toronto who had been working in New York City while he finished his PhD (SDUT 2010). A baby daughter soon arrived and family responsibilities were added to the research and teaching responsibilities that came with his position as Schwinger's assistant. The research project he undertook  was an investigation of the electromagnetic properties of mesons done in collaboration with fellow Schwinger post-doc Sidney Borowitz. His teaching consisted of an introductory physics course in the summers of 1949-1950 and a junior/senior level classical mechanics course in the summer of 1950.\,\footnote{An evaluation of Kohn written by Ms. Norine T. Casey provides insight into Kohn's teaching style and philosophy of physics. Ms. Casey was a 1949 Wellesley graduate pursuing an MA in teaching at Harvard. As part of her curriculum, she attended Kohn's summer 1950 class (devoted to introductory optics, electricity, magnetism, atomic physics, and nuclear physics) and wrote a four-page evaluation of her experience. Kohn received a copy of her report (Casey 1950), which states that
``Dr. Kohn's lectures were clear and concise. Demonstrations accompanied every lecture and were  given with great enthusiasm. . . . It was obvious from the beginning that [Dr. Kohn's] interest was not his own mastery of the mathematics, but in our understanding of the physics. . . . It was not infrequent that he read from source material giving direct quotes [such] as Newton's relating his discovery of the diffraction of light.''}

The second profound change in Kohn's life occurred through the good offices of John Van Vleck, the solid state theorist  he had rebuffed as a thesis supervisor. Van Vleck re-enters the story because Kohn supplemented his summer 1949 income by working for the Polaroid Corporation at their Cambridge, Massachusetts research laboratory. His job was to discover the mechanism whereby high-energy charged particles produce an image when they impinge on photographic plates.\,\footnote{This technique had recently been introduced to study cosmic rays using plates produced by another company and Polaroid wanted to enter the business (Powell and Occhialini 1947).} This task required a knowledge of solid state physics, which he acquired by  reading Frederick Seitz' {\it The  Modern Theory of Solids} (1940) and consulting with Van Vleck when necessary.\,\footnote{Frederick Seitz (1911-2008) was one of the creators and intellectual leaders of the American solid state physics community. He was Eugene Wigner's  first PhD student at Princeton and built influential research groups at four different universities between 1935 and 1965. Later, he served as president of Rockefeller University and president of the National Academy of Sciences. Seitz worked on a wide range of materials problems during World War II and was the author or co-author of more than 100 scientific papers (Slichter 2010).}

\begin{center}
\includegraphics[scale=.8]{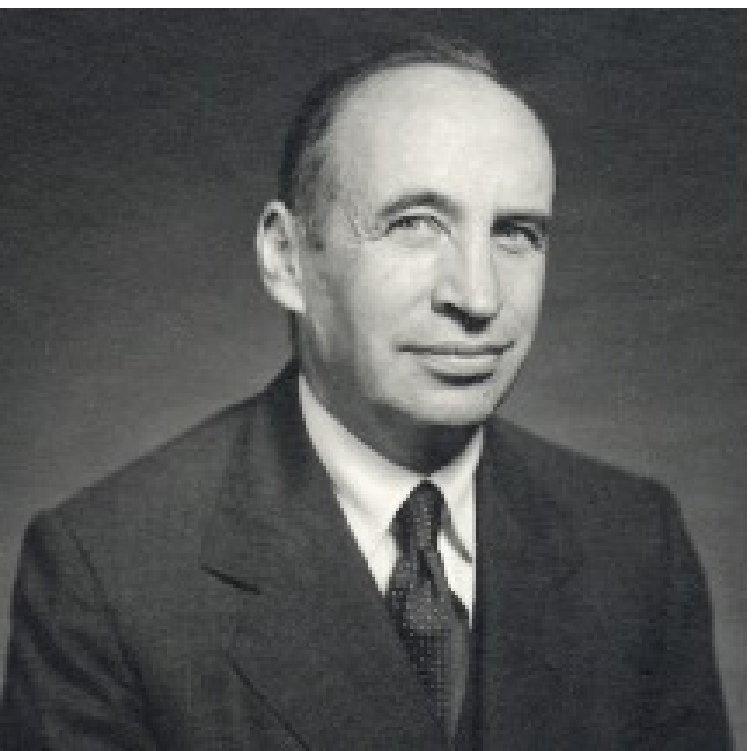}

\small Harvard's John Van Vleck facilitated Kohn's \\
transition from nuclear physics to solid state physics. \\ Courtesy of the UW-Madison Archives.
\end{center}

By this time, Kohn's first paper with Borowitz had appeared (Borowitz and Kohn 1949) and he had applied to the National Research Council for a Fellowship to spend the 1950-1951 academic year with Wolfgang Pauli at the Eidgen\"{o}ssische Technische Hochschule in Z\"{u}rich. Nevertheless, he was painfully aware that he had so far made only a ``very minor contribution'' to field-theory research. This fact, and the stunning quality of the latest achievements by Schwinger and Richard Feynman, made it easy for him  when (Kohn 1998)
\begin{quote}
Van Vleck explained to me that he was about to take a leave of absence and `since you are familiar with solid state physics', he asked me if I could teach a graduate course on this subject he had planned to offer  [for the fall 1949 semester]. This time, frustrated with my work on quantum field theory, I agreed.
\end{quote}
Kohn not only taught the solid-state physics course, he collaborated with Harvard graduate student Richard Allan Silverman to   find approximate numerical solutions of the Schr\"{o}dinger equation for the purpose of calculating the cohesive energy of metallic lithium (Silverman and Kohn 1950).\,\footnote{Silverman's PhD thesis  states that ``The author wishes to express his indebtedness to Professor Walter Kohn for suggesting the problem and for invaluable guidance during a large portion of the work (Silverman 1951). A generation of English-speaking solid-state physicists have Silverman to thank for his translation into English of {\it Methods of Quantum Field Theory in Statistical Physics} by A.A. Abrikosov, L.P. Gorkov, and I.E. Dzyaloshinski (1963).} Using the same numerical data, he and  Bloembergen estimated the Knight shift for lithium, the latter being a measure of the electron wave function amplitude at the atomic nucleus  accessible to experimenters  using magnetic resonance techniques (Kohn and Bloembergen 1950). It is notable that Silverman and Kohn conclude with the remark, ``One of us (W. Kohn) is investigating the cohesive energy by means of a variation iteration procedure based on the integral equation [Eq.~(\ref{four}) of this paper] and a Green function appropriate to a periodic lattice."

 Kohn, a  naturalized Canadian citizen since 1943, looked everywhere in Canada and the United States for an entry-level  academic position. Nothing turned up, but an early 1950 interview trip to the Westinghouse Research Laboratory in Pittsburgh, Pennsylvania  bore fruit even though his foreign citizenship precluded a job offer from Westinghouse. In Pittsburgh, Kohn stayed at the home of Alfred Schild, a friend from Toronto who had found a job teaching mathematics at the Carnegie Institute of Technology. Schild told him that the chairman of the  physics department, Frederick Seitz, had just resigned and was moving his solid state group to the  University of Illinois (Seitz 1994). Perhaps there was an opportunity at Carnegie Tech itself. The new chair, nuclear physicist Edward Creutz, interviewed Kohn and offered him a job as an Assistant Professor 48 hours later.\,\footnote{Edward Chester Creutz (1913-2009) earned his PhD in experimental nuclear physics  from the University of Wisconsin in 1938. He moved to Princeton as an Instructor and used  their cyclotron for nuclear physics projects until the Manhattan Project redirected his efforts to the synthesis of plutonium and the triggering of the plutonium bomb. After the war, Creutz  joined the faculty at Carnegie Tech to direct the construction of a proton synchrocyclotron and to build an experimental nuclear physics group. After nine years (seven as chair), Creutz moved to San Diego, California to help found the General Atomics division of General Dynamics Corporation. He spend 15 years there before concluding his career as Associate Director of the National Science Foundation (Hinman and Rose 2010).}  It turned out that Creutz needed someone who could teach solid state physics and mentor a few graduate students who had lost their advisors when the Seitz group left (Kohn 1998). Kohn was delighted to accept.

\section{Portrait of the Physicist as a Young Man}
Walter Kohn was thinking about the start of his academic career at Carnegie Tech when, in March 1950, the National Research Council approved his application for a fellowship to spend a year in Europe. Ed Creutz agreed to a one-year leave of absence, but only if Walter agreed to teach solid state physics for the fall 1950 semester. At the same time, Kohn was having  second thoughts about his original plan to work with Wolfgang Pauli. This led him to seek and secure the  approvals needed to switch the venue for his fellowship year from Z\"{u}rich to Niels Bohr's Institute for Theoretical Physics in Copenhagen. Accordingly, Kohn moved to Pittsburgh, taught his course, and left for Copenhagen at the end of the term.\,\footnote{One former Carnegie Tech graduate student remembers Kohn's solid state physics lectures as well-prepared, clearly delivered, and mathematically precise. He was surprised when the final exam avoided  mathematical issues and  focused entirely on qualitative aspects of the subject  (Arrott 2013).}
When he arrived at the Institute in January 1951, Walter  Kohn was an expert in scattering theory who had begun to think of himself as a solid state physicist. Unfortunately, ``no one in Copenhagen, including Niels Bohr, had even heard the expression ``solid state physics'" (Kohn 1998).

Kohn managed to publish two papers that year, but more important to his future was the fact that the Institute attracted a steady stream of short-term and long-term visitors from around the world from whom he could learn new physics.\,\footnote{Aage Bohr and Christian M{\o}ller were already fixtures at the Institute. Kohn's Harvard classmate Ben Mottelson arrived in the fall of 1950 and never left. A partial list of visitors who overlapped with Kohn for at least some time includes Hendrik Casimir, Freeman Dyson, Ugo Fano, Nicolas Kemmer, Louis Michel,  Wladyslaw Swiatecki, Jean Valatin, Nicolaas van Kampen, and Arthur Wightman (NBA 1951a).}
It was good luck for Kohn that post-war freedom of movement motivated Bohr to organize a meeting for all  foreign physicists who had ever worked at the Institute (Rozental 1967). The resulting   Conference on Problems in Quantum Physics (July 6-10 1951) was attended by an outstanding collection of theoretical physicists, many of whom Walter was able to incorporate into his expanding  professional network.\,\footnote{A partial list of attendees includes  Hans Bethe,  Homi Bhabha, L\'{e}on Brillouin, Richard Dalitz, Paul Dirac, Maria Mayer, Dirk ter Haar, Werner Heisenberg, Walter Heitler, L\'{e}on van Hove, Lamek Hulth\'{e}n, Egil Hylleraas, Hendrik Kramers, Ralph Kronig, Jens Lindhard, Lise Meitner, Wolfgang Pauli, Rudolf Peierls, L\'{e}on Rosenfeld, John Slater, Ernst Stueckelberg, Victor Weisskopf,  Gregor Wentzel, John Wheeler, and Gian-Carlo Wick (NBA 1951b).} A few weeks later, Kohn was tapped to lecture on solid state physics for two weeks at the first Summer School of Theoretical Physics organized by C\'{e}cile DeWitt  at Les Houches, near Chamonix in the French Alps.\,\footnote{Kohn was an emergency replacement for Mario Verde, a  nuclear physicist who had fallen ill (DeWitt 1951, 2013).}

At the end of 1951, Bohr wrote a formal evaluation which concluded  that (Bohr 1951)
 \begin{quote}
 Dr. Kohn has proved himself a highly qualified theoretical physicist with great knowledge of a wide field of problems. His ability to stimulate others in their work and his willingness to assist them with his knowledge has been of great value to the many members of our group.
 \end{quote}
This good opinion led Bohr to arrange a Rask {\O}rsted Foundation fellowship for Kohn so he could remain in Copenhagen through the summer of 1952. This was welcome news because Walter and his family enjoyed living in Denmark. Moreover, he had just begun a scientific collaboration with Res Jost, a Swiss mathematical physicist five years his senior who had lectured on quantum field theory at the Les Houches summer school.\,\footnote{Res Jost (1918-1990) wrote his PhD thesis under the supervision of Gregor Wentzel and  spent three years as the principal assistant to Wolfgang Pauli. He was a senior fellow at the Institute for Advanced Study in Princeton for six years (1949-1955)  before accepting a professorship at the Eidgen\"{o}ssische Technische Hochschule in Z\"{u}rich. Jost's research focused on  mathematical physics and quantum field theory, particularly axiomatic versions of the latter. (Kohn {\it et al.} 1992, Pais 1996).}
 Jost was interested in scattering theory and his ``predilection for mathematical rigor'' (Enz 2002) struck a responsive chord in Kohn.  Together, the two theorists completed three papers  (including an  `inverse scattering problem' where one deduces characteristics of the scattering potential from phase shift information) before Kohn  returned to Carnegie Tech to begin the 1952-1953 academic year (Jost and Kohn, 1952a, 1952b, 1953). On his own, Kohn studied the validity of the Born expansion for scattering (Kohn 1952a) and a non-Green function variational principle for electron waves in a periodic potential (Kohn 1952b).

Back in Pittsburgh, the Physics Department  had changed somewhat during Walter's absence. The senior experimentalist Immanuel Estermann had left to head the physics section of the Office of Naval Research (ONR) and his last PhD student, Simeon Friedberg, had taken over his low-temperature physics laboratory. A senior theorist, Gian-Carlo Wick,  had joined the faculty from Berkeley and  Roman Smoluchowski, an expert in the theory of defects in solids, had transferred to the Physics Department from the Metallurgy Department. A young experimenter, Jacob Goldman, and a young theorist, Paul Marcus, had joined the solid state physics group to complement the  senior experimentalist Emerson Pugh.   One familiar face was  Norman Rostoker, a Toronto native who had graduated from his hometown university as a physics major one year behind Kohn   and then received his PhD at Carnegie Tech under Pugh's supervision (Rostoker 2013).  Kohn and Rostoker  had become friends during the fall 1950 semester and Norman  was still working in the Physics Department as a post-graduate research scientist when Kohn returned from Europe.

 The fall 1952 semester  found Kohn teaching thermodynamics to  undergraduates and nuclear physics to graduate students. He was also named as co-principal investigator with Jack Goldman on an ONR contract to conduct solid state research.\,\footnote{For fifteen years after World War II ended, most solid state physics research in the United States was funded by the Office of Naval Research. Most nuclear physics research was supported by the United States  Atomic Energy Commission (Old 1961, Sapolsky 1990).}  His main research project  was to develop a Green function method to calculate the energy band structure for crystalline solids. In other words, he wanted to use  Eq.~(\ref{three}) to solve the Schr\"{o}dinger equation to find the energy eigenfunctions and eigenvalues for electrons moving in a periodic potential. A distraction  arose in the spring 1953 semester when Carnegie Tech learned that Walter had received job offers from the  Department of Mathematics at McGill University in Montreal and the Physical Research Department at Bell Telephone Laboratories in Murray Hill, New Jersey. Evidently, others besides  Niels Bohr had formed a  very positive impression of this new Assistant Professor. In the end, Carnegie Tech retained his services  by promoting him to Associate Professor (WKP 1953a) after only three semesters of academic service.

Kohn had brought to Carnegie Tech the germ of his Green function method to solve the electron band-structure problem. He  recruited Norman Rostoker to help with the numerical calculations and that activity continued  (part-time) while Walter was in Copenhagen.\,\footnote{The actual computing was performed by a `computress' named Alice Watson who operated a Friden  Model STW-1 Electro-Mechanical Calculator. Although her equipment changed to an IBM 650 digital computer in 1956, she continued to do computing tasks for Kohn the entire time he worked at Carnegie Tech  (Rostoker 2003, Young 2013).} The work accelerated when Kohn returned to Pittsburgh and he reported their still-unpublished results at two invited talks,  one at the June 1953  Summer Meeting of the American Physical Society and one at a July 1953 Gordon Research Conference devoted to the Physics and Chemistry of Metals. The latter was a particularly prestigious venue and it is notable that of the seven theorists invited to speak, the three  youngest (by far) were Walter Kohn, Jacques Friedel, and David Pines (WKP 1953b). Friedel, an expert on the theory of metals and alloys,  and Pines, an expert on electron-electron interactions in solids,  had both  published half a dozen papers in their fields by the time of the Gordon Conference.  It is an indication of Kohn's rising reputation that he had published only one full-length paper in solid state physics by this time. Kohn's  Green function paper (Kohn and Rostoker 1954) finally appeared in the June 1 1954 issue of the journal {\it Physical Review}. Therein, he and Rostoker (Kohn 1998)
\begin{quote}
 developed a theory for the energy band structure of electrons in solids harking back to my earlier experience with scattering, Green functions, and variational methods. We showed how to determine the band structure from a knowledge of purely geometric structure constants and a small number ( $\sim 3$) of scattering phase shifts of the potential in a single sphericalized cell.
\end{quote}
It happens that the same basic idea had been published several years earlier by the Dutch physicist Jan Korringa.  However, Korringa (1947) included no numerical applications and his paper went largely unnoticed.\,\footnote{Korringa (1994) relates that ``computers were rare in the Netherlands in 1946 and a cost estimate [for a numerical application] exceeded the annual research budget of our theory group."} Kohn and Rostoker illustrated their method by calculating the energy as a function of wave number for the 2s conduction band of lithium metal and comparing  their results with previous calculations in the literature. It is entirely characteristic of Kohn that he did {\it not} take his band structure formalism  and begin applying it to one material after another.\,\footnote{Kohn left the development and application of the Korringa-Kohn-Rostoker (KKR) method to others. It eventually became a standard method of band structure calculation (Zabloudil {\it et al}. 2005).} Instead, he made one use of his lithium results (Kohn 1954) and then proceeded to expand his personal research activities into other areas of   solid state physics. To understand  the choices he made, we interrupt our narrative briefly to survey the research agenda of solid state physics in the mid-1950's.

Wartime developments in computers, instrumentation, and materials processing had a profound effect on the issues addressed by solid state physicists at the mid-point of the twentieth century.\,\footnote{Hoddeson {\it et al.} (1992) is a history of solid state physics up to about 1960. The autobiography of Frederick Seitz (1994) provides a broad view from the perspective of a major  player in the development of solid state physics as the discipline matured through the 1950's and 1960's.} In June 1954, the National Science Foundation and the American Society for Engineering Education sponsored a meeting at  Carnegie Tech attended by representatives from  forty-five colleges and universities and several industrial and government laboratories. According to the conference co-chair, Professor Jack Goldman,\,\footnote{Jacob E. Goldman (1921-2011) was born in Brooklyn, New York and  studied physics at Yeshiva University and the University of Pennsylvania. His expertise in magnetism led him to the Westinghouse Research Laboratory in 1945  before he joined the faculty of the Carnegie Institute of Technology in 1951. He moved to the nascent Scientific Research Laboratory of the Ford Motor Company in 1955 and eventually became head of all Ford's corporate research and development. He joined the Xerox corporation in 1969 and one year later founded their Palo Alto Research Center (PARC). The first modern personal computer and the first graphical user interface were invented at PARC a few years later (Markoff 2011).} the purpose of the meeting was to ``make more definitive the state of knowledge of solid state physics and the levels at which various parts of it may be expected to be integrated into engineering education" (Goldman 1957). To this end, the  conferees identified six broad areas of active solid state physics research: the structure of crystalline matter, metals and alloys, surfaces, magnetism, semiconductors and dielectrics, and non-crystalline materials.

One needed to attend conferences reserved for specialists to learn the cutting-edge issues in each area. Happily, the same purpose was soon served by the articles published in {\it  Solid State Physics}, a series of volumes initiated by Frederick Seitz and David Turnbull in 1954 to provide  ``broad surveys of fields of advanced research that serve to inform and stimulate the more experienced investigator'' (Seitz and Turnbull 1955). The inaugural volume contained articles devoted to five issues: the band structure problem, the properties of valence semiconductors, electron-electron interactions, cohesion in solids, and the theory of order-disorder phase transitions. Kohn had already made a significant contribution to band structure theory and he now  added to his repertoire research projects devoted to semiconductors and to the effects of the electron-electron interaction (soon relabeled many-body physics). The total energy (cohesion) problem became a central concern when he developed  density functional theory a decade later.\,\footnote{Kohn never worked personally in the field of conventional thermal phase transitions. Nevertheless, he often supported a post-doctoral fellow trained in statistical mechanics to work on this class of problems.  See Domb (1996) for a history of  this subject.}

Kohn became interested in semiconductors because his flirtation with permanent employment at Bell Telephone Laboratories led to a summer consulting arrangement that lasted from 1953 to 1966.  His first summer project, a theoretical study of the damage done to germanium crystals after bombardment by energetic electrons, was motivated by experimental results obtained at Bell Labs by Walter Brown and Robert Fletcher  (Brown {\it et al.} 1953). These experiments, in turn, were part of an enormous in-house effort to investigate the properties of the elemental semiconductors germanium and silicon after  the  1947 invention of the transistor at Bell Labs by John Bardeen, Walter Brattain, and William Shockley  (Millman 1983).
Kohn was happy to return to Bell Labs  summer after summer, both to gain access to exciting experimental results and for the opportunity to interact with senior theorists on the Bell Labs staff like Conyers Herring and Gregory Wannier and  junior theorists closer to him in age like Peter Wolff and his Harvard classmate Philip Anderson.
He would later remark that  ``I owe this institution my growing up from amateur to professional" (Kohn 1998).

Walter's 1954  `summer vacation' at Bell Labs was particulary important because he began a long-lasting scientific collaboration with Joaquin Luttinger, another consultant to the semiconductor group.\,\footnote{Joaquin Mazdak Luttinger (1923-1997) earned his BS in physics at the Massachusetts Institute of Technology and remained there to complete a PhD thesis (1947) in statistical physics under the supervision of Lazslo Tisza.  He worked on quantum electrodynamics as the first American post-doctoral fellow of Wolfgang Pauli but reverted to  problems in many-body theory, solid state physics and statistical mechanics for the rest of his career as a professor, primarily at Columbia University (1960-1993) (Anderson {\it et al.} (1997).} They worked together to create a rigorous ``effective mass theory'' for the electronic energy levels produced when impurity atoms are purposely substituted for germanium or silicon atoms in pure crystals of the latter. The crucial importance of these impurities and their quantum mechanical states to the extraordinary electrical properties of semiconductors had been explained qualitatively by William Shockley in his seminal treatise, {\it Electrons and Holes in Semiconductors} (1950). However,  careful electron spin resonance and cyclotron resonance experiments at Bell Labs and elsewhere demanded a quantitative theory. Not for the first time and not for the last time, Kohn combined the creation of a novel and sophisticated theory with variational calculations designed to produce numbers for comparison with measurements for specific material systems. That fall, Kohn and Luttinger completed three substantial papers in  semiconductor physics  (Luttinger and Kohn 1955, Kohn and Luttinger 1955a,b) and thereby finished in a virtual dead heat with Berkeley  solid state theorist, Charles Kittel, who published similar work independently (Kittel and Mitchell 1954, Dresselhaus {\it et al.} 1955).\,\footnote{Sixty years later, the Bell Laboratories experimenter Robert Fletcher recalled that ``Walter was a very kind and thoughtful person to work with.  I
never had the impression he looked down on us experimenters as some
theorists are inclined to do" (Fletcher 2013). Another Bell Laboratories experimenter who published papers related to the Kohn-Luttinger theory was Walter's old friend Josef Eisinger (Eisinger and Feher 1958).}

\begin{center}
\includegraphics[scale=.35]{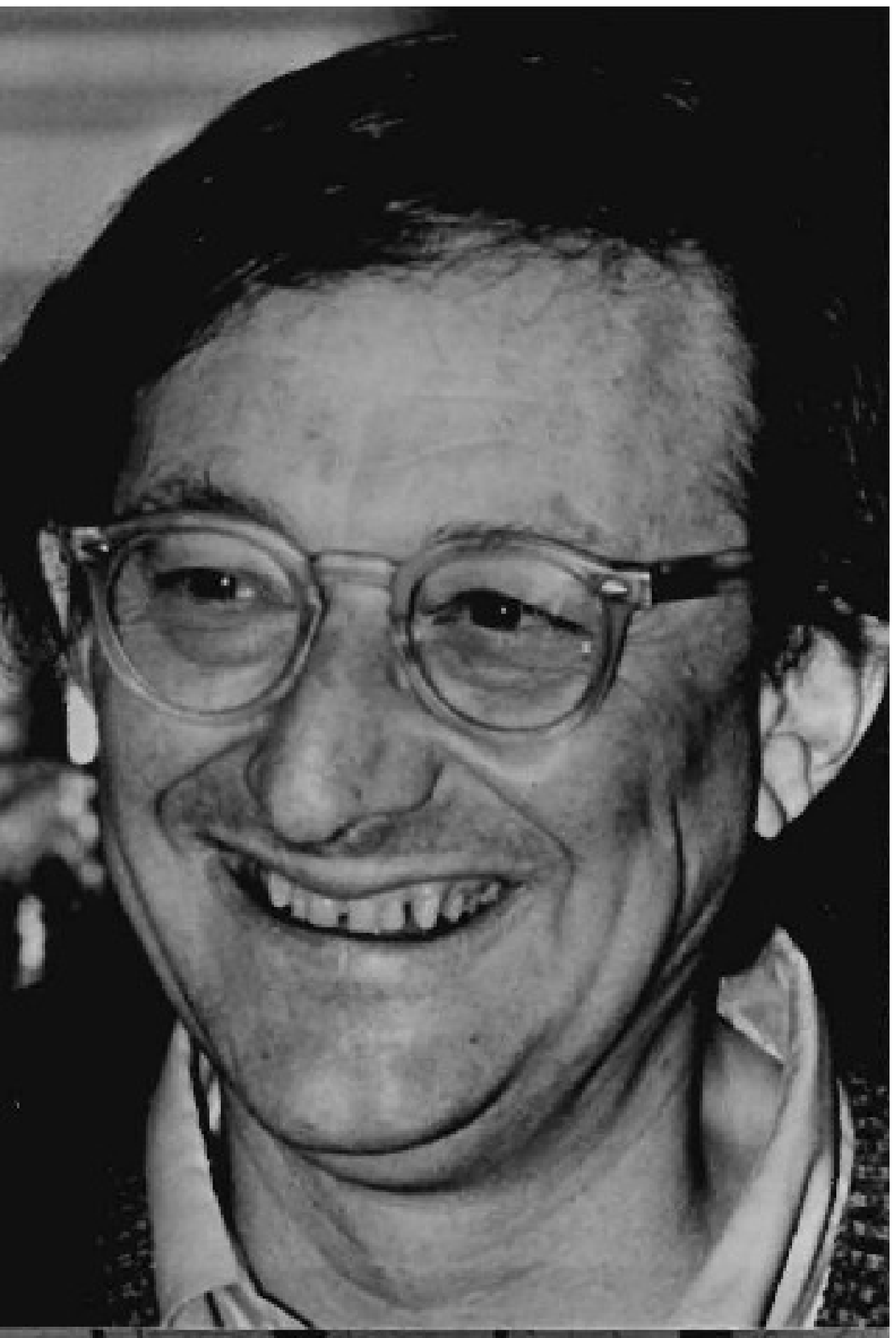}

\small Joaquin Luttinger was Kohn's principal  \\ scientific collaborator in the 1950's. \\ Courtesy of Walter Kohn.
\end{center}

In the spring of 1955, Walter worked hard to convince the British theoretical solid state physicist Harry Jones to accept an offer of a chaired position at Carnegie Tech. Jones had co-authored (with Nevill Mott) the influential book {\it The Theory of the Properties of Metals and Alloys} (1934) and he had spent the spring 1954 semester as a visiting professor in Kohn's department.  Kohn wrote to Jones and pointed out that ``all of us in solid state physics, as well as all the people in metallurgy, would be delighted to see you come here. With your field of interest, I honestly think that probably no other school in this country could offer you better opportunities for creative work along research and teaching lines" (WKP 1955).  Jones ultimately declined for personal reasons. The undeterred Kohn pursued the physics of semiconductors and metals  simultaneously and submitted two manuscripts to the {\it Physical Review}. The first paper, co-authored with his first  Carnegie Tech Ph.D. student, Daniel Schechter,  reported calculations for the wave functions and energy levels associated with shallow (weakly bound) impurity states in germanium.\,\footnote{Another graduate student, James Montague, never quite finished a thesis devoted to deep (strongly bound) impurity levels in semiconductors (Glasser 2013).} The second paper reported a Knight shift calculation for  metallic sodium with Terje Kjeldaas, a full-time employee of the Westinghouse Research Laboratories in East Pittsburgh (Kohn and Schechter 1955, Kjeldaas and Kohn 1956).\,\footnote{Kjeldaas pursued his Ph.D. part-time at the University of Pittsburgh. His 1959 thesis thanks Kohn and Westinghouse solid state theorist Theodore Holstein for acting as co-supervisors (Kjeldaas 1959). Westinghouse was a lively place for solid-state physics in the mid-1950's under the leadership of its Director of Science, Clarence Zener, himself a solid-state theorist. The theorists he recruited to complement Holstein included Edward Neufville Adams, Petros Argyres, William Mullins, and  Yako Yafet. The experimenters hired by Zener included Raymond Bowers, Robert Keyes, Colman Goldberg, and John Rayne. Walter Kohn had a consulting arrangement with Westinghouse. Holstein, Adams, and Yafet occasionally taught classes at Carnegie Tech (Ambegaokar 2013, Arrott 2013).}

For later reference, it is important to note that Kohn acted as an informal  consultant to the transition-metal magnetism groups of his faculty colleagues Emerson Pugh and  Jack Goldman. Goldman's PhD student Anthony Arrott recalls Kohn's surprise when Arrott successfully  used a simple energy band model to analyze his magnetic data for concentrated Cu-Ni alloys. At Arrott's  oral thesis defense, Kohn asked him a question that foreshadowed his motivation to invent density functional theory ten years later:  ``how can you use a band model when the potential felt by the electrons is not periodic'' (Arrott 2013)?

 As 1955 turned into 1956, Walter found himself thinking more and more about the effective mass equation he had derived with Luttinger for the energy levels of impurity states in silicon. Their ``one-particle'' method treated the impurity atom as unaware of its silicon host  except for whatever influence could be captured by two numbers: an effective mass $m^\ast$ which parameterized the  arrangement of atoms in the silicon crystal and an effective dielectric constant $\kappa^\ast$ which parameterized the ability of the silicon conduction electrons to ``screen'' or ``shield'' the Coulomb potential produced by a positively charge impurity embedded in the semiconductor.\,\footnote{By ``screening'' or ``shielding'', we mean that the electrons nearest to the positive charge are attracted to it and thereby partially neutralize the Coulomb force exerted by the positive charge on distant electrons and ions.}
 Why then did the energy levels calculated using the effective mass equation agree so very well with the energy levels measured in the laboratory? Surely, he reasoned, it must be that  ``this equation can be derived from some very general properties of the entire many-electron wave function without any recourse to the one-particle picture" (Kohn 1957). For the first time, Kohn attacked the quantum-mechanical  ``many-body problem'' in solid state physics where the repulsive Coulomb interaction between all pairs of electrons is taken seriously. Working alone, he managed to demonstrate his assertion for a hypothetical situation where the charge on the impurity nucleus exceeds that of the other nuclei by an infinitesimal amount. He announced this result in a comprehensive review of the Kohn-Luttinger theory written for  Seitz and Turnbull's {\it Solid State Physics} series (Kohn 1957a). A full account appeared later  (Kohn 1957b).

Meanwhile, back at Bell Labs, the resident theoretical physicists had successfully convinced the vice-president for research, William Oliver Baker,  to create (Anderson 2011)
 \begin{quote}
 a separate `super-department' for theorists . . .  with post-doctoral fellows, a rotating boss on whose identity we were consulted, sabbaticals, a travel budget under our control, and a spectacular summer visitor program. . . . One of the reasons for our success with management was the fact that for several years we had had Walter Kohn and Quin Luttinger as regular summer visitors and they had become so useful that our bosses desperately wanted to attract them permanently.
 \end{quote}
The advent of a `spectacular summer visitor program' meant that an unusually large number of theoretical physicists passed through and interacted with Kohn and Luttinger during their 1956 summer stay at the Labs.\,\footnote{Visitors to the theory group that summer included Elihu Abrahams, Kerson Huang, David Pines, J. Robert Schrieffer, and Philippe Nozi\`{e}res (Anderson 1978).} A hot topic was the effect on the properties of semiconductors when one systematically increased the number of impurities present. When the impurity concentration is low and the temperature is low, it was well known that electron scattering from impurities is the main source of a solid's electrical resistance. What happens when the concentration of impurities is high? Luttinger had been thinking about the  general subject of electric current flow in solids already in connection with his studies of the Hall effect in ferromagnets and it was not difficult to convince Kohn to work with him to produce as rigorous a theory of electrical conductivity as they could.\,\footnote{The Hall effect refers to a voltage that appears across a current-carrying sample when a magnetic field is applied in a direction perpendicular to the direction of current flow (Chien and Westgate 1980).}  After all, Luttinger's understanding of the experimental facts for the Hall effect came directly from a review paper written by Kohn's Carnegie Tech colleagues Emerson Pugh and Norman Rostoker (Pugh and Rostoker 1951). The fruits of that summer's labors were two long papers on the quantum theory of electrical transport in solids\,\footnote{Kohn and Luttinger did not ultimately  address the problem of the  effect of a large concentration of impurities on the electrical conductivity of a semiconductor. This was done by Phil Anderson (1958).  Simultaneous with Kohn and Luttinger's work on quantum transport, the Japanese physicist Ryogo Kubo proposed a theoretical approach to the same problem which ultimately became standard (Kubo 1957).} (Kohn and Luttinger 1957, Luttinger and Kohn 1958). In contrast to their effective mass theory work, which more reflected Kohn's style to address important physical questions with intuition, a good idea, and  mathematical elegance, the transport theory papers more reflected Luttinger's preference for general formalism and  mathematical rigor. In this way, the two young theorists enlarged each others' perspectives of their craft.

For Walter Kohn,  professor of physics,  the calendar change from 1956 to 1957 meant little more than a change in his teaching assignment from statistical mechanics  for physics majors to classical physics for engineers.  However, for Walter Kohn,  solid-state physicist,  the new year saw changes in his field that had a profound effect on his future research efforts.  In the words of Canadian physicist Allan Griffin, 1957 was
 a ``magic year'' when ``the way all theoretical physicists thought about interacting many-body systems underwent a revolution'' (Griffin 2007). The key event was the realization  that the  methods of quantum field theory could be applied with equal success to study the many-electron problem in solid-state physics.\,\footnote{There is no definitive history of this revolution. Different points of view can be found in  Pines (1961), Hoddeson {\it et al.} (1992), Gell-Mann (1996), Brueckner (2000), and Kaiser (2005). Percus (1963) is the  proceedings of a January 28-29 1957 meeting convened at the Stevens Institute of Technology  ``for the purpose of bringing together workers in the numerous rapidly moving fields of many-particle physics.'' The contributions to this volume (particularly the roundtable discussions) paint a vivid picture  of the first months of the revolution. The   Bardeen-Cooper-Schrieffer theory of superconductivity appeared later the same year, but Bardeen {\it et al.} (1957) makes no explicit use of field theoretic methods.} In particular, diagrammatic methods like those invented by Richard Feynman (1949) to study quantum electrodynamics made it  possible to define a perturbation theory that remained consistent as the number of particles in a system increased. Feynman diagrams posed no problem for a Harvard PhD like Kohn who was both well-trained in quantum mechanics and familiar with quantum field theory from  Julian Schwinger's lectures. He also had a ready-made problem:  his own desire to understand the success of the Kohn-Luttinger  effective mass equation from a many-body point of view.  A breakthrough paper by  Jeffrey Goldstone (1957) provided all  the technical details he needed.

 Walter became a naturalized citizen of the United States in 1957 and he had arranged a sabbatical leave from Carnegie Tech for the 1957-1958 academic year. He spent the fall of 1957 at the Physics Department of the University of Pennsylvania and it was there that he wrote up Kohn (1958), his first contribution to the many-body revolution--a diagrammatic analysis of the static dielectric constant of an insulator. At the end of this paper, Kohn acknowledges ``stimulating conversations'' with Keith Brueckner, a senior member of Penn's  Physics Department whose own thinking about the quantum many-body problem had stimulated Goldstone's work.\,\footnote{Keith Allen Brueckner (1924 - ~) studied mathematics at the University of Minnesota before earning his Ph.D in physics (1950) from the University of California (Berkeley) under the supervision of Robert Serber. As a professor at Indiana University and the University of Pennsylvania in the  1950's, Brueckner made many significant contributions to  the theories of nuclear matter and the electron gas.  In 1959, he  became the first member of the   physics department at the newly created University of California, San Diego. Brueckner divided his professional activities  between academia, industry, and the government until his retirement from UCSD in 1991.}  Brueckner, in turn, led an effort by his Penn colleagues to hire Kohn away from Carnegie Tech. A similar effort was mounted by the Physics Department of the University of Chicago  (WKP 1957). Kohn took these overtures seriously and Carnegie Tech responded by  awarding him tenure and promoting him to the rank of Professor with a substantial increase in salary (WKP 1958). Walter made his decision to return to Pittsburgh while completing his sabbatical and spending the spring 1958 semester with Harry Jones and his group at the Department of Mathematics of the Imperial College of Science and Technology in London.

The many-body revolution  introduced new ideas into solid state physics like the quasi-particles of Lev Landau (1956) and new objects for study like the one-particle and two-particle Green functions exploited  by Victor Galitskii and Arkday Migdal (1958).\,\footnote{The Green functions used in many-body theory are the ground state expectation values (averages over the ground state many-body wave function) of various quantum mechanical operators. The Green function for the Schr\"{o}dinger partial differential equation, Eq.~(\ref{two}), is related to some of the Green functions used in many-body theory when all the forces between the particles are turned off.} The physical and mathematical elegance of this subject caused some solid state theorists to  focus their attention exclusively on problems where many-body effects dominate the physics.\,\footnote{Samuel F.  Edwards, another Schwinger PhD student who switched from nuclear physics to solid state physics, has remarked that `the Green function formalism is very good to write down  solutions in abstract exact form, which gives unassailable answers when used in comparatively simple situations' (Edwards 1998).} Kohn did not follow this particular path because not every problem that interested him demanded a many-body analysis. For example, a theorem due to Felix Bloch (1928) demonstrated  that the eigenfunctions and energy eigenvalues of the Schr\"{o}dinger band for a spatially periodic system like a crystal have the form
\begin{equation}
\label{five}
\psi_{\bf k}({\bf r}) = \exp(i{\bf k}\cdot {\bf r})u_{\bf k}({\bf r})~~~~~~~~~~{\rm and}~~~~~~~~~~E({\bf k}),
\end{equation}
where the three quantum numbers collected in the vector  ${\bf k}=(k_x,k_y,k_z)$ are real numbers confined to a finite volume of the three-dimensional ${\bf k}$-space called the Brillouin zone, and the function $u_{\bf k}({\bf r})$ has the spatial periodicity of the crystal. While at Imperial College, Kohn performed an extensive study of the properties of the Bloch solutions  when  ${\bf k}$ is a complex-valued  vector. This allowed him to make precise statements about the exponential decay of a class of spatially localized wave functions first introduced by Wannier (1937).  In a separate project, Kohn analyzed the motion of Bloch electrons in a magnetic field with more rigor than had been done previously. He succeeded to show that an approximation for this problem first made by Peierls (1933) had a much broader range of validity than previously thought.

Kohn's Imperial College projects in mathematical physics did not disengage him from the more practical aspects of solid state physics.\,\footnote{Kohn became an Associate Editor of the {\it Journal of Mathematical Physics} in 1961.} For example, he was invited to the 1958 International Conference on Semiconductors in Rochester, New York, to report his many-body analysis of the effective mass equation for shallow impurity states. While there, he attended a session devoted to the calculation of energy bands and listened to a talk by  fellow-theorist James Phillips. Kohn asked Phillips whether he was doing ``physics or magic'' because Phillips' `pseudopotential method' reproduced the results of much more elaborate band structure calculations for silicon and germanium using only three parameters for each (Bassani and Tosi 1988). Similarly, Kohn paid careful attention when the {\it phonon spectrum} of a crystal was measured for the first time (Brockhouse and Stewart 1958) and also when the {\it Fermi surface} of a metal was measured for the first time (Pippard 1957, Gold, 1958).\,\footnote{A phonon is a quantized lattice vibration in  a crystal. The Fermi surface is the constant energy surface in the Bloch ${\bf k}$-space for the most energetic electrons in a metal.}

These experimental breakthroughs stimulated Walter's scientific imagination and he soon completed a simple and elegant analysis which predicted that the phonon spectrum of a metal possesses observable  ``anomalies'' which depend only on the existence and shape of the Fermi surface. More precisely, the Fermi surface locates a singularity in a linear  response function which describes the ability of the  conduction electrons to screen the ions which move during a lattice vibration. The short communication  which described what came to be known as ``Kohn anomalies'' was  one of four manuscripts he submitted for publication during the 1958-1959 academic year at Carnegie Tech (Kohn 1959a,b,c, Ambegaokar and Kohn 1959). Two longer papers described  the results of the projects begun at Imperial College and the paper co-authored by his PhD student Vinay Ambegaokar reported a new sum rule for insulators. The Kohn anomaly and Ambegaokar papers appeared in the same issue of {\it Physical Review Letters}, a journal spun-off from the {\it Physical Review} to provide ``speedy publication'' of ``new discoveries of major importance and for significant contributions to highly active and rapidly changing lines of research in basic physics" (Goudsmit and Trigg 1964).

Ambegaokar (2004, 2013) recalls that
\begin{quote}
Before going on leave [to Pennsylvania], Walter advised me to take a second course in quantum mechanics taught by
 Gian-Carlo Wick even though I had not finished a first course. . . . Upon his return [from England], he suggested a research project that was very much to my taste. We met for at least an hour a week and his supervision  was both precise and constructive. He thought hard during our meetings to keep the project moving along. . . . [Walter] could be formal as a person, but he opened up considerably with people he respected. He got me a summer job at Bell Labs and we played tennis together there frequently.
\end{quote}
Another student, James Langer,  was an undergraduate physics major at Carnegie Tech from 1951-1955. Langer  never took a formal course from Kohn, but in his senior year,  ``Walter somehow became my private instructor for a year-long supervised reading course. We went through the first edition of Leonard Schiff's classic text {\it Quantum Mechanics} essentially cover to cover''  (Langer 2003). Langer won a Marshall Scholarship to attend graduate school in Great Britain and Kohn directed him to Rudolf Peierls at the University of Birmingham.\,\footnote{In this way, Langer became the Peierls PhD student Kohn would have been if he had not gone to Harvard to work with Julian Schwinger (Kohn 2013).} Langer earned his PhD for a problem in nuclear physics and then returned to Carnegie Tech as an Instructor in  the fall of 1958. For the next year,  he and Seymour Vosko (a recent PhD student of Gian-Carlo Wick) functioned as post-doctoral fellows in Kohn's theoretical solid-state physics group.

The problem Kohn set for Langer and Vosko  was the shielding of a positively charged impurity embedded in a metal host.  This is  the analog  of the problem Walter had studied previously for the case of a non-metal host. Kohn's many-body perturbation theory calculation for the non-metal case confirmed the classical result that the Coulomb potential $q/r$ at distance $r$ from a point charge $q$ in vacuum changes to $q/\kappa r$ when the point charge is embedded in an insulator with dielectric constant $\kappa$. Nevill Mott (1936) studied the screening of a point charge in a metal in connection with a calculation of the electrical resistivity of dilute metal alloys. He used semi-classical Thomas-Fermi theory (March 1957) and showed that the potential $q/r$ in vacuum changes to $(q/r)\exp(-r/\ell)$ in a metal. The screening length $\ell$ depends on the density of conduction electrons and takes the value $1$-$2~ {\rm \AA}$ in a typical metal. Jacques Friedel (1958) revisited this problem using a  scattering theory method and found that the disturbance of the electronic charge density at a distance $r$ from the impurity charge, $\delta n(r)$, varied as
\begin{equation}
\label{fivehalf}
\delta n(r) \propto  {\cos(2k_F r + \Delta) \over r^3},
\end{equation}
where $k_F$ and $\Delta$ are two constants. The $1/r^3$ decay of this function falls off much more slowly with distance than the exponential variation predicted by Mott's theory and thus implies that the effect of isolated impurities might not be completely screened at the position of nearby atoms. The oscillatory behavior in  Eq.~(\ref{fivehalf}) is an intrinsically quantum effect. Kohn did not fully believe Friedel's result and therefore asked his post-docs to attack the problem themselves  (Kohn 2012b). Much to his surprise, Langer and Vosko (1959) fully confirmed Friedel's formula using diagrammatic many-body perturbation theory.\,\footnote{Langer and Vosko used a formulation of many-body perturbation theory due to John Hubbard (1957).}  Kohn  promptly applied their results to a quantitative calculation of the magnitude of the nuclear magnetic resonance signal in copper metal when small amounts of impurity atoms are introduced (Kohn and Vosko 1960). Confirmation of the Kohn-Vosko theory came from comparison with experiments performed by Theodore Rowland  at the Union Carbide Metals Company (Rowland 1960). Kohn knew Rowland from his Harvard days when Rowland was a PhD student of Nicolaas Bloembergen.

The summer of 1959 reunited Walter and Quin Luttinger at Bell Laboratories. Once again, the pair produced an interesting  paper (Kohn and Luttinger 1960) and once again, Kohn grappled with an offer from Keith Brueckner to leave Carnegie Tech. This time, however, Brueckner  was not soliciting on behalf of the University of Pennsylvania. He had  resigned from Penn a few months previously and his  mission now was to convince Walter to help him create the Physics Department at the soon-to-open University of California at La Jolla (later San Diego). Earlier in the year, Brueckner had flown Kohn to the beautiful site of the proposed campus to meet and hear the vision of its principal advocate, Roger Revelle,  the Director of the Scripps Institute for Oceanography. The salary was attractive and La Jolla seemed like an ideal place to relocate his wife and two elementary school-aged daughters.\,\footnote{The regents of the University of California committed unprecedented financial resources so UCSD could recruit  senior scientists like Kohn to its nascent faculty. In its first few years, 50 percent of new faculty hires at UCSD were made at the full professor level or above, as opposed to 15 percent for the  University of California system as a whole  (Kerr 2001).} Moreover, Ed Creutz,  the man who had hired Kohn at Carnegie Tech and who was now Vice-President of Research at the General Atomics division of the General Dynamics Corporation in San Diego,  had recently concluded a  consulting contract with him. This time, the allure was too great and Walter agreed to sign on. His only condition was that Keith Brueckner must serve as chair of the  new department (Brueckner 2013).

Walter's research group had grown to include three PhD students and four post-doctoral fellows by the fall of 1959 when he submitted his resignation to the President of Carnegie Tech. His  senior student, Vinay Ambegaokar, was one semester from graduation. His junior students, Larry Glasser and Edwin Woll, Jr. were not too far from the beginning of their research so  Walter invited them both to join him in La Jolla. Woll chose to accompany his advisor; Glasser remained  in Pittsburgh  and finished his degree with Assistant Professor J. Michael Radcliffe (Glasser 2013). Post-docs Hiroshi Hasegawa and Robert Howard had arrived the previous fall from Tokyo and Oxford, respectively, and worked together on a problem motivated by Kohn's experience with shallow donor states in semiconductors (Hasegawa 2004). Hasegawa accompanied Kohn to San Diego while Howard moved on to a permanent position at the National Bureau of Standards in Washington, D.C. Post-docs Emile Daniel and Anthony Houghton were former PhD students of Jacques Friedel in Paris and Geoffrey Chester (in the group of Rudolf Peierls) in Birmingham, respectively. Both began research projects in  metal alloy physics before moving to the west coast. Kohn himself taught a graduate course in advanced solid state physics and gave a talk on  ``The Electron Theory of Solids'' at a one-day  ``Solid State Symposium'' in New York City sponsored by the American Institute of Physics for the benefit of science writers from national magazines and newspapers.\,\footnote{The Symposium was organized by Conyers Herring from Bell Telephone Laboratories. Besides Kohn, the lecturers were John Fisher from the General Electric Research Laboratory, Jack Goldman from (by then) the Ford Motor Company Research Laboratory, and Frank Herman from  RCA Laboratories (WKP 1959a).}

Even before leaving Pittsburgh, Walter  worked hard to recruit faculty members to his new Physics Department in San Diego. He had immediate luck with three  Bell Laboratories scientists, the statistical mechanician Harry Suhl, the nuclear magnetic resonance experimenter George Feher, and the superconductivity experimenter Bernd Matthias.\,\footnote{Suhl and Feher arrived in San Diego in  1960. Matthias waited a year  because ``during the first year there will be too many administrative chores'' (Feher 2002).} In an October 26 1959 letter to Keith Brueckner, Kohn laments Quin Luttinger's decision to choose Columbia University over UCSD and suggests several solid state and/or low-temperature physicists whom Brueckner should approach (WKP 1959b). On the theoretical side, he proposed J. Robert Schrieffer, the junior author of a breakthrough paper on the theory of superconductivity published in 1957, Volker Heine, a specialist  in the electronic structure of metals from the University of  Cambridge, and Philippe Nozi\`{e}res, an expert in many-body theory trained at Princeton by David Pines. On the experimental side, he suggested his old Harvard friend and magnetic resonance practitioner, Charles Slichter, and the liquid helium experts William Vinen and Russell Donnelly. As it turned out, none of these people joined the UCSD faculty.\,\footnote{Offers were  also proffered to (and declined by) Kohn's old Harvard friend Ben Mottelson, a nuclear physicist at Bohr's Institute for Theoretical Physics, and the  French magnetic resonance expert, Anatole Abragam (WKP 1959c, Abragam 1989).}

Kohn's arrival in San Diego in January 1960 coincided with his election as a Fellow of the American Physical Society (APS), the professional organization of American physicists.\,\footnote{An APS Fellow is judged by his peers to have made ``exceptional contributions to the physics enterprise". In 1960, the total number of Fellows was 1653 out of a total Society membership of 16,157 (APS 2013b).} Resettlement and administrative issues dominated his time at first, so Kohn used his students, post-docs, and short-term visitors to pursue a research agenda now focused primarily on the physics of metals and alloys.\,\footnote{A visiting French scientist, Jacques des Cloizeaux,  arrived in the fall of 1960 and worked on a statistical mechanics problem.} A new post-doctoral fellow, Stephen Nettel, studied  the spatial arrangement of electron spins in the ground state of a  {\it homogeneous and non-interacting electron gas}, a much-studied hypothetical system composed of a collection of mobile electrons (with their mutual Coulomb interaction turned off) distributed throughout a uniform and static distribution of  electrically neutralizing positive charge (Kohn and Nettel 1960). Emile Daniel and former post-doc Seymour Vosko used many-body perturbation theory to study the sharpness of the Fermi surface for a fully interacting electron gas at zero temperature and Tony Houghton studied the specific heat and spin susceptibility of a dilute alloy (Daniel and Vosko 1960, Houghton 1961).\,\footnote{A ``sharp'' Fermi surface has the property that a quantum state labeled by the quantum number wave vector ${\bf k}$ is occupied by an electron if that wave vector lies inside the volume of the ${\bf k}$-space enclosed by the Fermi surface and unoccupied if ${\bf k}$ lies outside that volume.} The numerical calculations of Daniel and Vosko confirmed Quin Luttinger's analytic demonstration that electron-electron interactions do not destroy the sharpness of the Fermi surface of an electron gas. This result was reported by Luttinger at a (retrospectively famous) meeting attended by Kohn and Vosko on ``The Fermi Surface'' held at Cooperstown, New York on August 22-24, 1960. (Luttinger 1960, Harrison \& Webb 1960).

The La Jolla campus of the University of California opened for business in the fall of 1960. There were no undergraduates (until 1964), but sixteen physics graduate students arrived and began taking classes. Walter Kohn taught a course from 4:30 PM - 6:00 PM on Thursdays and from 9:30 AM - 11:00 AM on Saturdays using the textbooks {\it Thermodynamics} by Herbert Callen (1960) and {\it Elements of Statistical Mechanics} by  Dirk ter Haar (1954).\,\footnote{The unusual hours were chosen for ``the convenience of students employed in industry'' (UCSDA 1960).} The smallness of the new department and the setting just steps from the Pacific Ocean fostered considerable informality in the early days. Arnold Sherwood, a member of the first cohort of graduate students, recalled that ``Kohn had a charming European formality even when he was trying to be informal'' (Sherwood 2013). According to faculty member George Feher (2002):
\begin{quote}
Occasionally a student would come to class in a bathing suit or scantily dressed. We didn't mind it, except perhaps Walter Kohn, who was a bit more formal than the rest of us. But he couldn't tell the student off lest he be called a stuffy professor. He finally solved this problem by telling the students that he didn't mind their behavior but did not want them to acquire bad habits because some stuffy professor might take offense.
\end{quote}

Kohn flew to New York City on February 1 1961 to attend the annual joint meeting of the American Physical Society and the American Association of Physics Teachers. He was there to accept the 1961 Oliver Buckley Solid State Physics Prize  ``for having extended and elucidated the foundations of the electron theory of solids" (PT 1961).  This prize, endowed in 1952 by Bell Telephone Laboratories and named in honor its former president and board chairman, is awarded each year by the APS to ``recognize and encourage outstanding theoretical or experimental contributions to solid state physics''. Some measure of the esteem carried by this honor may be judged from the fact that four of the eight  persons who won the Buckley Prize before Kohn later won a Nobel Prize (William Shockley, John Bardeen, Clifford Shull, and Nicolaas Bloembergen).

A few months later, Kohn  submitted for publication his forty-eighth scientific paper in sixteen years. The first paragraph of this paper gives a good indication of his style, perspective, and level of engagement at this point in this career (Kohn 1961):
\begin{quote}
There has been considerable interest in recent months in the effects of the electron-electron interaction on the cyclotron resonance frequency and de Haas-van Alphen oscillations of a gas of electrons. As some of the theoretical treatments of these problems use very sophisticated methods, and others are based on incorrect qualitative reasoning, we wish here to present some simple considerations which we think shed some light on what has been a rather confusing situation.
\end{quote}
 The paper Walter cited as bringing ``very sophisticated methods'' to the issue at hand was written by his good friend and collaborator, Quin Luttinger (1961). A more precise statement would be that Luttinger's work employed very sophisticated {\it mathematical} methods. Kohn's paper exploited much simpler mathematics but rather sophisticated {\it physical} reasoning to reach the same conclusions.

 At UCSD, Kohn initiated a weekly Theoretical Solid State Lunch where faculty members, post-docs, and long-term visitors made presentations to anyone who brought a bag lunch and cared to listen. If a short-term visitor addressed the group, Kohn would take the visitor out to lunch and invite his personal research group to come along. After lunch,  everyone was invited to take a walk around the campus. Twenty-five years later, this behavior was parodied in a skit performed at a celebration to commemorate the anniversary of the founding of the UCSD Physics Department. The relevant dialog involves a visitor to the campus and a professor in the department (LJPS 1985):
\begin{quote}
Visitor: And what are those figures I see in the misty distance? It looks like a giant duck followed in a straight line by giant ducklings.

\noindent
Professor: Nothing too strange. That's Walter Kohn taking his students for a walk after lunch.
\end{quote}

 Sometime in the spring of 1961, Keith Brueckner announced that he was stepping down as the chair of the Physics Department after one year of service. He had accepted the position of vice-president and technical director of the Institute for Defense Analyses in Washington, D.C. (Brueckner 2013) Therefore, against his expressed desire and much to his chagrin, Walter Kohn found himself in the position of chair for the 1961 fall semester.
 As a department head at a start-up university, Kohn enjoyed opportunities and faced challenges that do not ordinarily arise for administrators at established universities. He was permitted to hire a dozen new faculty members (which doubled the size of his department), but the offices and laboratories he could offer new recruits occupied temporary space that would soon revert to the Scripps Institution of Oceanography.   One interesting hire was his friend Norman Rostoker, who had turned himself into a plasma physicist at General Atomics since he and Walter had collaborated at Carnegie Tech.
The senior faculty members recruited by Brueckner and Kohn had excellent research records, but many came from industrial or government laboratories with no teaching experience. It was Walter's responsibility to ensure that competent instructors staffed the courses offered to the first few classes of graduate students. At the same time, he maintained a research group of never less than five persons (graduate students, post-doctoral fellows, and visitors), served as an Associate Editor of the Journal of Mathematical Physics, team-taught a course on ``Advanced Solid State Physics'', presented a university-wide lecture on ``New Viewpoints in the Theory of Matter'', and submitted four research papers.\,\,\footnote{Kohn taught  semiconductor physics and the transport and optical properties of metals for five weeks in the fall of 1962 (Bruch 2013). His April 1962 campus-wide lecture was part of a series delivered by senior UCSD professors from various departments.}

Within his group, Edwin Woll, Jr., Kohn's PhD student from Carnegie Tech was making good progress with semi-quantitative calculations of Kohn's phonon anomaly in the metals aluminum, sodium, and lead. Walter also began working with Michael Greene and Max Luming, two PhD candidates from the 1960  crop of UCSD graduate students, and Chanchal Majumdar, a student from the 1961 class. Greene was tasked to use scattering theory to compute the resistivity of liquid alkali metals. Max Luming involved himself in  calculations of the orbital susceptibility of dilute metal alloys but switched to theoretical particle physics after the publication of Kohn and Luming (1963).\,\footnote{Luming (later Luming Ren) switched his PhD supervisor from Kohn to Assistant Professor David Wong because he felt that particle physics was ``more fundamental'' than solid state physics. He came to rue his decision when he was unable to find permanent employment as a particle physicist and made his career as a systems engineer for the Hughes Aircraft Company (Ren 2013).}  Majumdar began a project on the theory of positron annihilation in metals. Overall, Kohn acquired a reputation among potential theory students as a supervisor with very high standards who could assign a thesis problem that might take a very long time to complete (Feibelman 2012). This did not deter Michael Greene (PhD 1965), who joined Kohn's group (despite an initial lack of interest in solid state physics) because he was impressed by the thoughtful questions Walter asked at seminars. Greene wanted to learn to {\it think} like Kohn (Greene 2013).

  \begin{center}
\includegraphics[scale=.5]{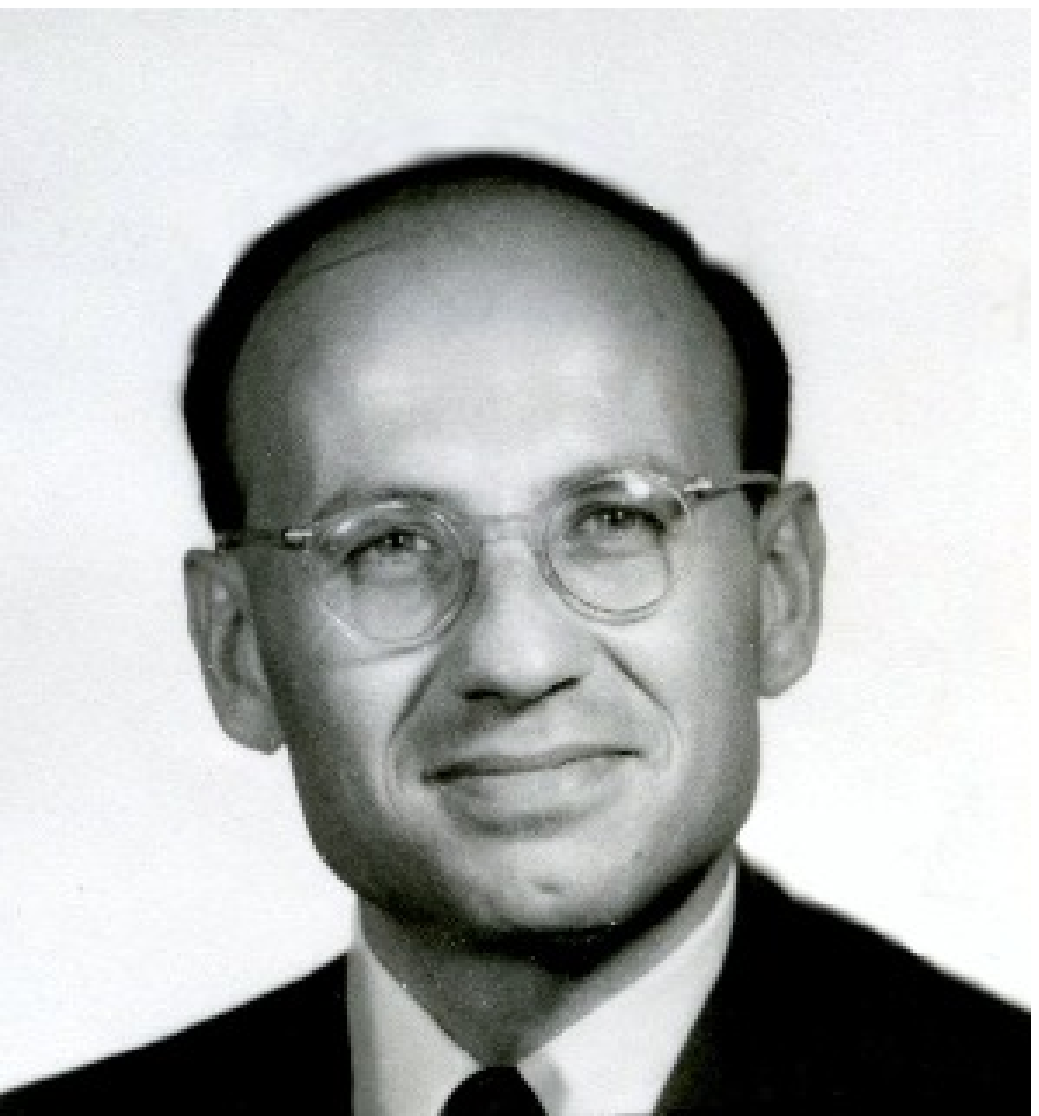}

\small Walter Kohn at age 39 (1962). \\ Courtesy of Walter Kohn and \\ the John Simon Guggenheim Foundation.
\end{center}

At the beginning of the fall 1962 semester, Walter learned that his colleague Norman Kroll would take over as Physics chair  beginning in the fall 1963 semester. Right away, he made an application to the John Simon Guggenheim Memorial Foundation for funds to support a recuperative leave  for the fall 1963 semester (Hohenberg, Kohn, and Sham 1990). His plan was to spend that time at the physics department of the \'{E}cole Normale Sup\'{e}rieure in Paris (Kohn, 1962). This was an ideal place to get back to full-time research. It was also an ideal place to renew his personal and scientific ties with three specialists in his own field of theoretical solid state physics: Jacques Friedel, Pierre-Gilles de Gennes, and Philippe Nozi\`{e}res. It was a bonus that the physics department was just a short walk from the Luxembourg Gardens, his favorite place in the whole world (Kohn 2001a).

Kohn's fellowship application proposed that he would study the interaction of electrons and phonons in metals (Kohn, 1962). This was a hot topic. The collisions between electrons and the particle-like phonons had long been recognized as important for a  proper description of  the electrical conductivity of metals (Ziman, 1960). However, only five years earlier, John Bardeen, Leon Cooper, and J. Robert Schrieffer had proposed a many-electron wave function for a {\it superconductor} based on a model where the electron-phonon interaction mediates an  effective attractive interaction between pairs of electrons with opposite spin (Bardeen {\it et al.} 1957). Moreover, barely a year earlier,  neutron scattering experiments had confirmed Kohn's own  prediction of anomalies in the phonon spectra of metals (Kohn, 1959c). Accordingly, Kohn proposed to spend the fall of 1963 generalizing the theory of  Kohn anomalies. For good measure,  he also proposed to develop a theory of the effect of electron-phonon interactions on the optical properties of metals.

The Guggenheim Foundation responded positively to Walter's application in the spring of 1963. This news must have been a great stimulant because he quickly completed a calculation which achieved ``a new and more comprehensive characterization of the insulating state of matter'' (Kohn 1964). The fundamental difference between the conducting state and the insulating state had been an issue for Kohn since his Bell Labs-inspired  work with Luttinger on electrical transport (Kohn and Luttinger 1957, Luttinger and Kohn 1958). Subsequent papers on the behavior of a point charge in a dielectric, the nature of Wannier's spatially localized states in solids, and the electromagnetic properties of insulators continued this theme (Kohn 1958, Kohn 1959a, Ambegaokar and Kohn 1960). His new work took seriously  a suggestion by Nevill Mott (1949) that the  many-body wave function in an insulator should be fundamentally different from the many-body wave function in a conductor. Kohn exploited a ground-breaking paper that focused attention on the gauge principle for the electromagnetic vector potential in quantum mechanics (Aharonov and Bohm 1959) and used a characteristically elegant method to calculate the electrical conductivity of a ring threaded by a line of magnetic flux. The result was a proof  that the spatial organization of the electrons in an insulator corresponds to a many-body wave function that breaks up into disconnected regions that do not overlap with one another. The published paper, Kohn (1964), has been called a ``a mine of ideas and methods'' by no less an expert than Walter's old Harvard classmate and Bell Laboratories colleague Philip Anderson (Anderson 2012).\,\footnote{Philip Warren Anderson (1923 - ~) is generally regarded as one of the preeminent theoretical physicists of the second half of the twentieth century. He attended Harvard University as an undergraduate, worked at the U.S. Naval Research Laboratory during World War II and returned to Harvard to earn his PhD in 1949 under the supervision of John van Vleck. Anderson spent a long and productive career at Bell Telephone Laboratories before moving to Princeton University in 1984. He shared the 1977 Nobel Prize in Physics with Nevill Mott and John Van Vleck for `fundamental theoretical investigations of the electronic structure of magnetic and disordered systems' (Anderson 1977).}

Now 40 years old, Walter Kohn was a  mature solid state physicist whose scientific talent and taste in problems had produced results that were highly valued by his peers. Two of those peers, David Pines and Charles Kittel,  highlighted Kohn's work four and eight times, respectively in their (now classic) 1963 graduate level textbooks {\it Elementary Excitations in Solids} and {\it Quantum Theory of Solids}. By the end of that summer, Walter's manuscript on the ``Theory of the Insulating State''  was ready for submission and  he had only to review some  professional correspondence before he could depart for Paris. In retrospect,  the most important letter on his desk came from Lu-Jeu Sham, a graduating PhD student from John Ziman's group at the University of Cambridge whom Kohn had earlier recruited to become a post-doctoral fellow. Kohn had written to Sham to inform him about his Paris sabbatical  and to urge him to come to San Diego as originally planned. Kohn proposed that Sham work on liquid metals with graduate student Mike Greene until he (Kohn) returned to campus. The return letter from Sham agreed to this plan (WKP 1963a).

\section{Alloys in Paris}
Walter Kohn's base of operations in Paris was the Ecole Normale Sup\'{e}rieure, one of the elite {\it grandes \'{e}coles} of the higher education system in France. His host was the 31-year old  Philippe Nozi\`{e}res, an expert in many-body theory who had just collaborated with Joaquin Luttinger to derive Landau's theory of the Fermi liquid using diagrammatic perturbation theory (Nozi\`{e}res and Luttinger, 1962, Luttinger and Nozi\`{e}res 1962).\,\footnote{Philippe Nozi\`{e}res (1932-~) graduated from the Ecole Normale Sup\'{e}rieure (ENS) in 1955 and earned his PhD two years later from the University of Paris, albeit under the supervision of David Pines at Princeton University. He spent a decade at the ENS before moving to the Institut Laue-Langevin (ILL) in Grenoble. He also lectures at the Coll\`{e}ge de France in his capacity (since 1983) as Professor of Statistical Physics (Nozi\`{e}res 2012a).} The setting inside the Physics Department at 24 rue Lhomond in the Latin Quarter was not typical. According to Nozi\`{e}res (Cheetham 1992, Nozi\`{e}res 2012b),
\begin{quote}
I had inherited a very magnificent office with an old desk made from an oak tree. It was gigantic, eight times the size of a normal desk,  and I put Walter on one side and me on the other side, facing each other.  The office had a very high ceiling, maybe 15 feet high, and an upper level balcony had been installed for a secretary, which I did not have. Instead, Pierre Hohenberg settled there, watching Walter and I from above.
\end{quote}
Pierre Hohenberg was a newly-arrived post-doc in Nozi\`{e}re's group. He places himself on the main floor of ``Philippe's own very large office . . . and I remember it to have been a general meeting place and thoroughfare, a little like trying to think deep thoughts in the middle of Times Square'' (Hohenberg 2003, 2012).

Kohn began his research activities, but he did not work on the electron-phonon interaction as he had proposed to the Guggenheim Foundation. Some months earlier,
 he had changed his mind and decided to think more deeply about the electronic structure of disordered metal alloys.  More precisely, he asked himself how one might best describe  the behavior of the electrons in a bulk metal composed of different types of atoms where there is at least some randomness in the identities of the atoms that occupy the sites of the underlying periodic lattice.\,\footnote{In July, Kohn had written to an editor at Academic Press confirming his interest to contribute to a book about ``Impurities in Metals'' and indicating that he would be ``working in this field'' during his stay in Paris (WKP 1963b).} Unlike most theoretical solid state physicists in the United States, Kohn had followed developments in alloy physics for more than a decade because of the intense experimental interest  in this subject by his faculty colleagues in the Physics and Metallurgy departments at Carnegie Tech. On the other hand, his personal contribution to the field consisted of only two published papers and both concerned  {\it dilute} alloys like ${\rm A}_x{\rm B}_{1-x}$  where the fraction $x$ of A-type  atoms dissolved in a host metal made of B-type atoms was  very small (Kohn and Vosko 1960, Kohn and Luming 1963). He now turned his attention to {\it concentrated} alloys where the populations of A-type atoms and B-type  atoms could be comparable. Philippe Nozi\`{e}res was not particulary interested in alloys (he was working on liquid helium at the time), but a short 30~km train ride took Kohn to  the suburban campus of the University of Paris in Orsay where his old friend  Jacques Friedel maintained his research group. Friedel was an acknowledged expert in the theory of metals and alloys.\,\footnote{Jacques Friedel (1921~-~) is a fourth-generation French scientist who was educated at the \'{E}cole Polytechnique (1944-46) and  the \'{E}cole Nationale Sup\'{e}rieure des Mines (1946-48) before earning his PhD in 1952 under the supervision of Nevill Mott at the University of Bristol. Friedel began his academic career at the Sorbonne, but moved in 1959 to the Orsay campus of the University of Paris,  now the University of Paris-Sud. Friedel's life-long interests in metallurgy and the physics of metals resulted in over 200 theoretical publications, most of them characterized by the use of simple models and elementary mathematics (Br\'{e}chet 2008). Kohn and Friedel met at a July 1953 Gordon Conference in Laconia, New Hampshire devoted to the Chemistry and Physics of Metals (Kohn 2012a).} Also present in  Orsay at the time were Andr\'{e} Guinier, an experimentalist renowned for his x-ray diffraction studies of alloys, and Pierre-Gilles de Gennes,  a theorist working on a set of problems he would soon collect and discuss in his book, {\it Superconductivity in Metals and Alloys} (1966).\,\footnote{Pierre-Gilles de Gennes (1932-2007) changed fields after the publication of his superconductivity book and won the 1991 Nobel Prize in Physics for his work on the statistical physics of liquid crystals and polymers.}

Kohn immersed himself in the literature of metals and alloys and soon discovered that two seemingly contradictory points of view dominated discussions of their electronic structure. I pause here to sketch the field as he found it, because his desire to reconcile these points of view was the immediate trigger for the creation of density functional theory. The fundamental problem was to calculate the eigenfunctions and energy eigenvalues for a binary alloy where A-type atoms replace a fraction of the atoms in a perfect B-type crystal. If the replaced B-type atoms are chosen randomly, the resulting structure is no longer periodic and the Bloch theorem which underlies conventional band structure theory is no longer valid.\,\footnote{Alloys of this kind are called {\it disordered}. The Bloch theorem remains valid for {\it ordered alloys} where the A-type atoms form a spatially periodic structure of their own.} By the end of the 1950's, approximate ways to analyze this situation had been proposed by Lothar Nordheim (1931), Harry Jones (1934), and Jacques Friedel (1954).  All of them acknowledge a debt  to the eminent English physical metallurgist William Hume-Rothery and his 1931 book, {\it The Metallic State}.

The first half of Hume-Rothery (1931) reviews years of experimental effort to systematize  the electric, thermo-electric, and thermionic properties of metals and alloys. The second half  reviews the classical and quantum mechanical theories that had been devised to explain some of these properties.  A typical result reported in {\it The Metallic State} was the observation that many disordered substitutional alloys ${\rm A}_x{\rm B}_{1-x}$  exhibit an  electrical resistivity that varies with the A-type atom concentration as $x(1-x)$. Nordheim (1931) explained this by replacing the real alloy, where dissimilar potentials $V_A(r)$ and $V_B(r)$ act on the  valence electrons near lattice sites occupied by A-type atoms and B-type atoms, respectively, by a fictitious {\it virtual crystal} where the valence electrons near every lattice site feel the same average potential, $\bar{V}(r)=xV_A(r)+(1-x)V_B(r)$.  By construction, the potential energy function for the virtual crystal is periodic and any band structure method becomes applicable to find the eigenfunctions and energy eigenvalues (Muto 1938).

Jones (1934) was concerned with some empirical `rules' deduced by Hume-Rothery which related the crystal structure of certain alloys to their  `electron concentration', {\it i.e.}, the ratio of the total number of valence electrons to the total number of atoms in the entire crystal.  Jones focused on the dilute limit and made two independent assumptions. First, he used a {\it nearly-free electron} description where the energy spectrum and wave functions of the host B-type metal was presumed to differ only slightly from the energy spectrum and wave functions of a collection of completely free electrons. Second, he made a {\it rigid-band} approximation which supposed that the sole effect of the A-type atoms was to contribute their valence electrons to the pre-existing  `sea' of valence electrons contributed by the B-type atoms. This implied that the electronic structure of the alloy was identical to the electronic structure of the host metal except that a few energy states were occupied (empty) in the alloy compared to the host if the valence of the A-type atoms was larger (smaller) than the valence of the B-type atoms.\,\footnote{This follows from the eigenstate occupation rules of  quantum mechanics which dictate that states are populated by electrons in order of increasing energy beginning with the lowest.} Using quantum mechanical perturbation theory, Jones estimated the change in total electronic energy when the Fermi surface of the rigid-band alloy contacts the Brillouin zone boundary for different crystal structures. In this way, he was able to rationalize the Hume-Rothery's electron concentration rules in a semi-quantitative way.

The virtual crystal and rigid-band approximations share a `delocalized' view of the electrons in a metal alloy. This means that each electron in the conduction band occupies an eigenstate whose Bloch-like wave function has a non-zero  amplitude on every atomic site of the alloy [see the left side of Eq. (\ref{five})]. This perspective gained popularity among practitioners because its successes were detailed  in the first two research monographs devoted exclusively to metal physics: {\it The Theory of Metals} (1936) by Alan Wilson and {\it The Theory of the Properties of Metals and Alloys} (1936) by Nevill Mott and Harry Jones.

\begin{center}
\includegraphics[scale=.3]{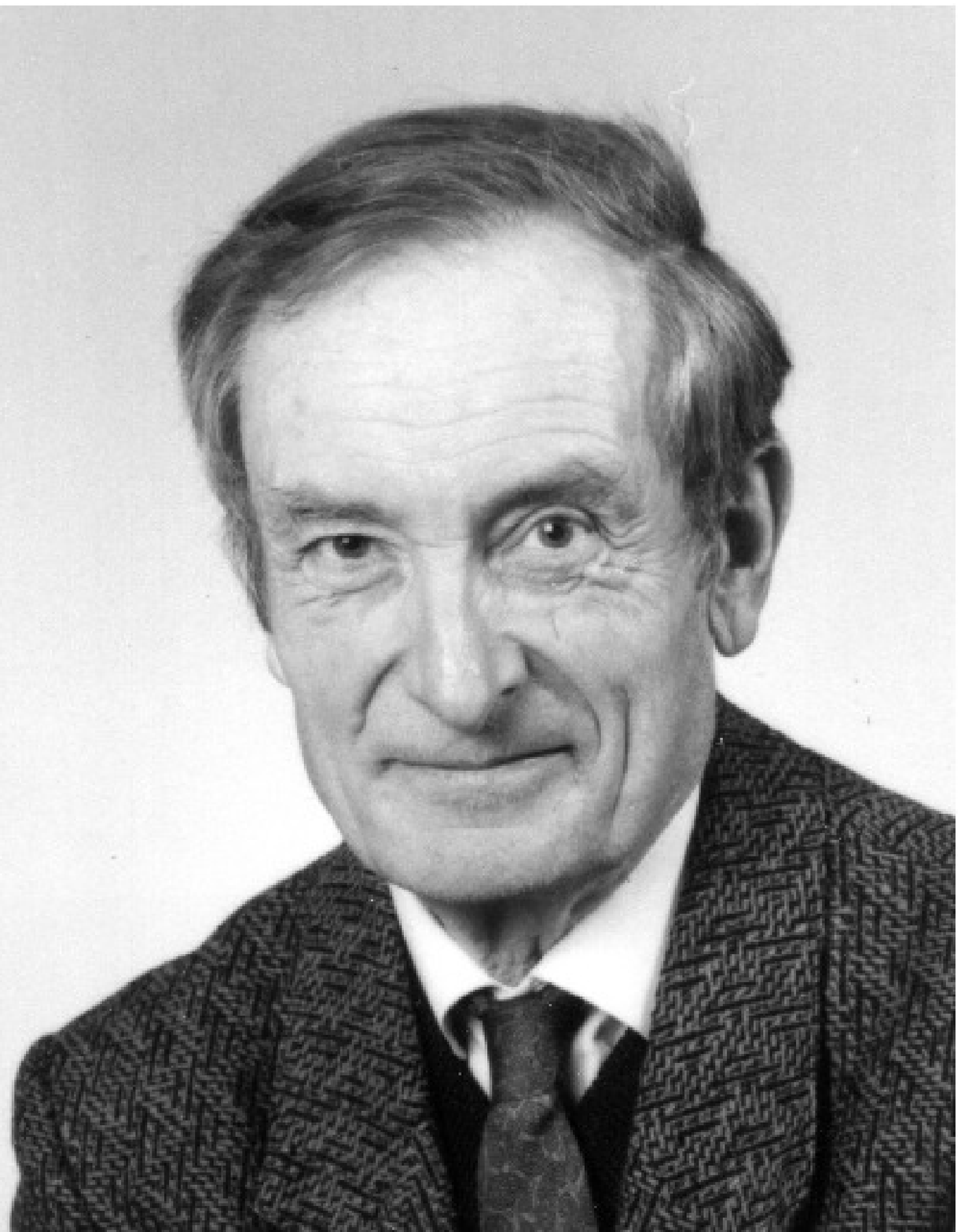}

Jacques Friedel pioneered a spatially local des-\\
cription of the electronic structure of alloys.\\
Courtesy of the AIP Emilio Segr\`{e} Visual Archive.
\end{center}

A rather  different, `localized' point of view was developed by Jacques Friedel (1954). He considered an alloy where $\Delta Z_V$ is the valence difference between the solute A-type atoms and the solvent B-type atoms and recognized that each A-type atom with its valence electrons removed amounts to a point-like impurity with charge $\Delta Z_V$ with respect to the host crystal. The screening of this impurity [discussed earlier in connection with the work of Langer and Vosko (1959)] by the entire sea of conduction electrons implies that electronic charge density accumulates in the immediate vicinity of the A-type atoms. The de-localized  electrons of the host scatter from these local charge accumulations  and Friedel used perturbation theory to show that the energy shift of each scattering state (with respect to the Fermi energy) was the same as the energy shift predicted by the rigid-band model. He went on to use this `local' perspective to  rationalize other experimental trends summarized in Hume-Rothery's book. An even more localized, covalent-bond approach to metals was advocated by Linus Pauling (1949).

The semi-quantitative nature of all existing theories of alloy electronic structure was criticized by John Slater at an October 1955 conference devoted to the theory of alloy phases.\,\footnote{This is the same John Slater whose book {\it Introduction to Chemical Physics} was purchased and read by Walter Kohn while he awaited release from internment in wartime Canada. See Section II.}  Speaking to an audience of physicists, physical chemists, metallurgists, and crystallographers, Slater opened the conference by remarking that (Slater 1955)
\begin{quote}
The metallurgist expects the physicist to be able to apply wave mechanics to the problem of the cohesive energy of metals. By this one means the energy of the metal as a function of the positions of the nuclei. . . . Unfortunately, the errors in our present calculations of the energies of the isolated atoms and of the atoms combined into a metallic crystal are both considerably larger than the energy difference between the two, which is the cohesive energy which we hope to find. . . . This theory is not yet in a position to make calculations of the accuracy which the metallurgists need and which they have been led to believe that they have been getting. Metallurgists have been understandably anxious to get real guidance from physicists regarding their problems. A few papers written by theoreticians have led them to think that this guidance could be actually given in a quantitatively satisfactory form. For instance, a large literature has grown up as a result of papers [published] in the middle 1930's by Jones . . . and Pauling has discussed metallic cohesion and ferromagnetism using methods that seem simple and quantitative. I wish to state my very firm opinion that these theories, as far as they pretend to be quantitative, are based on approximations which are not really justified. They may have qualitative truth in them, but they do not represent quantitative conclusions firmly based on fundamental theory.
\end{quote}
It is perhaps unsurprising that Slater's research group at the time was exploring the quantitative accuracy of his own  `augmented plane wave' method for band structure calculations (Slater 1953).

In Paris in the fall of 1963, Walter Kohn was in an excellent position to learn the latest developments in both the localized and de-localized approaches to alloy theory. The preceding January, Jacques Friedel and Andr\'{e} Guinier had completed editing {\it Metallic Solid Solutions}, a book which documented the proceedings of an international symposium on the electronic and atomic structure of alloys held in Orsay in July 1962.\,\footnote{The unedited proceedings of this conference were also published in the October 1962 issue of {\it Journal de Physique et le Radium}. Reading them may have been the trigger for Kohn to change the subject of his Paris research from the electron-phonon interaction to alloy physics.} In one invited paper, Stanley Raimes of Imperial College noted that Fermi surface measurements for the noble metals invalidated the nearly-free electron assumption used by Jones (1934), but did not quash the rigid-band approximation itself  (Raimes 1963). Similarly, an invited paper by Frank Blatt of Michigan State University asserted that a broad range of transport measurements showed  that ``the rigid-band model is a good approximation to the electronic structure of dilute alloys'' (Blatt 1963). On the other hand, Blatt continued,  ``it is difficult to overstate the importance'' of Friedel's local screening model for the interpretation of not only resistivity and thermoelectric  data for alloys, but also for data obtained from measurements of impurity diffusion,  positron annihilation, and the Knight shift.'' Friedel himself reported an extension of his previous work to the case of transition-metal atom impurities where screening occurs by the occupation of atomic-like orbitals spread out in energy into `virtual bound state resonances' (Friedel 1963).

The most general conclusion to be drawn from the papers collected in {\it Metallic Solid Solutions} was that some observations  were best understood assuming that the conduction electrons of an alloy are delocalized through the volume of the crystal  while other observations were best understood assuming that the most relevant electrons are localized in the immediate vicinity of the solute atoms. At least two prominent metallurgists regarded these  points of view as complementary rather than contradictory. In the $4^{\rm th}$ edition of their {\it Structure of Metals and Alloys}, William Hume-Rothery and Geoffrey Vincent Raynor write (Hume-Rothery and Raynor 1962)
\begin{quote}
The covalency interpretation and the Brillouin zone picture each express a part of the truth. In the case of the diamond structure, for example, the covalency theory . . . gives the more correct picture of the probable cloud density of valence electrons in the crystal. This concept by itself ignores the fact that electrons are free to move in the crystal, and this freedom is emphasized by the zone theories, which in their turn ignore the variation of the electron-cloud density in space. Both concepts are required to express the whole truth.
\end{quote}
The complementarity expressed by this paragraph was well-known to many solid-state physicists, and particularly so to Walter Kohn who had published a fundamental paper on the relationship between the delocalized Bloch functions and the localized Wannier functions (Kohn 1959a).  Kohn had also kept up with band structure calculations and could well appreciate the point John Slater had made about the absence of quantitative calculations in alloy theory (Callaway and Kohn 1962).

Kohn has written about his 1963 survey of the alloy literature on several occasions. His 1990 account, which is closest in time to the actual events,  recalls  the ``rough and ready'' rigid-band model and  then recounts a calculation of the total energy of a disordered alloy  (Hohenberg, Kohn, and Sham 1990). This is interesting because the total energy of a real metal was not a focus of research at the time. Indeed, the only paper in {\it Metallic Solid Solutions} concerned  with total energy begins with the statement (Cohen 1963),
\begin{quote}
In the early history of the theory of metals, calculation of the cohesive energy was a central concern. Apart from very considerable development of the theory of the electron gas, recent effort, both experimental and theoretical, has been focused primarily on properties of one-electron character. In the present work, we have returned to the study of the cohesive energy of metals.
\end{quote}
The remainder of this paper, written by Walter's friend Morrel Cohen from the University of Chicago, displays an  exact formula for the total energy of a  uniform (constant density) electron gas due to  Nozi\`{e}res and Pines  and then generalizes it to the case of an arbitrary ``non-uniform system.'' Both formulae involve an integral over a parameter $\lambda$ with the  quantities in the integrand computed assuming a charge on the electron of $\lambda e$. The limits of the integral extend from $\lambda=0$ (the non-interacting electron gas) to $\lambda=1$ (the fully interacting electron gas).

In his 1990 reconstruction, Kohn wrote a similar integral to compute  $\Delta E$, the {\it difference} in total energy between a real ${\rm A}_x{\rm B}_{1-x}$ alloy and the  same alloy in the virtual crystal approximation. If  $Z^{\rm A}$ and $Z^{\rm B}$ are the atomic numbers of the  A-type and B-type atoms, every ion in the virtual crystal has nuclear charge $\bar{Z}=xZ^{\rm A}+(1-x)Z^{\rm B}$. Then, replacing  $\Delta Z=Z^{\rm A}-Z^{\rm B}$ by $\lambda \Delta Z$, Kohn evaluated the integral of $dE/d\lambda$ from $\lambda=0$ to $\lambda=1$ to second order in $\Delta Z$ (keeping $\bar{Z}$ fixed). The resulting expression for  $\Delta E$  depended on two quantities only: the electron density distribution of the virtual crystal,  $\bar{n}({\bf r})$, and the electron density distribution of the real alloy,  $n({\bf r})$. At this point, ``the question occurred to Kohn whether a knowledge of $n({\bf r})$ alone determined--at least in principle--the total energy'' (Hohenberg, Kohn, and Sham 1990). In other words, could the many qualitative successes of the Friedel point of view,  which put great emphasis on the space-varying electronic charge density inside an alloy,  be elevated to show that the exact charge density uniquely predicts the exact total energy?\,\footnote{In later recollections, Kohn emphasizes the concept of charge transfer in alloys. His Nobel Prize autobiography reports that he read some  ``metallurgical literature in which the concept of an effective charge $e^\ast$ of an atom in an alloy was prominent, which characterized in a rough way the transfer of charge between atomic cells'' (Kohn, 1998). His Nobel Prize lecture notes similarly that   ``there is a transfer of charge between . . . unit cells on account of their chemical differences. The electrostatic interaction energies of these charges is an important part of the total energy. Thus in considering the energetics of this system there was a natural emphasis [in the literature] on the electron density distribution $n({\bf r})$" (Kohn 1999). Despite these remarks, I have been unable to find any significant discussion of charge transfer or ``effective charge'' in the pre-1963 literature of metallurgy or metal physics. Indeed, if they mention this type of charge transfer at all, review articles of the period consistently refer to the same two papers---one by  Nevill Mott (1937) which concerns ordered alloys and one by John Henry Oliver Varley (1954) which demonstrates that the electrostatic energy associated with charge transfer is negligible in  disordered alloys. On the other hand, less than  ten years after the events narrated here, charge transfer became an important variable in two proposed theories of binary alloy formation (Hodges and Stott 1972, Miedema, de Boer, and de Chatel 1973). I have found no literature of the time that emphasizes the spatially-varying electron density distribution $n({\bf r})$ in alloys except in the most qualitative terms (see the passage by Hume-Rothery and Raynor quoted earlier in this section).}

It is important to understand the revolutionary nature of this question.
Notwithstanding the  work of Cohen mentioned above, most solid state theorists in the early 1960's thought about the total energy of a solid  along the lines laid out by Frederick Seitz in his widely admired textbook, {\it The Modern Theory of Solids} (1940). One starts  with the  nuclear charge $eZ_k$ and the fixed position ${\bf R}_k$ of each of the $M$ nuclei in the system.  The Coulomb potential energy of interaction between pairs of nuclei is a classical quantity which poses no problem to compute. The remaining energy terms all contribute to the Schr\"{o}dinger equation for the $N$-electron wave function:  the  kinetic energy of the electrons, the Coulomb potential energy of interaction between every pair of electrons, and the  Coulomb potential energy of interaction between every electron and every nucleus.

It will be convenient to define $v({\bf r})$ as the potential energy of interaction between an electron at the point ${\bf r}$ and all the fixed nuclei:
 \begin{equation}
\label{six}
v({\bf r})=-e^2\sum\limits_{k=1}^M {Z_k \over |{\bf r}-{\bf R}_k|}.
\end{equation}
This potential energy appears in the Schr\"{o}dinger equation, as do the operators for the electron kinetic energy and the electron-electron potential energy,
\begin{equation}
\label{seven}
\hat{T}=- {\hbar^2 \over 2m}\sum\limits_{k=1}^N \nabla_k^2~~~~{\rm and}~~~~\hat{U}= {1\over 2}\sum\limits_{k=1}^N \sum\limits_{m=1}^N {e^2 \over |{\bf r}_k -{\bf r}_m|}.
\end{equation}
The Schr\"{o}dinger equation determines the  $N$-electron wave function, $\psi({\bf r}_1, {\bf r}_2, \dots, {\bf r}_N)$, and then the electron density distribution,
\begin{equation}
\label{eight}
n({\bf r})=\int \psi^\ast({\bf r}, {\bf r}_2, \dots, {\bf r}_N)\, \psi({\bf r}, {\bf r}_2, \dots, {\bf r}_N)\, d{\bf r}_2 \cdots d{\bf r}_N.
\end{equation}
If we omit the classical ion-ion energy, which does not involve the electrons, the total energy is
\begin{equation}
\label{nine}
E = \int n({\bf r})\,v({\bf r})\, d{\bf r} + \langle \psi |\hat{T} |\psi  \rangle + \langle \psi |\hat{U}|\psi \rangle,
\end{equation}
where the last two terms are expressed as averages (expectation values) with respect to the ground state wave function.

The conventional perspective outlined in the previous paragraph shows that the external potential, $v({\bf r})$, is the only term in the Schr\"{o}dinger equation  which distinguishes one alloy from another. Therefore, $v({\bf r})$ determines the wave function from the Schr\"{o}dinger equation,  which in turn  determines the  electron density $n({\bf r})$ from Eq.~(\ref{eight}) and the total energy $E$ from Eq.~(\ref{nine}). From this point of view, the energy amounts to a {\it functional} of the external potential. Of course, the accuracy of the computed energy depends on the quality of the choice made for the form of the $N$-electron wave function  used to solve the Schr\"{o}dinger equation.

Kohn now contemplated a radical inversion of this procedure.  Was it possible that the total energy depended only on the electron  density $n({\bf r})$?  That is, could the total energy be a functional of the density alone? If so, knowledge of $n({\bf r})$ was sufficient to determine (implicitly) the external potential, the $N$-particle wave function, and all ground state properties, including the Green functions of  many-body theory!  This was a very deep question. Walter realized he wasn't doing alloy theory anymore (Kohn 2012b).

\section{The work of Hohenberg and Kohn}

Kohn's proposition that the total energy of an electron system was a functional of the density seemed preposterous on the face of it. How could knowledge of the electron density, which is a function of three Cartesian variables, be sufficient to compute the total energy when the last two terms in Eq.~(\ref{nine}) depend explicitly on the many-electron wave function, which is a function of $3N$ Cartesian variables? Looking for support, Walter asked himself if he knew  {\it any} examples where complete knowledge of $n({\bf r})$ implied complete knowledge of $v({\bf r})$ (Hohenberg, Kohn, and Sham 1990). One such example was the elementary Schr\"{o}dinger equation,
\begin{equation}
\label{eleven}
-{\hbar^2 \over 2m}\nabla^2 \psi+ \left [v({\bf r}) -E\right ] \psi({\bf r})=0.
\end{equation} \vspace{0.05em}

\noindent One can always choose $\psi({\bf r})$ real for this equation. Therefore, the electron density is $n({\bf r})=\psi^2({\bf r})$ and the inversion we seek is a matter of algebra:
\begin{equation}
\label{twelve}
v({\bf r})=E+{\hbar^2 \over 2m}{\nabla^2 \sqrt{n({\bf r})} \over \sqrt{n({\bf r})}}.
\end{equation}

Unfortunately, Eq.~(\ref{eleven}) applies only to a one-particle system and thus does not shed any light on the many-body problem. A more relevant  example known to Kohn was
the Thomas-Fermi method, a semi-classical but self-consistent approximation to the quantum theory of a many-electron system with a non-uniform density $n({\bf r})$ (March 1975, Zangwill 2013). Nevill Mott had used this method to calculate the  screening of a point charge by the conduction electrons in a metal. For the problem considered here, Thomas-Fermi theory method shows that $n({\bf r})$ determines  $v({\bf r})$ through the relation\,\footnote{The coefficient of the kinetic energy term in Eq.~(\ref{fifteen}) includes the electron mass $m$ and Planck's constant $h$.}
\begin{equation}
\label{fifteen}
 v({\bf r})=E-{1\over 2m}\left({3h^3\over 8\pi}\right)^{\hspace{-.25em} 2/3}\hspace{-1em} n^{2/3}({\bf r}) - e^2\int d{\bf r'} {n({\bf r'})\over |{\bf r}-{\bf r'}|}. \vspace{0.15in}
\end{equation}

Encouraged by these examples, Kohn sought to  prove  that {\it the ground state electron density $n({\bf r})$ uniquely determines the external potential $v({\bf r})$.} Such a simple result (if true) could not  depend on the details of the many-electron wave function. Therefore, he looked for a very general idea to exploit and found it with  the Rayleigh-Ritz variational principle  (Hohenberg, Kohn, and Sham 1990). The reader will recall that this was the first tool  Walter had placed in his theoretical toolbox while still a student at the University of Toronto. The {\it reductio ad absurdum} proof is ``disarmingly simple'' (Parr and Yang 1989).

Assume that a Hamiltonian (energy) operator $\hat{H}=\hat{v}+\hat{T}+\hat{U}$ produces a ground state energy $E_0$, a  ground state wave function $\psi({\bf r}_1, \dots, {\bf r}_N)$ and an electron density $n({\bf r})$. Contrary to what we wish to prove, assume that another Hamiltonian  $\hat{H'}=\hat{v}'+\hat{T}+\hat{U}$ produces a  ground state energy $E'_0$, a ground state  wave function $\psi'({\bf r}_1, \dots, {\bf r}_N)$, but the {\it same} electron density $n({\bf r})$.  The idea is to use $\psi'$ as a trial function in the variational principle for $\hat{H}$ in Eq.~(\ref{one}). Ignoring the possibility of degeneracy, this gives the strict inequality\,\footnote{Later analysis showed that removing the assumption of no degeneracy does not invalidate the final result  (Parr and Yang 1989).}
\begin{widetext}
\begin{equation}
\label{sixteen}
E_0 < \langle \psi' | \hat{H} | \psi' \rangle = \langle \psi' | \hat{H}' |\psi' \rangle + \langle \psi' |\hat{H}-\hat{H}' |\psi' \rangle =E'_0 + \int n({\bf r}) [v({\bf r})-v'({\bf r})] \, d{\bf r}.
\end{equation}
\end{widetext}
Similarly, we may use  $\psi$ as the trial function in the variational principle for $H'$. The result is the same as Eq.~(\ref{sixteen}) with the primed and unprimed variables exchanged:
\begin{widetext}
\begin{equation}
\label{seventeen}
E'_0 < \langle \psi | \hat{H}' | \psi \rangle = \langle \psi | \hat{H} |\psi \rangle + \langle \psi |\hat{H}'-\hat{H} |\psi \rangle =E_0 + \int n({\bf r}) [v'({\bf r})-v({\bf r})] \, d{\bf r}.
\end{equation}
\end{widetext}
Adding Eq.~(\ref{sixteen}) to Eq.~(\ref{seventeen}) yields $E_0 + E'_0 < E'_0 + E_0$, which is impossible. Hence, our assumption that two external potentials correspond to the same electron density is false.

\begin{center}
\includegraphics[scale=.65]{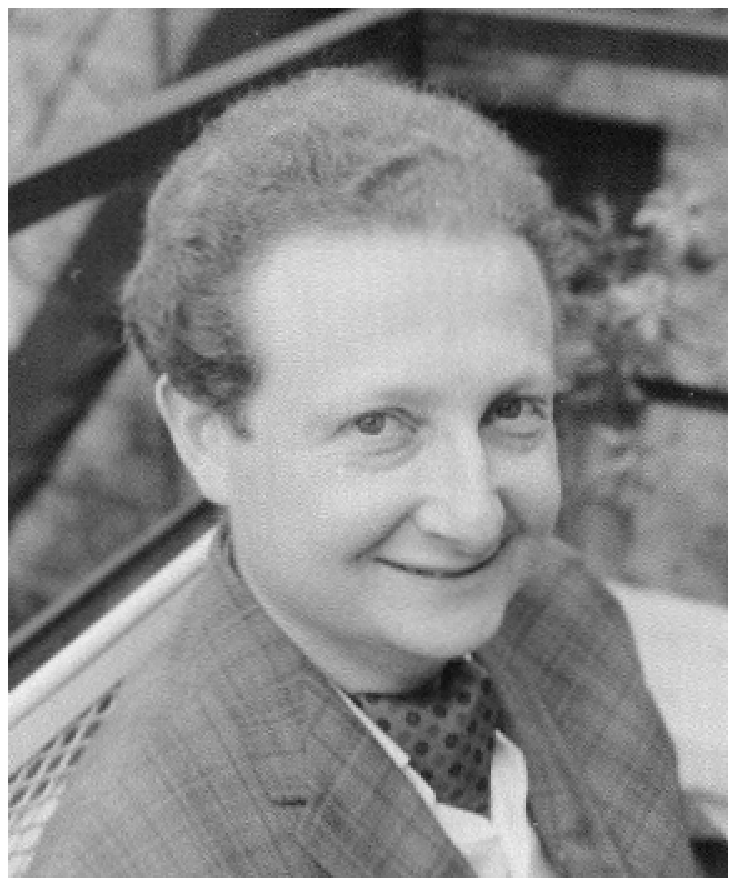}

Pierre Hohenberg in 1965, soon after his work  \\
with Kohn on  density functional theory.  \\ Courtesy of Pierre Hohenberg.
\end{center}

Kohn was exhilarated by his proof, but ``it seemed such a remarkable result that I did not trust myself'' (Kohn 1998). He looked around for help and found it in the person of  Pierre Hohenberg, the recently arrived post-doctoral fellow who was also ensconced in Philippe Nozi\`{e}res' enormous office. Hohenberg had just completed a post-doctoral year doing many-body theory with Alexei Abrikosov and Lev Gor'kov in Moscow, but  was having some trouble getting the attention of Nozi\`{e}res, his  new post-doctoral supervisor (Hohenberg 2012). Walter proposed that they work together and Pierre agreed. Like Kohn's previous collaborators, Res Jost and Quin Luttinger, Hohenberg was
  a rather formal theoretical physicist. He had been trained at Harvard by Paul Martin and his 1962 PhD thesis, ``Excitations in a dilute condensed Bose gas'', used density matrix and Green function methods to provide microscopic justification for phenomenological theories that had been offered  by Lev Landau and others to explain the properties of superfluid helium.

Hohenberg's first task was a literature search to discover if the theorem Kohn had proved was already known. Apparently not. However, it had been known for about a decade that a computation of  the total energy of a system of nuclei and electrons did not really require complete knowledge of the entire $N$-particle wave function (L\"{o}wdin 1959, McWeeny 1960, Coleman 1963). It was sufficient to specify a quantity derived from the wave function called the {\it second-order density matrix}.\,\footnote{Another necessary quantity, the first-order density matrix, is derivable from the second-order density matrix.}
In principle, there exists a variational principle for the energy where the second-order density matrix is varied rather than the  wave function.  Unfortunately, it was not known what properties the density matrix had to possess to ensure that it was  derivable from an $N$-particle wave function. This became known as the `$N$-representability problem'.

 In a similar way, Kohn's theorem reminded Hohenberg  of ``formal work on stationary entropy and renormalization which had just been completed by Paul Martin and Cyrano de Dominicis, also working together in Paris'' (Hohenberg, Kohn, and Sham 1990). These authors  had studied the grand partition function of an arbitrary many-body system and used the mathematical technique of Legendre transformations to effect the desired  `renormalizations'. The latter eliminated the functional dependence  on any one-body potential (like the external potential above) in favor of functional dependence on a one-particle distribution function. It also eliminated the functional dependence on any two-particle potential (like the Coulomb interaction between pairs of electrons) in favor of functional dependence on a  two particle distribution function (de Dominicis 1963, de Dominicis and Martin 1964). Fifty years later, Hohenberg and Kohn differ slightly in their recollection of whether Legendre transformations played any role in their work together. Kohn recalls that he and Hohenberg recognized that his theorem could be interpreted as a Legendre transformation only after the fact and notes that the final published paper makes no mention of it (Kohn 2013b). Hohenberg (2012) recalls that
 \begin{quote}
we were certainly thinking in the language of Legendre transformations, but we did not need that idea in the end. It was characteristic of Walter's style to introduce [in print] only those theoretical ideas needed to solve the problem at hand.
\end{quote}

Martin had returned to Harvard, but de Dominicis worked at the Centre d'Etudes Nucl\'{e}aires in nearby Saclay. According to Hohenberg, ``it took a number of intense but informative discussions with de Dominicis and his colleague Roger Balian to convince them that the procedure worked with the density rather than the distribution function (Hohenberg, Kohn, and Sham 1990). Philippe Nozi\`{e}res finally joined the conversation when he learned of the larger theoretical issue that now engaged Hohenberg and Kohn. As he later recalled (Nozi\`{e}res 2012b),
\begin{quote}
The three of us discussed it a lot. But I was not fully convinced. In my opinion, putting all the emphasis on the density did not account properly for exchange and correlations. I did not share the enthusiasm of Walter and Pierre and I stayed aside.
\end{quote}
Accordingly, Hohenberg and Kohn proceeded on their own to the natural next step: a reformulation
of the Rayleigh-Ritz variational principle in terms of the density rather than the many-body wave function.

Kohn's theorem implies that the many-body wave function is a functional of the ground state electron density $n({\bf r})$. The same is true of  the exact kinetic energy $T[n]$ and the exact electron-electron potential energy $U[n]$. Therefore, if
\begin{equation}
\label{exactT}
F[n]=T[n] + U[n]
\end{equation}
is the sum of these two, the total energy in Eq.~(\ref{nine}) is
\begin{equation}
\label{eighteen}
E[n] = \int n({\bf r})\,v({\bf r})\, d{\bf r} + F[n].
\end{equation}\vspace{0.15em}

\noindent Now let $n_T({\bf r})\ge 0$ be a {\it trial density} which produces the correct total number of electrons, $N=\int n_T({\bf r})\, d{\bf r}$. By Kohn's theorem, this density determines its own  external potential, Hamiltonian, and ground state  wave function $\psi_T$. In that case, the usual wave function variational principle expressed by Eq.~(\ref{one}) tells us that
\begin{widetext}
\begin{equation}
\label{nineteen}
\langle \psi_T |\hat{H} | \psi_T \rangle = \int n_T({\bf r})\,v({\bf r})\, d{\bf r} + F[n_T]=E[n_T] \ge E[n].
\end{equation}
\end{widetext}
The variational principle in Eq.~(\ref{nineteen}) establishes that the energy in Eq.~(\ref{eighteen}) evaluated with the true ground state density is minimal with respect to all other density functions for the same number of particles.

The proofs given just above appear early in Hohenberg and Kohn (1964), the first foundational paper of density functional theory.  In connection with Eq.~(\ref{nineteen}), Hohenberg and Kohn (HK) use a  footnote to warn the reader of a possible `v-representability problem' because ``we cannot prove whether an arbitrary positive definite distribution $n({\bf r})$ which satisfies $\int n({\bf r})\, d{\bf r}=N$ can be realized for {\it some} external potential.'' They  express confidence that all ``except some pathological distributions'' will have this property but the mere fact that they draw attention to this point demonstrates how carefully the authors looked for holes in their arguments. \,\footnote{The  term `v-representability problem' was coined in 1975. See Parr and Yang (1989) and Kryachko and Lude$\tilde{\rm n}$a (1990) for extensive discussion of this issue.}

HK make a point to remark that the functional $F[n]$ in Eq.~(\ref{exactT}) is ``universal'' in the sense that it is valid for any number of particles and any external potential. Even more significant for later work, they define yet another functional $G[n]$ by extracting from $F[n]$ the classical Coulomb energy of a system with charge density $en({\bf r})$. Doing this renders Eq.~(\ref{eighteen}) in the form
\begin{widetext}
\begin{equation}
\label{twenty}
E[n] = \int n({\bf r})\,v({\bf r})\, d{\bf r} + {e^2\over 2}\int
\int {n({\bf r})\, n({\bf r'}) \over
|{\bf r}-{\bf r'}|}\,d{\bf r}\, d{\bf r'} + G[n].
\end{equation}
\end{widetext}
The functional $G[n]$ in Eq.~(\ref{twenty}) includes the exact kinetic energy of the electron system and  all the potential energy associated with the electron-electron interaction that is {\it not} already counted by the classical electrostatic energy. Needless to say, HK could not write down $G[n]$. Doing so would imply that they had solved the entire many-electron problem.

The approach Hohenberg and Kohn took to analyze $G[n]$ reflects the way Kohn chose to frame the final published paper. There is no mention of the alloy problem or even of any desire to re-formulate the electronic structure problem for solids.  Instead, the title of the HK paper is simply ``Inhomogeneous electron gas'' and the first line of the abstract announces that ``this paper deals with the ground state of an interacting electron gas in an external potential $v({\bf r})$.'' The Introduction goes on to note that ``during the last decade there has been considerable progress in understanding the properties of a homogeneous, interacting electron gas.''A footnote refers the reader to  David Pines' book {\it Elementary Excitation in Solids} (1963) for the details. HK then remind the reader about the Thomas-Fermi method,
\begin{quote}
in which the electronic density $n({\bf r})$ plays a central role. . . This approach has been useful, up to now, for simple though crude descriptions of inhomogeneous systems like atoms and impurities in metals. Lately, there have been some important advances along this second line of approach. . . . The present paper represents a contribution in the same area.
\end{quote}

HK  confirmed that the Thomas-Fermi model of electronic structure follows from  Eq.~(\ref{twenty}) by using an expression for  $G[n]$ which accounts approximately for the kinetic energy of the electrons but takes no account of the non-classical electron-electron potential energy.  Specifically,  the kinetic energy density at a  point ${\bf r}$ of the real system is set equal to the kinetic energy density of an infinite gas of {\it non-interacting} electrons with a uniform density $n=n({\bf r})$. The latter is computed using elementary statistical mechanics and one finds (March 1975, Zangwill 2013)
\begin{equation}
\label{twentyonehalf}
G_{\rm TF}[n]= {3 \over 10m}\left({3h^3\over 8\pi}\right)^{2/3}\int n^{5/3}({\bf r})\, d{\bf r}.
\end{equation}
After inserting  Eq.~(\ref{twentyonehalf}) into Eq.~(\ref{twenty}) to produce $E_{\rm TF}[n]$, the variational principle in Eq.~(\ref{nineteen}) directs us to minimize $E_{\rm TF}[n]$ with respect to density. This is done using Lagrange's method to ensure that the total particle number, $N=\int n({\bf r})\, d{\bf r}$, remains constant. The final result is the Thomas-Fermi expression in Eq.~(\ref{fifteen}).

The Thomas-Fermi model never went out of fashion as a quick and easy way to gain qualitative information about atoms, molecules, solids, and plasmas. By 1957,  the British solid state physicist Norman March was able to publish a 100-page review article surveying the successes and failures of the model and its generalizations  (March 1957). Earlier, I  labeled 1957 as the  `magic year' when many-body Green functions and diagrammatic perturbation theory transformed the study of many-electron systems. Therefore, it is not surprising that several physicists--including Kohn's old post-doctoral colleague Sidney Borowitz--applied these methods with the aim to systematically generalize the Thomas-Fermi model to include the effects of electron correlation and spatial inhomogeneity  (Baraff and Borowitz 1961, DuBois and Kivelson 1962). These papers are among the  ``important advances'' noted by HK in the passage quoted just above. \,\footnote{It is curious that Hohenberg and Kohn do not refer to papers by Hubbard (1958), Bellemans and de Leener (1961), and Jones, March, and Sampanthar (1962), all of whom studied the energy of an electron gas in the presence of a lattice of positive charges using methods superior to the Thomas-Fermi approximation.}

As trained solid-state physicists, HK knew that the entire history of research on the quantum mechanical many-electron problem could be interpreted as attempts to identify and quantify the physical effects described by $G[n]$. For example,
many years of approximate quantum mechanical calculations for atoms and molecules had established that the phenomenon of
{\it exchange}---a consequence of the Pauli exclusion principle---contributes significantly to the potential energy part of  $G[n]$. Exchange reduces the Coulomb potential energy of the system by tending to keep electrons with parallel spin spatially separated. The remaining potential energy part of $G[n]$ takes account of short-range {\it correlation} effects. Correlation also reduces the Coulomb potential  energy by tending to keep {\it all} pairs of electrons spatially separated. The effect of correlation is largest for electrons with anti-parallel spins because these pairs are not kept apart at all by the exchange interaction.  I note for future reference that the venerable Hartree-Fock approximation takes account of the  kinetic energy and  the exchange energy exactly but (by definition) takes no account  of the correlation energy (Seitz 1940, L\"{o}wdin 1959, Slater 1963).

HK devoted the remainder of their time together to studying $G[n]$ for two cases: (i) an interacting electron gas with a nearly constant  density; and (ii) an interacting electron gas with a slowly-varying  density. For the nearly constant density case, HK wrote  $n({\bf r}) = n_0 + \tilde{n}({\bf r})$ with  $\tilde{n}({\bf r})/n_0 \ll 1$ and pointed out that $G[n]$ admits a formal expansion in powers of $\tilde{n}({\bf r})$:
\begin{equation}
\label{twentytwo}
G[n]= G[n_0] + \int K(|{\bf r}-{\bf r'}|) \tilde{n}({\bf r})\tilde{n}({\bf r'}) \, d{\bf r} \,d{\bf r'}  + \cdots
\end{equation}
$K(|{\bf r}-{\bf r'}|)$ is a  linear  response function for the uniform and interacting electron gas which had been studied intensely by experts in many-body theory  (Pines 1963). In particular, the derivative of its  Fourier transform $\tilde{K}(q)$ was known to  diverge at a certain value of $q$ (Pines 1963).  Walter knew this divergence well: it was responsible for the oscillatory algebraic form of the Friedel density disturbance formula in Eq.~(\ref{fivehalf}).  It was also  responsible for the ``Kohn anomalies'' in the phonon spectrum of metals. HK point this out and remark in passing  that  ``the density oscillations in atoms which correspond to shell structure . . . are of the same general origin.''

 The divergence in $\tilde{K}(q)$ disappears if one retains only a finite number of terms in its power series expansion. The  corresponding expansion of $K(|{\bf r}-{\bf r'}|)$ reduces Eq.~(\ref{twentytwo}) to  a {\it gradient expansion},  \begin{equation}
 \label{nodiverge}
 G[n]=G[n_0] +a\int \tilde{n}({\bf r})\, d{\bf r} + b\int |\nabla \tilde{n}({\bf r})|^2 \, d{\bf r} + \cdots,
 \end{equation}
 where $a$ and $b$ are constants related to  $\tilde{K}(q)$. Therefore, as HK  pointed out, any `quantum oscillations'  produced by the divergence in $\tilde{K}(q)$ cannot be captured by any low-order gradient expansion of $G[n]$ like Eq.~(\ref{nodiverge}). This explained the failures of the many generalizations of the Thomas-Fermi approximation which involved adding gradient terms.

Walter and Pierre took some time off in the first week of December 1963 to attend an All-Union Conference on Solid State Theory in Moscow.  Kohn joined John Bardeen and  George Vineyard (Brookhaven National Laboratory) as the only senior Americans in attendance. Hohenberg had left Moscow in July and he relished the opportunity to visit his old friends there. The topics discussed at the meeting were superconductivity, the theory of metals, semiconductors and dielectrics, lasers, the magnetic properties of rare earths, neutron scattering, the M\"{o}ssbauer effect, phonons,  dislocation theory, and the theory of radiation damage in crystals (Agranovich 1964).

Upon their return to Paris, Hohenberg and Kohn focused on $G[n]$ for a system with a slowly-varying density function. This assumption precluded variations with short (spatial) wavelengths, but allowed for the possibility of substantial variations in the overall magnitude of the density. For this case, the appropriate gradient-type expansion is
\begin{equation}
\label{twentythree}
G[n] = \int g_0(n({\bf r})) \, d{\bf r} + \int g_2(n({\bf r}))\, |\nabla n({\bf r})|^2\, d{\bf r} + \cdots,
\end{equation}
where $g_0$ and $g_2$ are functions (not functionals) of $n({\bf r})$. By specializing  Eq.~(\ref{twentythree}) to the previously studied case of a nearly uniform electron gas, HK were able to express these functions in terms of the properties of the uniform and interacting electron gas. For example, $g_0(n)$ is the sum of the kinetic energy density, the exchange energy density, and the correlation energy density for an interacting electron gas with uniform density $n$. With this information, HK performed a partial (infinite) summation of the entire  gradient expansion in Eq.~(\ref{twentythree}). Their final expression had the great virtue of recovering the singularity in $\tilde{K}(q)$ needed to describe quantum oscillations.

Kohn left Paris in January 1964 and visited physicists in London, Cambridge, and Oxford before returning to California. He wrote up a first draft of a manuscript and sent it to Hohenberg for his  review. Pierre made several suggestions, all of which were incorporated into the final version (WKP 1964). The paper was sent off to the {\it Physical Review} in the second week of June 1964 and published in the November 9 issue.  It is notable  that the  `Concluding Remarks' section does not remind the reader of the two theorem's proved at the beginning of the paper. Instead the authors remark that  they ``have developed a theory of the electronic ground state which is exact in two limiting cases.'' The importance Hohenberg and Kohn ascribed to the ability of their theory to capture quantum oscillations may be judged from the fact that their paper ends with the statement  that  `the most promising formulation of the theory . . . appears to be that obtained by partial summation of the gradient expansion''. They say this despite warning the reader in the previous sentence that ``actual electronic systems'' have neither nearly constant densities nor slowly-varying densities, {\it i.e.}, the situations where the partial summation was expected to be valid.

Any reader of Hohenberg and Kohn (1964)  cannot help but be struck by its  understated and rather formal tone. The Introduction is succinct,  the basic theorems are proved quickly,  most of the paper is taken up with the gradient expansions, and no applications are discussed or proposed. Earlier, I quoted Kohn to the effect that he understood that the basic theorem which drives the paper was ``remarkable''.  If so, he and Hohenberg made a conscious decision not to  ballyhoo the truly revolutionary idea at its core: that the ground state electron density, in principle, determines all the properties of an electronic system. That being said, we learn from their 1990 reminiscence that the authors spent at least a little time talking about applications (Hohenberg, Kohn, and Sham 1990):
\begin{quote}
The question arose as to what the method might be good for, and Kohn suggested that one could try using it to improve current techniques for calculating the band structure of solids. Hohenberg's immediate reaction was to say, ``But band structure calculations are horribly complicated, isn't that the sort of stuff better left to professionals?'' To this, Kohn simply replied, ``Young man, I am the Kohn of Kohn and Rostoker (1954) !''
\end{quote}

A recommendation from Kohn helped Hohenberg win a job at Bell Telephone laboratories in the fall of 1964. Recently, he recalled that ``I gave a talk on my Paris work during my first few months at Bell Labs. Phil Anderson, Bill McMillan, and Phil Platzman were in the audience and there was no enthusiasm. They correctly understood that our results would not help them solve the difficult many-body problems they were struggling with" (Hohenberg 2012). In the event, Hohenberg turned to hydrodynamics and phase transitions as topics for research and worked productively in those areas over a thirty-year career at Bell. He was elected to the US National Academy of Sciences in 1989 and moved to Yale University in 1995 to accept the position of Deputy Provost for Science and Technology. Since 2004, he has served at New York  University as a professor of physics and Vice Provost for Research.

\section{The Work of Kohn and Sham}

Kohn returned to La Jolla in early 1964 ``to find a group of postdocs and visitors eagerly awaiting a first-hand account of the work'' (Hohenberg, Kohn, and Sham 1990). According to one of those post-docs, Vittorio Celli (2013),
\begin{quote}
Walter gave a full departmental colloquium rather than just a technical seminar after his return from Paris. Keith Brueckner (who had returned from Washington and was Dean at the time)  said that colloquia were usually reserved for ``foreign stars'' but that today we have ``our own star'' to give a talk. I remember thinking that the theory with Hohenberg was cute but would not have many consequences. I certainly did not think it compared in significance with the many-body calculations for the electron gas that had been obtained by Brueckner and his collaborators.
\end{quote}
Kohn proposed to another post-doc, N. David Mermin, that he exploit his knowledge of statistical mechanics to generalize the Hohenberg-Kohn results to non-zero temperature. Mermin realized almost immediately that ``a strange variational principle for the free energy that I had formulated for  an utterly unrelated purpose'' was perfectly suited for the job. As a result,  ``it took me less than an hour to check that the HK proof went through'' almost without change (Mermin 2004). Kohn was skeptical, but the simplicity of the proof could not be assailed. On the other hand, the significance of Mermin's work was not very clear and he was disinclined to write it up. It took six months, his imminent departure from La Jolla, and  Kohn's insistence that ``it someday might be important'' for the manuscript of Mermin (1965) to get written (Mermin 2013).

Kohn's colloquium featuring the Hohenberg-Kohn theorems generated skepticism from several of the many-body theorists who were in La Jolla to work with Keith Brueckner.  Most notably, ``Nobuyuki Fukuda,  visiting from the University of Tokyo, constructed counter-example after counter-example purporting to demonstrate the non-uniqueness of the potential given a density distribution. The job of resolving these fell to Lu Jeu Sham'' (Hohenberg, Kohn, and Sham 1990).

Lu Sham was the graduating PhD student to whom Kohn had written requesting that he begin his post-doctoral fellowship at UCSD while Kohn was still in Paris. Sham, a native of Hong Kong, arrived in San Diego by way of Imperial College and Cambridge University where he had earned, respectively, an undergraduate degree in Mathematics and a graduate degree in physics. His 1963 PhD thesis, ``The electron-phonon interaction'', was supervised by John Ziman, a distinguished theorist whose books, {\it Electrons and Phonons} (1960) and {\it Principles of the Theory of Solids} (1964), helped train a generation of solid-state physicists. In La Jolla, Sham used the months before Kohn returned to write a paper on the phonon spectrum of sodium metal.  This quantitative calculation used a pseudo-potential for the sodium ionic potential and a self-consistent modification of the Hartree-Fock method to take approximate account of correlation effects which act to screen (reduce) the exchange interaction (Sham 1965).

\begin{center}
\includegraphics[scale=1.6]{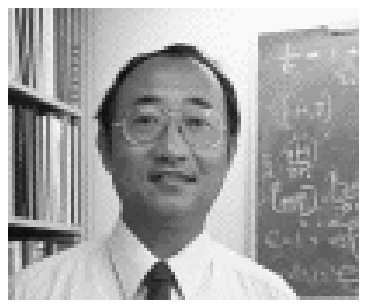}

Lu-Jeu Sham (circa early 1980's) was a  \\
post-doc with Walter Kohn from 1963-1966. \\ Courtesy of Lu-Jeu Sham.
\end{center}

Quantum density oscillations were much on Walter's mind when he related the details of his Paris experience  to Sham. He was acutely  aware that no oscillations would  emerge from Eq.~(\ref{nodiverge}), the most natural choice for $G[n]$ when the external potential was slowly-varying function of position. Therefore, in the same letter where Walter informed Pierre Hohenberg that the HK manuscript had been submitted for publication, he related that ``Lu Sham and I have started looking at situations like a heavy atom where one has a localized and rapidly-varying potential'' (WKP 1964).
In principle, their goal was to develop a general theory of quantum density oscillations for use in situations where the electron density is strongly non-uniform. In practice, they developed a method to find the leading quantum corrections to the Thomas-Fermi electron density for a collection of non-interacting electrons moving in a one-dimensional potential.  The published paper, Kohn and Sham (1965a), reports the results of an elegant Green function calculation which related the electron density (which did exhibit the desired quantum oscillations) to the potential  in a way which generalized Eq.~(\ref{fifteen}).  Unfortunately, the extension of their method to three-dimensional periodic potentials presented a daunting numerical challenge which did not lend itself to practical calculations for real solids. Accordingly, Kohn and Sham dropped this line of investigation and moved in a different direction, albeit one still motivated by the basic results obtained by Hohenberg and Kohn.\,\footnote{The Web of Science database (accessed January 2014) lists 230 citations to the quantum density oscillations paper,  Kohn and Sham (1965a). A review of these citations shows that the vast majority of the citing papers $(>90\%)$ seem  unaware of its actual content. They incorrectly  cite it along with Hohenberg and Kohn (1964) and Kohn and Sham (1965b) as one of the foundational papers of density functional theory.}

By the late fall of 1964,  Kohn was thinking about alternative ways to transform the theory he and Hohenberg had developed into a practical scheme for atomic, molecular, and solid state calculations.  Happily, he was very well acquainted with an approximate approach to the many-electron problem that was notably superior to the Thomas-Fermi method, at least for the case of atoms. This was a theory proposed by Douglas Hartree in 1923 which exploited the then just-published Schr\"{o}dinger equation in a heuristic way to calculate the orbital wave functions $\phi_k({\bf r})$, the electron binding energies $\epsilon_k$, and the charge density $n({\bf r})$ of  an $N$-electron atom (Park 2009, Zangwill 2013). Hartree's theory transcended Thomas-Fermi theory primarily by its use of the exact quantum-mechanical expression for the  kinetic energy of independent  electrons. The {\it Hartree equations} which define the theory for an atom with nuclear charge $Z=N$ are
\begin{equation}
\label{H1}
-{\hbar^2 \over 2m}\nabla^2 \phi_k+ \left [v_{\rm eff}({\bf r}) -\epsilon_k\right] \phi_k({\bf r})=0,~~~~~k=1,\dots N,
\end{equation}
where
\begin{equation}
\label{H2}
v_{\rm eff}({\bf r})= v({\bf r})+e^2\int d{\bf r'} {n({\bf r'})\over |{\bf r}-{\bf r'}|}
\end{equation}
and $v({\bf r})=-Ze^2/r$.  The electron density $n({\bf r})$
 in Eq.~(\ref{H2}) is calculated assuming that the electrons occupy thet $N$ lowest energy eigenfunctions of the Schr\"{o}dinger equation in Eq.~(\ref{H1}). Therefore, if $k=1,2,\dots, N$ labels these $N$ lowest energy orbitals,
\begin{equation}
\label{H3}
n({\bf r})=\sum_{k=1}^N |\phi_k({\bf r})|^2.
\end{equation}
The effective potential energy function in Eq.~(\ref{H2}) shows that every electron interacts with the charge of the  `external' nucleus and with the charge of the entire atomic electron cloud taken as a whole. \,\footnote{In the actual equations written by Hartree (1923), each electron feels the  classical electrostatic potential produced by every electron except itself. Therefore,  in contrast to Eq.~(\ref{H2}), each electron in Hartree's theory feels a slightly different electrostatic potential.} Hartree stressed that these equations must be solved {\it self-consistently}. That is,
an iterative numerical method is required to  ensure that the $\phi_k({\bf r})$ generated by Eq.~(\ref{H1}) are the same as the $\phi_k({\bf r})$ used  in Eq.~(\ref{H3}) to construct
the particle density $n({\bf r})$.

Slater (1930) and Fock (1930) had provided a rigorous derivation of the Hartree equations. They used the variational principle in Eq.~(\ref{one}) and  evaluated the total energy using a many-electron wave function of the form $\psi({\bf r}_1, {\bf r}_2, \dots, {\bf r}_N)=\phi_1({\bf r}_1)\phi_2({\bf r}_2)\cdots \phi_N({\bf r}_N)$. Minimizing this energy with respect to different choices for the $\phi_k$ functions generates the Hartree equations.  Slater and Fock also  evaluated the total energy using a more sophisticated many-body  wave function (called a Slater determinant) which combines the same $N$ orbital functions $\phi_k$ in such a way that the Pauli exclusion principle is obeyed automatically.  With this choice, minimizing the total energy with respect to choices for the $\phi_k$ functions generates what are called the  Hartree-Fock equations. Hartree-Fock theory is superior to Hartree theory because  the kinetic energy {\it and} the exchange energy are treated exactly. Unfortunately, the Hartree-Fock equations are significantly harder to solve  than the Hartree equations.

Kohn suggested to Sham that he try to derive the Hartree equations from the Hohenberg-Kohn formalism.
Walter had good reason to believe this could be done (Kohn 2001a, Kohn 2012b). On the one hand, his work with Hohenberg had established  the central role of the electron density $n({\bf r})$ for a complete description of any electronic system. On the other hand, the  Hartree equations could be read as a self-consistent scheme to deduce an approximate expression for  $n({\bf r})$. Therefore, it should be possible to derive the Hartree equations as an {\it example} of the HK variational principle. Specifically, the variational minimization of some approximate form of the total energy functional $E[n]$ should lead to Hartree's equations. Sham set to work with enthusiasm. His state of mind at that particular moment was noted by Philip Taylor, a fellow former-graduate student of John Ziman's at Cambridge University. Taylor had taken a job at the Case Institute of Technology in Cleveland, Ohio and had recently published a quantitative analysis of Kohn anomalies in the phonon spectrum of metals (Taylor 1963). It was during a visit to La Jolla to consult with Kohn and visit with Sham that Sham remarked to Taylor that he was ``thrilled'' by the research project Kohn had given to him (Taylor 2013).

Kohn and Sham recognized that the Hartree method regards each electron as moving independently in an effective  potential $v_{\rm eff}({\bf r})$ which does not recognize the individual identity of the other electrons. Consistent with this, the kinetic energy implied by Eq.~(\ref{H1}) is correct only for independent and non-interacting electrons. This was the key to progress because the Hohenberg-Kohn analysis implied that the kinetic energy of a strictly non-interacting system of electrons is also a functional of the density. If we call this functional $T_S[n]$, ordinary quantum mechanics specifies
that\,\footnote{The correctness of Eq.~(\ref{H4}) requires that the functions $\phi_i$ and $\phi_j$ be orthonormal, which means that  the integral $\int \phi^*_i({\bf r})\, \phi_j({\bf r}) \, d{\bf r}$ is one when $i=j$ and zero when $i\neq j$.}
\begin{equation}
\label{H4}
T_S[n]=\sum_{k=1}^N \int \phi_k^\ast({\bf r})\,\left[-{\hbar^2 \over 2m}\nabla^2 \right] \phi_k({\bf r}) \, d{\bf r}.
\end{equation}

The path was now open for Sham to derive the Hartree equations from a density functional point of view (Kohn 2001a). He chose the approximate total energy functional
\begin{widetext}
\begin{equation}
\label{H6}
E_H[n] = T_S[n] + \int v({\bf r}) \, n({\bf r}) \, d{\bf r} +  {e^2\over 2}\int
\int {n({\bf r})\, n({\bf r'}) \over
|{\bf r}-{\bf r'}|}\,d{\bf r}\, d{\bf r'},
\end{equation}
\end{widetext}
and minimized it with respect to a density assumed to have the form in Eq.~(\ref{H3}). The latter transforms the density  variation of the first term in Eq.~(\ref{H6}) into variations with respect to the $\phi_k$ functions. The density variation of the remaining terms in Eq.~(\ref{H6}) is straightforward and the final result is exactly the Hartree equations (\ref{H1}) and (\ref{H2}).

Kohn and Sham (KS) now knew how to move forward with the general many-electron problem. Motivated by Eq.~(\ref{H6}),
they {\it defined} a functional $E_{xc}[n]$ by the partition $G[n]=T_S[n] +E_{xc}[n]$. This puts the exact total energy functional in the form
 \begin{equation}
 \label{E2}
 E[n]=E_H[n] + E_{xc}[n]. \vspace{0.4em}
 \end{equation}
 The great virtue of Eq.~(\ref{E2}) is that it has exactly the same structure as Eq.~(\ref{H6}), even if we revert to Eq.~(\ref{six}) for $v({\bf r})$. Therefore,  the interacting electron density $n({\bf r})$ which minimizes the original total energy in Eq.~(\ref{nine}) is precisely equal to the non-interacting electron density $n({\bf r})$ which minimizes  Eq.~(\ref{E2}). Carrying out the latter minimization explicitly produces the  {\it Kohn-Sham equations}, which are identical to the Hartree equations  Eqs.~(\ref{H1}) and (\ref{H3}) with Eq.~(\ref{H2}) replaced by
\begin{equation}
\label{KS1}
v_{\rm eff}({\bf r}) = v({\bf r})+e^2\int d{\bf r'} {n({\bf r'})\over |{\bf r}-{\bf r'}|} + v_{xc}({\bf r})
\end{equation}
where
\begin{equation}
\label{KS2}
v_{xc}({\bf r})={\delta E_{xc}[n]\over \delta n({\bf r})}.
\end{equation}
The exchange-correlation potential energy, $v_{xc}({\bf r})$, obtained in Eq.~(\ref{KS2}) from the functional derivative of $E_{xc}[n]$, is a  function of ${\bf r}$ and not a functional of $n({\bf r})$. Therefore, according to KS, a numerical procedure no more difficult that Hartree's original method is sufficient to compute the  ground state electron density and thus the ground state total energy of an arbitrary many-electron system subject to an external potential. If $E_{xc}[n]$ was known exactly, one could calculate $n({\bf r})$ and $E[n]$ exactly as well.

The  ``exchange-correlation'' energy functional $E_{xc}[n]$ in Eq.~(\ref{E2}) is similar to  $G[n]$ in Eq.~(\ref{twenty}) in the sense that it  accounts for all the energy associated with the Coulomb interaction between electrons {\it not} already counted by the classical Coulomb self-energy. However, while  $G[n]$ had also to account for the total kinetic energy of the real interacting electron system [called $T[n]$ in Eq.~(\ref{exactT})], $E_{xc}[n]$ has only to account for the {\it difference} between the  kinetic energy of an interacting electron system and the kinetic energy of  a non-interacting electron system with exactly  the same density $n({\bf r})$. Of course, the exact and universal functional $E_{xc}[n]$ is no better known that $G[n]$ for the interacting electron problem.

Unlike the Slater-Fock methodology sketched earlier, the foregoing derivation of the Hartree-like Kohn-Sham equations {\it does not introduce a many-electron wave function at any stage.} Instead, Kohn and Sham  replace the true interacting electron system with a non-interacting electron reference system which has  exactly the same ground state electron density. The wave function of the reference system is unambiguously Hartree-like, so it is correct to use the exact kinetic energy for non-interacting electrons in Eq.~(\ref{H4}) and represent the density function as in Eq.~(\ref{H3}). On the other hand, the eigenfunctions $\phi_k({\bf r})$ and the eigenvalues $\epsilon_k$ in Eq.~(\ref{H1}) have {\it no} direct physical meaning for the true interacting electron system.

Finally, Kohn and Sham proposed an approximation for $E_{xc}[n]$ which has come to be known as the  {\it local density approximation} (LDA). Namely,
\begin{equation}
\label{KS3}
E_{xc}[n]=\int n({\bf r})\, \epsilon_{xc}(n({\bf r}))\, d{\bf r},
\end{equation}
where $\epsilon_{xc}(n)$ is the exchange and correlation energy per electron of a fully interacting electron gas with uniform  density $n$. Hohenberg and Kohn had introduced a similar approximation for $G[n]$ in their paper and it was reasonable for Kohn and Sham to ``regard $\epsilon_{xc}(n)$ as known from theories of the homogeneous electron gas.''\,\footnote{The local density approximation for $G[n]$ retains only the first term on the right hand side of Eq.~(\ref{twentythree}).}  When Eq.~(\ref{KS3}) is used for $E_{xc}[n]$, the exchange-correlation potential in Eq.~(\ref{KS2}) becomes
\begin{equation}
\label{KS4}
v_{xc}({\bf r}) = {d\over dn}\left[n\epsilon_{xc}(n)\right].
\end{equation}


The foregoing results were reported in a short manuscript, ``Exchange and correlation effects in an inhomogeneous gas'' which Kohn and Sham submitted to {\it Physical Review Letters} in May of 1965.
Samuel Goudsmit, one of the editors of {\it Physical Review Letters} at that time, informed Sham by letter that the Kohn-Sham manuscript ``deserves publication as an Article in the {\it Physical Review}, but it is not of such urgency to warrant speedy publication in {\it Physical Review Letters}'' (Goudsmit 1965). The authors responded by withdrawing the manuscript and, three weeks later, submitted to the {\it Physical Review} a longer and more detailed paper  with a new title, ``Self-consistent equations including exchange and correlation effects''. The published version, Kohn and Sham (1965b) is  the second foundational paper of density functional theory. It is also one of the most highly cited papers in the history of physics.\,\footnote{The Web of Science  database (accessed January  2014) lists  21,372 citations to Kohn and Sham (1965b). Only four physics-related papers have more citations and all four of them are density functional papers which owe their existence to the Kohn-Sham paper.} Interestingly, it was only at the page-proof stage of the longer paper  that the authors realized that their Hartree-like equations with Eq.~(\ref{KS2}) represented a formally exact statement of the complete many-body problem (Sham 2014). For that reason,  Eq.~(\ref{KS2}) appears only in a ``Note Added in Proof'' while Eq.~(\ref{KS4}) appears in the main exposition.

Kohn and Sham knew it was straightforward to use Eq.~(\ref{KS4}) and write down an explicit and analytic formula for $v_{xc}({\bf r})$ and incorporate it seamlessly  into existing computer programs to calculate the electronic structure of atoms, molecules, and solids. Briefly, the separation   $\epsilon_{xc}(n)=\epsilon_x(n) + \epsilon_c(n)$ known for the interacting and uniform electron gas implies that the exchange-correlation potential  Eq.~(\ref{KS4}) separates similarly into $ v_{xc}({\bf r})=v_x({\bf r}) + v_c({\bf r})$.
The exact exchange energy density,  $\epsilon_x(n)$,  had been calculated years earlier by Dirac (1930) for the purpose of improving the Thomas-Fermi approximation. Using Dirac's formula, KS reported their result for the LDA exchange potential:\,\footnote{Kohn and Sham were unaware that the Hungarian physicist  Rezs\"{o} G\'{a}sp\'{a}r had derived Eq.~(\ref{x}) ten years earlier by similarly computing the variational derivative of Dirac's exchange energy with a local density approximation (G\'{a}sp\'{a}r 1954).}
  \begin{equation}
 \label{x}
 v_{x,{\rm LDA}}({\bf r}) = -e^2\left[{3\over \pi} n({\bf r})\right]^{1/3}.
 \end{equation}
This expression was consequential at the time because, when  Eq.~(\ref{x}) replaces  $v_{xc}({\bf r})$ in Eq.~(\ref{KS1}), the Kohn-Sham equations become almost identical to a set of equations John Slater had proposed in 1951 as a local approximation to the non-local Hartree-Fock equations. I say `almost' identical because the {\it ad hoc} local exchange potential proposed by Slater was
\begin{equation}
\label{Slater}
v_{x,{\rm Slater}}({\bf r})=-{3\over 2}e^2\left[{3\over \pi} n({\bf r})\right]^{1/3}.
\end{equation}
KS argue for the correctness of their proposed exchange potential  and it is notable that the abstract of Kohn and Sham (1965b) devotes a sentence to announcing the factor of $3/2$ difference between Eqs.~(\ref{x}) and (\ref{Slater}). The authors' motivation to do this was surely their awareness that the  so-called `Hartree-Fock-Slater' method was in  wide use by physicists performing band structure calculations for real solids (Callaway 1958, Herman 1964).

The Kohn-Sham equations [with and without the local density approximation for $v_{xc}({\bf r})$] are the reason for the enduring importance of Kohn and Sham (1965b). The paper itself differs in tone from Hohenberg and Kohn (1964) in the sense that the abstract notion of an ``inhomogeneous electron gas'' disappears from the title and from most of the text. Instead, there is the practical promise of  ``self-consistent equations'' appropriate to ``real systems (atoms, molecules, solids, etc.) [where] the electronic density is nonuniform." The Introduction is even more specific and makes the point that
\begin{quote}
most theoretical many-body studies have been concerned with elementary excitations and as a result there has been little progress in the theory of cohesive energies, elastic constants, etc. of real metals and alloys. The methods proposed here offer the hope of new progress in the latter area.
\end{quote}
That being said, KS did not themselves report any calculations of the cohesive energy or elastic constants (or any other measurable quantity) for any `real' electronic system. Indeed, they did not even bother to write down an explicit form for the correlation part of  $v_{xc}({\bf r})$ in the local density approximation. This was a straightforward exercise for anyone familiar with the electron gas literature. As for the LDA itself, KS remark that it should ``give a good representation of exchange and correlation effects . . for metals, alloys, and small-gap insulators.'' On the other hand, they warn the reader that the LDA should have  ``no validity [at] the `surface' of atoms and the overlap regions of molecules. . . We do not expect an accurate description of chemical binding.''

Walter left for his annual visit to Bell Labs after the June 1965 submission of the longer  Kohn-Sham manuscript. He collaborated with Quin Luttinger as usual and their efforts produced a prediction for a new mechanism for superconductivity based  on a presumed  oscillatory interaction between pairs of electrons (Kohn and Luttinger 1965).  Meanwhile, back  in La Jolla, Lu Sham began work on two density functional projects. One of these, which became the  final paper he and Kohn would publish together, examined the one-body Green function of many-body theory (Sham and Kohn 1966). It was important for them to study this quantity because its properties determine the energy, lifetime,  and spatial extent of single-particle-like excitations out of the ground state of a many-body system. At the same time, the  Hohenberg-Kohn theory implies that the Green function is as  a functional of the ground state electron density.

Sham, who did most of the calculations, demanded that the Green function satisfy the requirements of particle-number conservation and charge neutrality and deduced thereby  that an electron at a point ${\bf r}$ in an atom, molecule, or solid  responds to the  electrostatic potential at that point and  to exchange and correlation effects which depend  on the electron density distribution in the immediate vicinity of ${\bf r}$ only. For a slowly-varying density, this conclusion justifies a local density approximation for the Green function, which in turn provides an approximation for the energy spectrum and an independent justification for using the LDA with the Kohn-Sham equations to calculate the ground state electron density $n({\bf r})$.

Sham's second post-Kohn-Sham density functional project was done in collaboration with Bok Yin Tong, a thirty-year old graduate student who had begun to work with Kohn. Their goal was to solve the Kohn-Sham equations numerically for several atoms and ions.   Luckily for them, Frank Herman and Sherwood Skillman had just published a computer program which solved the Hartee-Fock-Slater equations for atoms (Herman and Skillman 1963). Tong and Sham needed only to replace Eq.~(\ref{Slater}) by Eq.~(\ref{x}) in the program and add some code  for the correlation part of $v_{xc}({\bf r})$ in the LDA. For this they used an interpolation formula for the correlation energy derived from the information given in Pines (1963). The published paper, Tong and Sham (1966), focused on total energies, total energy differences, and charge densities. The final results with correlation omitted were gratifying,  giving ``slightly better results for energies and substantially better results for densities that Slater's method." The correlation correction worsened the results,  ``presumably because the electronic density in atoms has too rapid a spatial variation.''

Lu Sham published three articles unrelated to density functional theory before beginning an Assistant Professorship at the  University of California at Irvine  in the fall of 1966. Two years later, he accepted an offer to return to La Jolla as an Associate Professor. He was promoted to Professor in 1975, served as a Dean from 1985-1989, and is currently Emeritus Professor of Physics. He was elected to the US National Academy of Sciences in 1998.  At UCSD, Sham developed a broad research program  with a particular expertise in the theory of the electronic and optical properties of semiconductor heterostructures.   However, the twenty papers he published on density functional theory over the years show that he never completely abandoned the main subject of his post-doctoral work.

Walter Kohn remained at San Diego until 1979, when he accepted the position as founding Director of the Institute for Theoretical Physics (ITP), a research facility established and supported by the US National Science Foundation at the University of California at Santa Barbara. He served as  Director for five years and continued as a Professor of Physics at UCSB until 1991 when he gained Emeritus status. Kohn and his collaborators published over 150 papers between 1965 and 2006. One third of these explore some aspect of density functional theory, particulary its application to solid surfaces. An equal number of papers concern Kohn's pre-DFT interests including disordered states of matter, superconductivity, Bloch and Wannier functions, scattering theory, and the transition between the conducting and insulating states of matter. Kohn was elected to the National Academy of Sciences (1969) and was a recipient of a National Medal of Science (1998) before winning a share of the 1998 Nobel Prize in Chemistry.

\section{Discussion and Conclusion}
In March 2001, Kohn addressed a symposium on ``The History of the Electronic Structure of Atoms, Molecules, and Solids'' at a meeting of the American Physical Society. With an audience well-schooled to interpret the lower case Greek letter $\epsilon$ as an infinitesimally small quantity, Walter was understood immediately when he characterized the initial reception of
 density functional theory as ``$+\,\epsilon$ by  theoretical physicists and zero by theoretical chemists'' (Kohn 2001a).
The small reaction by physicists reflected the fact that Kohn's theory did not directly address the ``big'' issues that occupied  many solid-state physicists at the time: superconductivity, the Kondo effect, superfluid helium, disorder-induced localization, the metal-insulator transition, and quasi-one-dimensional conductors.  The positive response was limited mostly to  band structure theorists who were well-positioned to carry out the numerical work needed to solve the Kohn-Sham equations for real materials.\,\footnote{Despite the warning in Sham and Kohn (1966) that the $\epsilon_k$ parameters should not be interpreted as one-electron energies, ``the temptation to use [them] as band structures in solids proved irresistible'' (Hohenberg, Kohn and Sham 1990).} The lack of response (or negative response) from chemists came mostly from their near-universal belief that no theory of electronic structure based on the particle density alone could possibly be correct. In a future publication, I will trace the evolving response of both the physics and chemistry communities to DFT. For the present,  it suffices to sketch very briefly the path which led from virtually no response by chemists to a Nobel Prize in Chemistry.

An important point mentioned earlier is that a computer program written to solve the local exchange potential equations of the Hartree-Fock-Slater (HFS) method (Slater 1951) was easily adapted to solve the Kohn-Sham equations in the local density approximation. Therefore, after the 1965 publication of the Kohn-Sham paper, systematic calculations for atoms began to reveal that Eq.~(\ref{x}) was superior to Eq.~(\ref{Slater}) as a local approximation to the exact, non-local exchange potential (Herman, Van Dyke, and Oretenburger 1969). HFS and LDA calculations for molecules and solids were more difficult to evaluate because computational exigencies encouraged the use of a `muffin-tin approximation'' where the effective potential in Eq.~(\ref{KS1}) was replaced by its spherical average inside a set of touching spheres centered at the atoms. A constant potential was used outside the spheres.

 Beginning in the 1970's, a small group  of scientists  committed themselves to carrying out HFS and LDA calculations for small molecules without imposing the muffin-tin (or any other) constraint on the self-consistent potentials (Baerends, Ellis, and Ros 1973, Gunnarsson, Harris, and Jones 1977, Becke 1982).  Over time, HFS calculations disappeared in favor of LDA calculations and the basic conclusion was that one obtained reasonable binding energies and predictions for molecular structures that agreed well with experiment (Jones 2012). On the other hand, the LDA was {\it not} capable of the kind of  `chemical accuracy' ($\pm 2~ {\rm kcal/mole}$) that the best {\it ab initio} methods of traditional quantum chemistry could achieve. Then, in the 1980's, efforts to go beyond the local density approximation led to the proposal and testing of various non-local approximations for the exchange and correlation functional (Langreth and Mehl 1983, Perdew 1986, Becke 1988, Lee, Yang, and Parr 1988). These, so-called ``generalized gradient approximations'' (GGA) replaced the electron gas exchange-correlation energy density in Eq.~(\ref{KS3}) with much more complicated functions of both the local density $n({\bf r})$ and the local density gradient $\nabla n({\bf r})$.\,\footnote{The generalized gradient approximations go far beyond the simple gradient expansions analyzed by HK and KS.}
Systematic Kohn-Sham calculations for atoms and molecules using GGA, particulary a hybrid approach introduced by Axel Becke (1993), quickly began to approach chemical accuracy.

A turning point occurred in 1991 at the VII$^{\rm th}$ International Congress of Quantum Chemistry in Menton, France (Kohn 2001a). John Pople gave the final talk and summarized the achievements of `G2' theory, his most comprehensive {\it ab initio} attempt to improve the Hartree-Fock approximation using perturbation theory (Curtiss, Raghavachari, Trucks, and Pople 1991).  On the other hand, earlier in the week,  the Congress had  given its triennial ``outstanding young scientist'' award to Axel Becke for his manifestly non-{\it ab-initio} work with DFT. Specifically,  ``for unique advances in numerical methods in density functional theory as applied to molecules, and for important developments in the understanding of the exchange-correlation functional that enters density functional theory'' (IAQMS 2014).
Pople  made a point in his talk to remark that he found Becke's results ``stimulating and intriguing'' (Pople 1991).

Barely a year later, Pople's group published a systematic comparison of the best quantum chemical calculations with  DFT calculations performed using a variety of exchange-correlation potentials for 32 molecules. They concluded that the most sophisticated non-local functionals ``outperformed correlated {\it ab initio} methods, which are computationally more expensive. Good agreement with experiment was obtained with a small basis set'' (Johnson, Gill, and Pople 1992).  DFT was promptly incorporated into Pople's  widely-used GAUSSIAN computer program and, with this endorsement, the popularity of DFT calculations among chemists  began to grow exponentially (see Fig. 2). Accordingly, when the  Nobel Chemistry Committee decided it was time to honor quantum chemistry with a Prize, it was not difficult for them to split the award between John Pople and Walter Kohn.

At the beginning of this paper, I suggested that density functional theory might  be unknown today if Walter Kohn had not created it in the mid-1960's. Such a  counter-factual claim can never be proved. However, it is interesting to examine the  evidence that supports it. We do this with full awareness of a long tradition which  examines great scientific discoveries from the {\it personalistic} and {\it naturalistic} points of view (Boring 1950). The former focuses on the specific attributes of an individual whose ``exceptional insight may lead to an original discovery which has not been anticipated by others and which is relatively independent of the times.'' The latter posits that ``the {\it zeitgeist} (scientific climate of the time) determines the great discovery and that he who makes the discovery is great merely because the times employed him." Here, I  begin with the zeitgeist of electronic structure theory in Kohn's lifetime and then turn to his particularism for the case of DFT.

\begin{center}
\includegraphics[scale=.55]{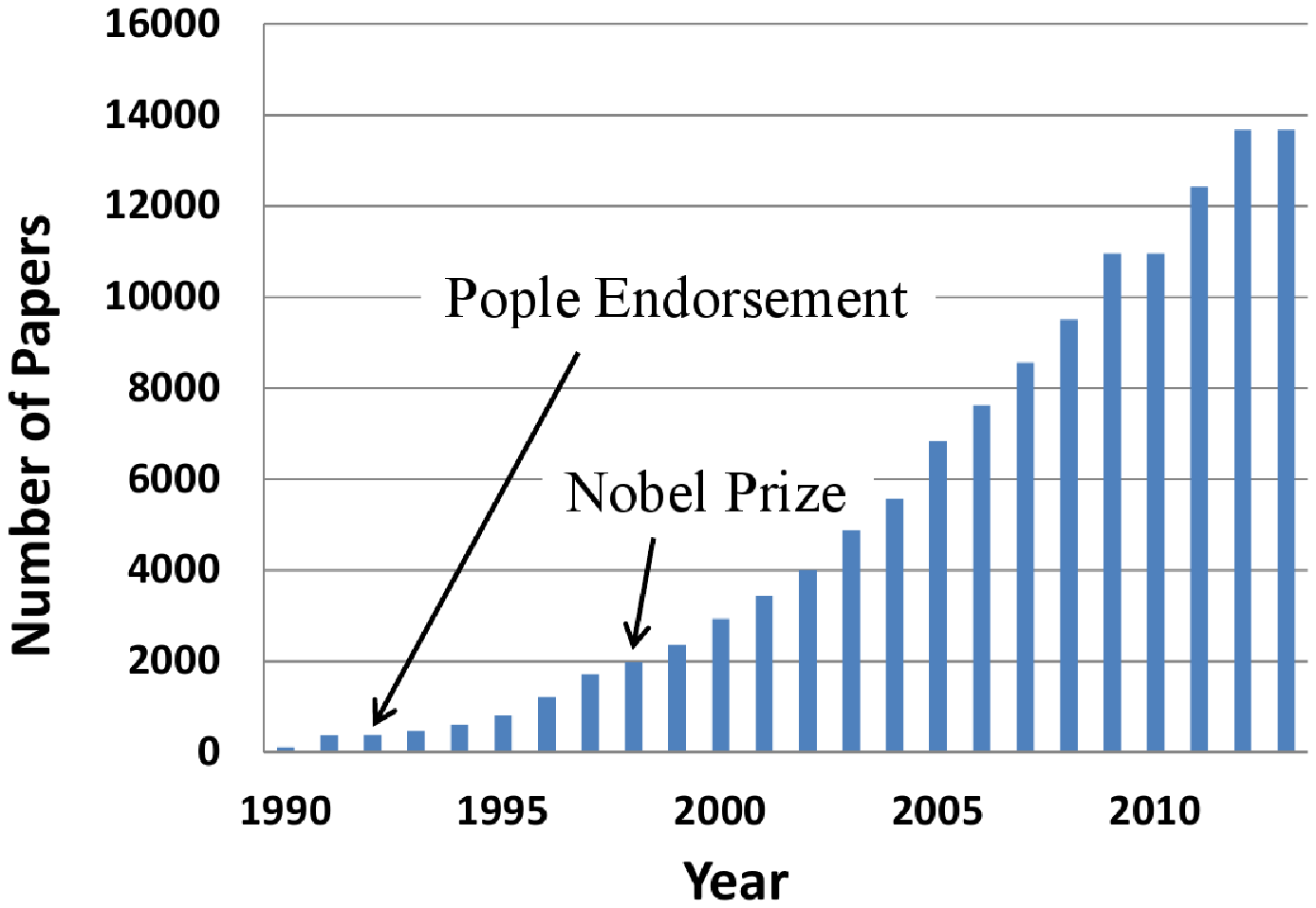}

\small Fig. 2. Numbers of papers that mention \\
       DFT as found by the Web of Science. \\
\end{center}

The modern concept of the electronic structure of atoms, molecules, and solids began when the old quantum theory of Bohr gave way to the new quantum theory of Schr\"{o}dinger (Jammer 1966). From 1925 to 1960, no fewer than 1000 papers in the scientific literature concerned themselves with the  `many-electron' problem.\,\footnote{Data collected from Google Scholar in January 2014. The true number of papers is higher because not all authors used the term `many-electron' in their writing.}  No fewer than  1000 more papers focused on this topic between 1961 and 1965. The abstracts of these papers reveal five principal approaches to this issue:  the Thomas-Fermi method, electron gas models, many-electron wave function methods, density matrix methods, and quantum field theory methods. Of these, only the approximate Thomas-Fermi method singles out the ground state  electron density $n({\bf r})$ as the primary quantity for study.  Attempts to build ``quantum corrections'' into the Thomas-Fermi model were current in the early 1960's, but none of these suggested that the theory could be ``exactified'' to reveal the density as a truly fundamental quantity. A paper inspired by Friedel's alloy work which ignores correlation and derives a formal perturbation series for the density in term of the external potential is perhaps closest in this regard (March and Murray 1961).

In the late 1950's and early 1960's, the ground state electron density was not very interesting to  many  theoretical solid state physicists. They focused instead on the excited states of solids and the powerful new methods of quantum field theory which made their study possible (Hoddeson {\it et al.} 1992). Those who did appreciate the general importance of the charge density---primarily the practitioners of band structure calculations---saw no reason and had no motivation to elevate it to the lofty status of the many-electron wave function (Herman 1958, Pincherle 1960). The latter attitude was  shared by the quantum chemistry community who devoted enormous efforts to solving the Schr\"{o}dinger equation for molecules with greater and greater accuracy  (Barden and Schaefer 2000, Gavroglu and Sim\~{o}es 2012). An interesting exception is  the Canadian Richard Bader, perhaps the  greatest champion of $n({\bf r})$ among theoretical chemists. Just a year before  the publication of Hohenberg and Kohn (1964), he wrote (Bader and Jones 1963):
\begin{quote}
The manner in which the electron density is disposed in a molecule has not received the attention its importance would seem to merit. Unlike the energy of a molecular system, which requires a knowledge of the second-order density matrix for its evaluation, many of the observable properties of a molecule are determined in whole or in part by the simple three-dimensional electron density distribution.
\end{quote}
Despite his fondness for the density, even Bader could not deny the primacy of the second-order density matrix (L\"{o}wdin 1959). This was the rock-solid  quantum chemical view that Hohenberg had discovered in Paris when he reviewed the literature of many-electron theory.

To my knowledge, the only work in the pre-1964 electronic structure literature where the electron density plays a  fundamental role is a one-page paper by the distinguished theoretical chemist, E. Bright Wilson, Jr. (Wilson 1962). Wilson asks the rhetorical question, ``Does there exist some procedure for calculating $n({\bf r})$ [for an $N$-electron system] which avoids altogether the use of $3N$ dimensional space?'' He then uses a device mentioned earlier (see Section IV) and defines $n({\bf r},\lambda)$ to be the exact ground state electron density of a many-electron system where the charge of every electron is taken to be $\lambda e$ rather than $e$. Using just a few lines of calculation, Wilson shows that the total energy $E$ in Eq.~(\ref{nine}) can be written as an integral from $\lambda=0$ to $\lambda=1$ of the sum of the classical Coulomb potentials produced by $n({\bf r},\lambda)$ at the positions of all the nuclei.\,\footnote{According to Musher (1966), Wilson's  method  goes back to Pauli.}

From the foregoing, I conclude that no scientists before Walter Kohn in 1963 were even vaguely thinking about using the ground state charge density as a fundamental quantity from which to build an exact theory of a many-electron system. The idea was not ``in the air'' and all eyes were riveted either on the many-electron wave function, the first- and second-order density matrices, or the Green functions of many-body field theory. The zeitgeist of electronic structure theory was simply not moving in the direction of the electron density function for some particulary well-prepared scientist to exploit and earn the accolades of discovery.

I now turn to Kohn himself. Section~IV detailed how Walter chose to focus his fall 1963 sabbatical leave on a problem that was {\it not} under active investigation by many of his theoretical colleagues. Namely, how might one calculate the electronic structure of a three-dimensional disordered metal alloy, a system with no underlying spatial periodicity?\,\footnote{Early work on this problem published by Korringa (1958) and Beeby (1964) blossomed into a full-scale theory of the electronic structure of disordered alloys in the late 1960's and early 1970's (Ehrenreich and Schwartz 1976).} His study of the metallurgy and metal physics literature inspired him to ask whether the electron density $n({\bf r})$ was sufficient to completely characterize a many-electron system. It must be admitted that Kohn was a product of his scientific milieu  as much as any other electronic structure  theorist working at the time. Therefore, the mere fact that this question came to his mind {\it and he took it seriously} must be regarded as a legitimate `eureka moment' which few are privileged to experience. That being said, I wish to argue further that his particular history,  style of research, and scientific tastes made him unusually well-suited to exploit  this insight and create from it the edifice of density functional theory.

Two aspects of Kohn's pre-college years (surveyed in Section II) bear on the narrow question of his future life as a physicist. First, the cataclysm of the {\it Anschluss} put the budding classics scholar  into contact with Emil Nohel and Victor Sabbath, two high school teachers whose passion  for their subjects converted him to an enthusiastic student of  mathematics and physics.  Second, the camp schools Kohn attended while interned in Canada exposed him to sophisticated one-on-one instruction from professional scientists. A pedagogical experience of this kind is barely imaginable at a conventional high school, then or now.  In a normal setting,  there is little chance that the teen-aged Kohn  would have encountered (much less devoured) books like  Hardy's  {\it A Course in Pure Mathematics} and Slater's  {\it Introduction to  Chemical Physics}.\,\footnote{It is impossible to know how the frightful loss of his home and parents to Nazi terror motivated Kohn to succeed in later life. A statistical study of Viennese children who had similar experiences during World War II and then emigrated to America shows that those who entered the sciences achieved success (by conventional measures)  more than twice as often as native-born American scientists of the same generation. The study quotes several participants who said ``they felt a great responsibility to make the most of their lives because their survival was such a rare and unlikely event.'' (Sonnert and Holton 2006)}

Walter's undergraduate and master's level classroom experiences  at the University of Toronto  were probably typical of first-rate academic institutions at the time. What was not typical was the unusually high calibre of the individuals who  mentored him  and who (because of their own professional interests) repeatedly emphasized variational  principles for both general proofs and for detailed numerical calculations. It is true that all well-trained theoretical physicists learn about variational principles and many use them occasionally in their professional work. However, very few physicists who learn about them as undergraduates go on to work with a doctoral supervisor like Julian Schwinger who attacked almost every problem from a variational point of view, and then write a thesis where variational principles are used (again) both to prove a general result and to obtain numerical results for a specific quantum situation. Fundamental  and numerical variational calculations appear over and over again in Kohn's solid state work at Carnegie Tech and Bell Labs through the 1950's.  It is little wonder that he turned quickly to this powerful tool when he sought to prove the first Hohenberg-Kohn theorem, which states that the many-body wave function and everything calculable from it are functionals of the ground state electron density $n({\bf r})$. The  second Hohenberg-Kohn theorem, which states  that the  total energy  takes its minimum value when $n({\bf r})$ is the exact ground state density, is explicitly a variational result.

The Hohenberg-Kohn paper is austere, elegant, and deep. Like several other papers in his oeuvre, it demonstrates a characteristic of his work that a Kohn-watcher of fifty years tenure summarized in this way (Langer 2003):
\begin{quote}
He always has loved mathematical elegance, but he reserves it for situations where it is truly necessary. His emphasis [is always] on the most important physical questions and the ways in which they could be answered with insight and confidence.
\end{quote}
This observation is interesting and important,  but it does not distinguish Kohn from a number of other theoretical physicists with a taste for proving theorems. However, I believe it is  unlikely that any of them would have had either the interest or the inclination to derive the Kohn-Sham equations and suggest the local density approximation for practical calculations. To make this case, I have surveyed the publications of the most active theoretical solid state physicists working between 1950 and 1980. As a matter of personal taste, three broad  activities engage them: formal calculations and proofs of theorems, analyses of model Hamiltonians, and numerical calculations for specific materials systems. If an individual was active in more than one of these, it was most often the first and second activities or the second and third activities. Kohn is unusual among his peers simply because he followed up a paper which asks and answers a  deep theoretical question with a paper which  constructs a practical tool to perform  calculations for specific systems.

Kohn's own early research history demonstrates a willingness to compute actual numbers for direct comparison with experiment. The Kohn-Sham equations are the vehicle for this activity in the context of electronic structure theory. In practice, Kohn turned over much of the explicit numerical work on DFT  to his students and post-docs, but there is complete agreement among this cohort that he never regarded this activity as a less important part of his group's research. This is the reason that a senior quantum chemist could remark that Kohn is ``the least arrogant of the deep physicists'' in the sense that he does not ``give a lower standing to those parts of physics that deal with the complexities of phenomena governed by known laws'' (Baerends 2003). The `complexities' mentioned here are discovered  only by carrying out numerical computations for specific systems using the `known laws' described by the Kohn-Sham formalism.

In summary, Walter Kohn earned one-half of the 1998 Nobel Prize in Chemistry by asking himself a simple (yet deep) scientific  question about the electronic structure of matter. He answered that question in an elegant and thought-provoking manner and then exploited his result to re-formulate the quantum many-electron problem in a manner which made calculations for real systems computationally cheap and surprisingly accurate. These things may have been achieved by someone else if Kohn and his post-doctoral associates had not done so in the years 1963-1965,  but that person would probably look very much like Walter Kohn himself.

\begin{center}
\includegraphics[scale=.4]{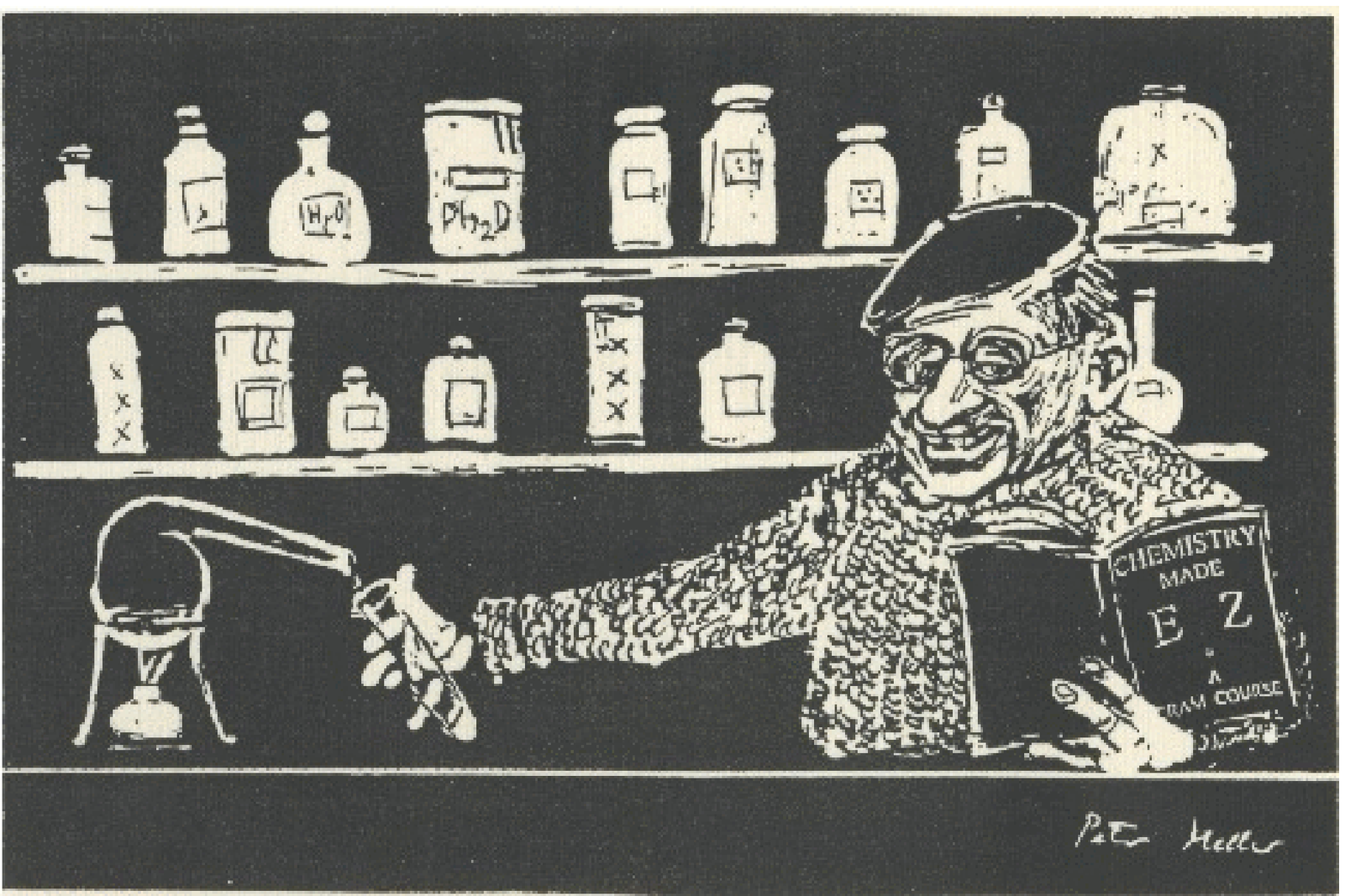}

\vspace{1em}
 The physicist Walter Kohn learns a new trade after \\
winning one-half the 1998 Nobel Prize in Chemistry.  \\ Drawing by Peter Meller. Courtesy of Walter Kohn.
\end{center}

\section{Acknowledgments}
First and foremost, I thank Walter Kohn, who graciously granted me two interviews,
 answered my follow-up questions by email, and gave me access to several private documents and photographs. Prof. Kohn's administrative assistant,  Ms. Chris Seaton,  deserves special mention for her unfailing help. I am also grateful to   Pierre Hohenberg and Lu Jeu Sham, who  shared with me both their memories and their personal photographs. Dozens of people communicated with me about their experiences with Kohn and his work and I collectively thank all of them here.  Ms. Pia Otte provided excellent translations from German into English and I acknowledge Andre Bernard (John Simon Guggenheim Foundation),  Carlo Siochi (University of Toronto), Gregory Giannakis (McGill  University), and Felicity Pors (Niels Bohr Archive)  for sending me copies of original documents. The staffs of the Department of Special Collections of the Library of the University of California at Santa Barbara and of the Vancouver Holocaust Education Centre provided essential help. Special thanks go to my Georgia Tech colleague, Glenn Smith, who read every word and offered many helpful suggestions.

\section{References}


\noindent Abragam, A. 1989. {\it Time Reversal}. Oxford: Clarendon

Press.

\noindent Abrikosov, A.A., L.P. Gorkov, and I.E. Dzyaloshinski.

1963. {\it Methods of Quantum Field Theory in Statistical

Physics} Englewood Cliffs, N.J.: Prentice-Hall.

\noindent Adhikari, S.K. 1998. {\it Variational Principles and the

Numerical Solution of Scattering Problems}. New York:

Wiley.

\noindent Agranovich, V. 1964. At the all-union conference on solid

state theory. {\it Soviet Atomic Energy} 17: 863-865.

\noindent Aharonov, Y. and D. Bohm. 1959. Significance of

electromagnetic potentials in the quantum theory.

{\it Physical Review} 115: 485-491.

\noindent Allin, E.J. 1981. {\it Physics at the University of Toronto

1843-1980}. Toronto: University Press.

\noindent Allis, W.P. and M.A. Herlin. 1952. {\it Thermodynamics

and Statistical Mechanics}. New York:  McGraw Hill.

\noindent Ambegaokar, V. 2013. May 12 2013 telephone interview

with the author.

\noindent Ambegaokar, V. and W. Kohn. 1959. Connection

between true effective mass and optical absorption

in insulators. {\it Physical Review Letters} 2: 385.

\noindent Ambegaokar, V. and W. Kohn. 1960. Electromagnetic

properties of insulators. {\it Physical Review} 117: 423-431.

\noindent Anderson, P.W. 1958. Absence of diffusion in certain

random lattices. {\it Physical Review} 109: 1492-1505.

\noindent Anderson, P.W. 1977. Nobel Prize autobiography.

http://www.nobelprize.org/nobel\textunderscore\ prizes/physics/

laureates/1977/anderson-bio.html

\noindent Anderson, P.W. 1978. Local moments and localized

states. {\it Reviews of Modern Physics}  50: 191-201.

\noindent Anderson, P.W. 1987. John Hasbrouk Van Vleck.

{\it Biographical Memoir} Washington, D.C.: National

Academy of Sciences Press.

\noindent Anderson, P.W. 1999. Interview of Philip Anderson by

Alexei Kojevnikov on May 29 1999.  Niels Bohr Library

and Archives, American Institute of Physics, College

Park, MD, USA,  http://www.aip.org/history/ohilist/

23362\textunderscore\ \!2.html

\noindent Anderson, P.W. 2011. Physics at Bell Labs, 1949-1984.

pp. 73-80 in: {\it More and Different}. Hackensack, New

Jersey: World Scientific.

\noindent Anderson, P.W. 2012. August 7 2012 correspondence

with the author.

\noindent Anderson, P.W., R.M. Friedberg, and W. Kohn. 1997.

Joaquin M. Luttinger. {\it Physics Today} 50 (12) 89-90.

\noindent APS 2013a. American Physical Society. Prizes.

Available at http://www.aps.org/programs/honors/

prizes/index.cfm

\noindent APS 2013b. American Physical Society. Membership

growth, 1899-2013. Available at http://www.aps.org/

membership/statistics/upload/APSMembership

Growth\textunderscore\ \!2013.pdf

\noindent Aronszajn N. and A. Weinstein. 1943. On the unified

theory of eigenvalues of plates and membranes.

{\it American Journal of Mathematics} 64: 623-645.

\noindent Arrott, A. 2013. July 19 2013 telephone interview

with the author and subsequent correspondence.

\noindent Auger, M.F. 2005. {\it Prisoners on the Home Front}.

Vancouver: UBC press.

\noindent Avery, D.H. 1998. {\it The Science of War: Canadian

Scientists and Allied Military Technology during the

Second World War}. Toronto: University Press.

\noindent Bader, R.F.W. and G.A. Jones. 1963. Electron density

distributions in hydride molecules. The ammonia

molecule. {\it Journal of Chemical Physics} 38: 2791-

2802.

\noindent Baerends, E.J. Walter Kohn, the chemist. pp. 20-22 in:

M. Scheffler  and P. Weinberger (eds.) {\it Walter Kohn--

Personal Stories and Anecdotes}. Berlin: Springer.

\noindent Baerends, E.J., Ellis, D.E., and P. Ros. 1973. Self-

consistent molecular Hartree-Fock-Slater calculations

I. the computational procedure. {\it Chemical Physics}

2: 41-51.

\noindent Baraff, G.A. and S. Borowitz. 1961. Green's function

method for quantum corrections to the Thomas-Fermi

model of the atom. {\it Physical Review} 121: 1704-1713.

\noindent Bardeen, J., L.N. Cooper, and J.R. Schrieffer. 1957.

Theory of superconductivity. {\it Physical Review} 108:

1175-1204.

\noindent Barden, C.J. and H.F. Schaeffer III. 2000. Quantum

chemistry in the $21^{\rm st}$ century. {\it Pure and Applied

Chemistry} 72: 1405-1423.

\noindent Bassani, F. and M. Tosi. 1988. Theoretical research

in the physics of solids. pp. 129-136: in G. Giuliani

(ed.) {\it The Origins of Solid State Physics in Italy:

1945-1960}. Bologna: Italian Physical Society.

\noindent Baym, G. 1969. {\it Lectures on Quantum Mechanics}.

Menlo Park: Benjamin Cummings.

\noindent Becke, A.D. 1982. Numerical Hartree-Fock-Slater

calculations on diatomic molecules. {\it Journal of

Chemical Physics} 76: 6037-6045.

\noindent Becke, A.D. 1988. Density functional exchange energy

approximation with correct asymptotic behavior.

{\it Physical Review A} 38: 3098-3100.

\noindent Becke, A.D. 1993. Density functional thermochemistry.

III. The role of exact exchange. {\it Journal of Chemical

Physics} 98: 5648-5652.

\noindent Beeby, J.L. 1964. Electronic structure of alloys.

{\it Physical Review} 135: A130-A143.

\noindent Bellemans A. and M. De Leener. 1961. Ground-state

energy of an electron gas in a lattice of positive point

charges. {\it Physical Review Letters} 6: 603-604.

\noindent Beller, S. 1989. {\it Vienna and the Jews: 1867-1938}.

Cambridge: University Press.

\noindent Berkley, G.E. 1988. {\it Vienna and its Jews}. Cambridge,

MA: Abt Books.

\noindent Blatt, F.J. 1963. Transport properties in dilute alloys.

pp. II-1-11 in: J. Friedel and A. Guinier (eds.)

{\it Metallic Solid Solutions}. New York: Benjamin.

\noindent Bloch, F. 1928.  \"{U}ber die Quantenmechanik der

Elektronen in Kristallgittern. {\it Zeitschrift f\"{u}r Physik}

52: 555-600.

\noindent Bohr, N. 1951. Evaluation of Walter Kohn. Niels Bohr

Archive. Copenhagen.

\noindent Boring, E.G. 1950. Great men and scientific progress.

{\it Proceedings of the American Philosophical Society}

94: 339-351.

\noindent Borowitz, S. and W. Kohn. 1949. On the electro-

magnetic properties of nucleons. {\it Physical Review}

76:818-827.

\noindent Br\'{e}chet, Y. 2008. Presentation of Professor Jacques

Friedel. `Leonardo da Vinci Award of the European

Academy of Sciences'. http://www.eurasc.org/

davinci/davinci2010.asp

\noindent Brockhouse, B.N. and A.T. Stewart. 1958. Normal

modes of aluminum by neutron spectroscopy. {\it Reviews

of Modern Physics} 30: 236-249.

\noindent Brook, A.G. and W.A.E. McBryde. 2007. {\it Historical

Distillates: Chemistry at the University of Toronto

since 1843}. Toronto: Dundurn Group.

\noindent Brown, W.L., R.C. Fletcher, and K.A. Wright. 1953.

Annealing of bombardment damage in germanium:

experimental. {\it Physical Review} 92: 591-596.

\noindent Brueckner, K. 2000. Highlights of many-body physics.

pp. 25-29 in: R.F. Bishop, K.A. Gernoth, N.R. Walet,

and Y. Xian (eds.)  {\it Recent Progress in Many-Body

Theories}. Singapore: World Scientific.

\noindent Brueckner, K. 2013. Keith Brueckner and the

founding of UCSD. {\it Chronicles: Newsletter of the

UCSD Emeriti Association} 12 (4): 6-7.

\noindent Bruch, L.W. 2013. April 10 2013 correspondence with

the author.

\noindent Buckingham, A.D. 2006. Sir John Anthony Pople.

{\it Biographical Memoirs of Members of the Royal Society}

52: 299-314.

\noindent Callaway, J. 1958. Energy bands in solids. pp: 99-212

in: F. Seitz and D. Turnbull (eds.) {\it Solid State Physics},

volume 7, New York: Academic.

\noindent Callaway, J. and W. Kohn. 1962. Electron wave func-

tions in metallic lithium. {\it Physical Review} 127: 1913.

\noindent Callen, H.B. 1960. {\it Thermodynamics}. New York: John

Wiley.

\noindent Casey, N.T. 1950. A critical analysis of Physics S-1b.

Walter Kohn Papers. Box 23, Folder 4. UArch FAcP

34. Department of Special Collections. University

of California, Santa Barbara.

\noindent Celli, V. 2013. February 8 2013 telephone interview with

the author.

\noindent Cesarini, D. and T. Kushner. 1993. {\it The Internment of

Aliens in Twentieth Century  Britain}. London: Frank

Cass \& Co. Ltd.

\noindent Cheetham, A. 1992. Video interview of Walter Kohn and

Philippe Nozi\`{e}res for the Vega Science Trust.

Available at http://vega.org.uk/video/programme/134

\noindent Chien, C.L. and C.R. Westgate. 1980. {\it The Hall Effect

and Its Applications}. New York: Plenum.

\noindent Cohen, M.H. 1963. Interatomic interactions in metals.

pp. XI-1-9 in: J. Friedel and A. Guinier (eds.) {\it Metallic

Solid Solutions}. New York: Benjamin.

\noindent Coleman, A.J. 1963. Structure of fermion density

matrices. {\it Reviews of Modern Physics} 35: 668-687.

\noindent Crawford, T.D, Wesolowski, S.S., Valeev, E.F., King,

R.A., Leininger, M.L., and H.F Schaefer III. 2001. The

past, present, and future of quantum chemistry. pp.

219-246 in:  E. Keinan and I. Schechter (eds.) {\it Chemis-

try for the $21^{\rm st}$ Century}. Weinheim: Wiley-VCH.

\noindent Curio, C. 2004. Invisible children: the selection and

integration strategies of relief organizations. {\it Shofar}

23: 41-56.

\noindent Curtiss, L.A, Raghavachari, K., Trucks, G.W., and

J.A. Pople. 1991. Gaussian-2 theory for molecular

energies of first- and second-row compounds.

{\it Journal of Chemical Physics} 94:7221-7230.

\noindent Daniel, E. and S.H. Vosko. 1960. Momentum distri-

bution of an interaction electron gas. {\it Physical

Review} 120:2041-2044.

\noindent DeWitt, B.S. 1951. Theoretical Physics. {\it Physics Today}

4(12): 22-23.

\noindent DeWitt, C.M. 2013. Program of the 1951 Summer School

of Theoretical Physics, Les Houches, France. Courtesy

of Prof. C\'{e}cile deWitt, University of Texas.

\noindent Diaz, J.B. 1978. {\it Alexander Weinstein Selecta}. London:

Pitman.

\noindent Dirac, P.A.M. 1930. Note on exchange phenomena in the

Thomas atom. {\it Proceedings of the Cambridge

Philosophical Society} 26: 376-385.

\noindent Dirac, P.A.M. 1935. {\it The Principles of Quantum

Mechanics} $2^{\rm nd}$ edition. Oxford: Clarendon Press.

\noindent Domb, C. 1996. {\it The Critical Point: a Historical

Introduction to the Modern Theory of Critical

Phenomena}. London: Taylor \& Francis.

\noindent de Dominicis, C. 1963. Variational statistical

mechanics in terms of `observables' for normal and

superfluid systems. {\it Journal of Mathematical

Physics} 4: 255-265.

\noindent de Dominicis, C. and P.C. Martin. Stationary entropy

principle and renormalization in normal and superfluid

systems. {\it Journal of Mathematical Physics} 5: 14-30.

\noindent DuBois, D.F. and M.G. Kivelson. 1962. Quasi-classical

theory of electron correlation in atoms. {\it Physical

Review} 127: 1182-1192.

\noindent Dresselhaus, G., A.F. Kip, and C. Kittel. 1955.

Cyclotron resonance of electrons and holes in silicon

and germanium crystals. {\it Physical Review} 98: 368-384.

\noindent Duff, G.F.D. 1969. Arthur Francis Chesterfield Steven-

son. 1899-1968. {\it Proceedings of the Royal Society of

Canada}. Series IV. 8: 104-107.

%

\noindent Edwards, S.F. 1998. Reminiscences. pp. 203-204 in:

E.A. David (ed.) {\it Nevill Mott, Reminiscences and

Appreciations}. London: Taylor and Francis.

\noindent Ehrenreich, H. and L.M. Schwartz. 1976. The electronic

structure of alloys. pp. 149-286 in: H. Ehrenreich,

F. Seitz, and D. Turnbull (eds.) {\it Solids State Physics},

volume 31, New York: Academic.

\noindent Eisinger, J. and G. Feher. 1958. Hfs anomaly of Sb$^{121}$

and Sb$^{123}$ determined by the electron nuclear double

resonance technique. {\it Physical Review} 109: 1172-1183.

\noindent Eisinger, J. 2003. For Rappa on his $80^{\rm th}$ Birthday from

Terry. pp. 63-65 in:  M. Scheffler  and P. Weinberger

(eds.) {\it Walter Kohn--Personal Stories and Anecdotes}.

Berlin: Springer.

\noindent Eisinger, J. 2011. {\it Einstein on the Road}. Amherst, New

York: Prometheus Books.

\noindent Eisinger, J. 2013. July 24 2013 telephone interview by

the author and subsequent correspondence.

\noindent Ehrlich, G. 2003. Reunion in history. pp. 60-62 in: M.

Scheffler and P. Weinberger (eds.) {\it Walter Kohn--

Personal Stories and Anecdotes}. Berlin: Springer.

\noindent Faddeev, L. 1965. {\it Mathematical Aspects of the

Three Body Problem in Quantum Scattering Theory}.

Jerusalem: Israel Program for Scientific Translations.

\noindent Fast, V.K. 2011. {\it Children's Exodus: a History of the

Kindertransport}. London: I.B. Tauris.

\noindent Feher, G. 2002. The creation of the physics department.

{\it Chronicles: Newsletter of the UCSD Emeriti

Association} 2 (2): 6-8.

\noindent Feibelman, P.J. 2012. June 28 2012 telephone interview

with the author. Peter Feibelman was a PhD student

of Keith Brueckner at UCSD.

\noindent Feldberg, W.S. 1960. Bruno Mendel. 1897-1959.

{\it Biographical Memoirs of Fellows of the Royal

Society} 6: 190-199.

\noindent Fernandez, B. 2013. {\it Unravelling the Mystery of the

Atomic Nucleus} New York: Springer.

\noindent Feynman, R.P. 1949. Space-time approach to quantum

electrodynamics. {\it Physical Review} 76: 769-789.

\noindent Fletcher, R.C. 2013. October 30 2013 correspondence

with the author.

\noindent Florides, P.S. 2008. John Lighton Synge. 1897-1995.

{\it Biographical Memoirs of Fellows of the Royal Society}

54: 401-424.

\noindent Fock, V. 1930.  Approximate methods for the solution

 of the quantum mechanical many-body problem (in

 German). {\it Zeitschrift f\"{u}r Physik} 61: 126-148.

\noindent Ford, J. 2013. June 19 2013 correspondence between the

author and Jon Ford, Head Teacher, Imberhorne Sec

ondary School, East Grinstead, England.

\noindent Frank, P. 1947. {\it Einstein, his Life and Times}. New York:

Alfred A. Knopf.

\noindent Friedel, J. 1954. Electronic structure of primary solid

solutions. {\it Advances in Physics} 3: 446-507.

\noindent Friedel, J. 1958. Metallic alloys. {\it Supplemento del

Nuovo Cimento} 7: 287-311.

\noindent Friedel, J. 1963. The concept of the virtual bound state.

pp. XIX-1-22 in: J. Friedel and A. Guinier (eds.)

{\it Metallic Solid Solutions}. New York: Benjamin.

\noindent Frisch, O.R. 1979. {\it What Little I Remember}. Cambridge:

 University Press.

\noindent Galitskii, V.M. and A.B. Migdal. 1958. Application of

quantum field theory methods to the many-body

problem. {\it Journal of Experimental and Theoretical

Physics} 34: 139-150.

\noindent G\'{a}sp\'{a}r, R. 1954. About an approximation to the

Hartree-Fock potential through a universal potential

function (in German). {\it Acta Physica Hungarica} 3: 263-

286.

\noindent Gavroglu, K. and A. Sim\~{o}es. 2012. {\it Neither Physics nor

Chemistry} Cambridge, MA: MIT Press.

\noindent Gell-Mann, M. 1996. Reminiscences. {\it Philosophical

Magazine} 74: 431-434.

\noindent de Gennes, P.G. 1966. {\it Superconductivity in Metals and

Alloys}. Reading, MA: Benjamin.

\noindent Geoffroy, P.R. 1946. {\it Report on the Magnetometer Survey

of the property of Dante Red Lake Gold Mines Ltd.}

Available as file 52N04N9958.pdf from the Ontario

Ministry of Northern Development and  Mines.

\noindent Giannakis, G. 2013. August 6 2013 correspondence

between the author and G. Giannakis, a member of

the reference staff of the McGill University Archives.

\noindent Gilman, P. and L. Gilman. 1980. {\it Collar the Lot: How

Britain Interned and Expelled its Wartime Refugees}.

London: Quartet Books.

\noindent Glasser, M.L. 2013. July 12 2013 telephone interview by

the author and subsequent correspondence with former

Carnegie Tech graduate student Larry Glasser.

\noindent Gold, A.V. 1958. An experimental determination of the

Fermi surface in lead. {\it Philosophical Transactions of

the Royal Society A} 251: 85-112.

\noindent Goldman, J.E. 1957. {\it The Science of Engineering

Materials}. New York: John Wiley.

\noindent Goldstone, J. 1957. Derivation of Brueckner many-body

theory. {\it Proceedings of the Royal Society A} 239:

267-279.

\noindent Goudsmit, S.A. and G.L.Trigg. 1964. October 1964

Memorandum from the Editors of {\it Physical Review

Letters}.

\noindent Goudsmit, S.A. 1965. May 28 letter to L.J. Sham.

Courtesy of L.J. Sham.

\noindent Greene, M.P. 2013. March 30 2013 telephone interview

with the author.

\noindent Griffin, A. 2007. Many-body physics in the 1960's: a

golden age. A talk given at the symposium, `Fifty

Years of Condensed Matter Physics'. June 16 2007.

Available at http://www.lassp.cornell.edu/

lassp\textunderscore\,data/LASSP/50YearsCMP-VA2007.pdf

\noindent Gunnarsson, O., Harris, J. and R.O. Jones. 1977.

Density functional theory and molecular bonding. I.

first-row diatomic molecules. {\it Journal of Chemical

Physics} 67: 3970-3979.

\noindent ter Haar, D. 1954. {\it Elements of Statistical Mechanics}

Rinehart: New York.

\noindent Hanta, K. 1999. From exile to excellence: an interview

with Nobel prize laureate Walter Kohn.

{\it Austria Kultur} 9 (1). Available online at

http://www.auslandsdienst.at/de/projekt/

pressearchiv-1999/austria-kultur-vol-9-no1-

januaryfebruary-1999-exile-excellence-interview

\noindent Hardy, G.H. 1938. {\it A Course in Pure Mathematics}.

Seventh edition. Cambridge: University Press.

\noindent Harrison, W.A. and M.B. Webb. 1960. {\it The Fermi

Surface} John Wiley: New York.

\noindent Hasegawa, H. 2004. Walter Kohn in Japan. pp. 93-95

in:  M. Scheffler and P. Weinberger (eds.) {\it Walter

Kohn--Personal Stories and Anecdotes}. Berlin:

Springer.

\noindent Herman, F. 1958. Theoretical investigation of the

electronic energy band structure of solids. {\it Reviews of

Modern Physics} 30: 102-121.

\noindent Herman, F. 1964. Recent progress in energy band theory.

pp. 3-22 in: Hulin, M. (ed.) {\it Physics of Semi-

conductors}. Paris: Dunod.

\noindent Herman, F. and S. Skillman. 1963. {\it Atomic Structure

Calculations}. Englewood Cliffs, New Jersey:

Prentice-Hall.

\noindent Herman, F., Van Dyke, J.P., and I.B. Ortenburger.

1969. {\it Physical Review Letters} 22: 807-811.

\noindent Hinman, G. and D. Rose. 2010. Edward Chester Creutz.

{\it Biographical Memoir}  Washington, D.C.: National

Academy of Sciences Press.

\noindent Hoddeson, L., E. Braun, J. Teichmann, and S. Weart.

 1992. {\it Out of the Crystal Maze}. New York: Oxford

University Press.

\noindent Hoddeson, L., H. Schubert, S.J. Heims, and G. Baym.

1992. Collective Phenomena. pp. 489-616: in

Hoddeson, L., E. Braun, J. Teichmann, and S. Weart

(eds.) {\it Out of the Crystal Maze}. New York:

Oxford University Press.

\noindent Hodges, C.H. and Stott, M.J. 1972. Theory of electro-

chemical effects in alloys. {\it Philosophical Magazine}

26: 375-392.

\noindent Hohenberg, P.C. 2003. A personal tribute to Walter

Kohn on his $80^{\rm th}$ birthday. pp. 99-102  in: M.

Scheffler and P. Weinberger (eds.) {\it Walter Kohn--

Personal Stories and Anecdotes}. Berlin: Springer.

\noindent Hohenberg, P.C. 2012. July 19 2012 interview with the

author.

\noindent Hohenberg, P. and Kohn, W. 1964. Inhomogeneous

electron gas. {\it Physical Review} 136: B864-B871.

\noindent Hohenberg, P.C., W. Kohn, and L.J. Sham. 1990. The

beginnings and some thoughts on the future.

{\it Advances in Quantum Chemistry} 21: 7-26.

\noindent Hollander, N. 2000. Interview with Walter Kohn. Nobel

Voices Video History Project. Archives Center,

Smithsonian National Museum of American History.

\noindent Horsley, A. 2013. June 25 2013 telephone interview by

the author with Allen Horsley (son of Caperton B.

Horsley, the founder of the  Sutton-Horsley Co.) and

subsequent email correspondence with his sisters,

Rose Shelton Horsley Cruz and Lucile Horsley

Blanchard.

 \noindent Houghton, A. 1961. Specific heat and spin susceptibility

 of dilute alloys. {\it Journal  of the Physics and  Chemistry

  of Solids} 20: 289-293.

\noindent Hubbard, J. 1957. The description of collective motions

in terms of many-body perturbation theory.

{\it Proceedings of the Royal Society (London)

A} 239: 539-560.

\noindent Hubbard, J. 1958. The description of collective motions

in terms of many-body perturbation theory III. The

extension of the theory to the non-uniform gas.

{\it Proceedings of the Royal Society (London) A} 244:

199-211.

\noindent Hulth\'{e}n, L. 1946. The variational principle for

continuous spectra. pp. 201-206: in {\it Dixi\`{e}me

Congre\`{e}s des Math\'{e}maticiens Scandinaves}

Copenhagen: Julius Gjellerups Forlag.

\noindent Hume-Rothery, W. 1931. {\it The Metallic State}. Oxford:

 Clarendon Press.

\noindent Hume-Rothery, W. and G.V. Raynor. 1962. {\it The

Structure of Metals and Alloys}. $4^{\rm th}$ edition.

London: Institute of Metals.

\noindent IAQMS. 2014. Awards website of the International

Academy of Quantum Molecular Science.

www.iaqms.org/awards.php

\noindent Infeld, L. 1978. {\it Why I Left Canada}. $2^{\rm nd}$ edition.

Montreal: McGill-Queen's University Press.

\noindent Jammer, M. 1966. {\it The Conceptual Development

of Quantum Mechanics}. New York: McGraw-Hill.

\noindent Johnson, B.G., Gill, P.M.W., and J.A. Pople. 1992.

{\it Journal of Chemical Physics} 97: 7846-7848.

\noindent Jones, H. 1934. The theory of alloys in the $\gamma$-phase.

{\it Proceedings of the Royal Society of London A} 144:

225-234.

\noindent Jones, R.O. 2012. Density functional theory: a personal

view. pp. 1-28: in M. Avella and F. Mancini (eds.):

{\it Strongly Correlated Systems: Theoretical Methods}.

Berlin: Springer-Verlag.

\noindent Jones, T. 1988. {\it Both Sides of the Wire}. Fredericton,

New Brunswick, Canada: New Ireland Press.

\noindent Jones, W., March, N.H., and S. Sampanthar. 1961.

The energy and the Dirac density matrix of a non-

uniform electron gas. {\it Physics Letters} 1: 303-304.

\noindent Jost, R. and W. Kohn. 1952a. Construction of a

potential from a phase shift. {\it Physical Review} 87:

977-992.

\noindent Jost, R. and W. Kohn. 1952b. Equivalent potentials.

{\it Physical Review} 88: 382-385.

\noindent Jost, R. and W. Kohn. 1953. On the relation between

phase shift energy levels and the potential. {\it Det

Kongelige Danske Videnskarbernes Selskab

Matematisk-fysiske Meddelelser} 27: 1-19.

\noindent Kaiser, D. 2005. {\it Drawing Theories Apart}. Chicago:

University Press.

\noindent Kerr, C. 2001. {\it The Gold and the Blue: a Personal

Memoir of the University of California, 1949-1967}.

Volume  1. Academic Triumphs. Berkeley: University

of California Press.

\noindent Kittel, C. and A.H. Mitchell. 1954. Theory of donor and

acceptor states in silicon and germanium. {\it Physical

Review} 96: 1488-1493.

\noindent Kittel, C. 1963 {\it Quantum Theory of Solids} New York:

Wiley.

\noindent Kjeldaas, T. and W. Kohn. 1956. Interaction of con-

duction electrons and nuclear magnetic moments in

metallic sodium. {\it Physical Review} 101: 66-67.

\noindent Kjeldaas, T. 1959. Theory of Ultrasonic Cyclotron Reso-

nance in Metals at Low Temperature. Ph.D. Thesis.

University of Pittsburgh.

\noindent Klein, F. and A. Sommerfeld. 1898. {\it \"{U}ber die Theorie

des Kreisels} Leipzig: Teubner.

\noindent Koch, E. 1980. {\it Deemed Suspect}. Toronto: Methuen.

\noindent Kohn, W. 1945. The spherical gyrocompass. {\it Quarterly

of Applied Mathematics} 3: 87-88.

\noindent Kohn, W. 1946. Contour integration in the theory of the

spherical pendulum and the heavy symmetrical top.

{\it Transactions of the American Mathematical

Society} 59: 107-131.

\noindent Kohn, W. 1947. Two applications of the variational

method to quantum mechanics. {\it Physical Review} 71:

635-637.

\noindent Kohn, W. 1948. Variational methods in nuclear collision

problems. {\it Physical Review} 74: 1763-1772.

\noindent Kohn, W. 1952a. Validity of Born expansions.

 {\it Physical Review} 87: 539-540.

\noindent Kohn, W. 1952b. Variational methods for periodic

lattices. {\it Physical Review} 87: 472-481.

\noindent Kohn, W. 1954. Interaction of conduction electrons and

nuclear magnetic moments in metallic lithium.

{\it Physical Review} 96: 590-592.

\noindent Kohn, W. 1957a. Shallow impurity states in Si and Ge.

{\it Solid State Physics}. F. Seitz and D. Turnbull (eds.)

pp:257-320.

\noindent Kohn, W. 1957b. Effective mass theory in solids from a

many-particle point of view. {\it Physical Review} 105:

509-516.

\noindent Kohn, W. 1958. Interaction of charged particles in a

dielectric. {\it Physical Review} 110: 857-864.

\noindent Kohn, W. 1959a. Analytic properties of Bloch waves and

Wannier functions. {\it Physical Review} 115: 809-821.

\noindent Kohn, W. 1959b. Theory of Bloch electrons in a magnetic

field: the effective Hamiltonian. {\it Physical Review}

 115: 1460-1478.

\noindent Kohn, W. 1959c. Image of the Fermi surface in the

vibration spectrum of a metal. {\it Physical Review Letters}

2: 393-394.

\noindent Kohn, W. 1961. Cyclotron resonance and de Haas-

van Alphen oscillations of an interacting electron gas.

{\it Physical Review} 123: 1242-1244.

\noindent  Kohn, W. 1962. John Simon Guggenheim Memorial

Foundation Application. Used by permission of Walter

Kohn and the Guggenheim Memorial Foundation.

\noindent Kohn, W. 1964. Theory of the insulating state. {\it Physical

Review} 133: A171-A181.

\noindent Kohn, W. 1996. Tribute to Julian Schwinger. pp. 61-64:

in Y. Jack Ng (editor): {\it Julian Schwinger: the

Physicist, the Teacher, and the Man}. Singapore.

World Scientific.

\noindent Kohn, W. 1998. Nobel Prize autobiography. Available

online at http://www.nobelprize.org/nobel\textunderscore\,prizes/

chemistry/laureates/1998/kohn-autobio.html.

\noindent Kohn, W. 1999. Nobel lecture: electronic structure of

matter--wave functions and density functionals.

{\it Reviews of Modern Physics} 71: 1253-1266.

\noindent Kohn, W. 2000. Letter of August 23 2000 from

Walter Kohn to the Principal of the Imberhorne

Secondary School, East Grinstead, England.

\noindent Kohn, W. 2001a. A personal account of the history of

density functional theory. Session S3 (History of

electronic structure theory in atoms, molecules, and

solids), American Physical Society annual meeting,

March 14 2001, Seattle, WA. Sound  recording AV

2001-374z used with the permission of the Niels Bohr

Library \& Archives, American Institute of  Physics

One Physics Ellipse, College Park, MD 20740, USA.

\noindent Kohn, W. 2001b. Graduate School Days. p. 6 in the fall

2002 Newsletter of the American Physical Society

Forum on Graduate Student Affairs. One Physics

Ellipse, College Park, Maryland.

\noindent Kohn, W. 2003. A Fireside Chat with Nobel Laureate

Professor Walter Kohn. An August 15 2003 broadcast

by the University of California Television. Online at

http://www.youtube.com/watch?v=VDNNiKdtyhg.

\noindent Kohn, W. 2004. My honored teachers in Vienna. pp.

43-50 in: F. Stadler (editor): {\it \"{O}sterreichs Umgang

mit dem Nationalsozialismus}. Wien: Springer.

\noindent Kohn, W. 2012a. Congratulation. {\it Journal of Super-

conductivity and Novel Magnetism} 25: 551.

\noindent Kohn, W. 2012b. December 18 2012 interview conducted

by the author.

\noindent Kohn, W. 2013a. March 3 2013 interview conducted by

Nina Krieger for the Internment Project of the Van-

couver Holocaust Education Centre (VHEC). The

author thanks archivist Elizabeth Shaffer of the

VHEC for making a videotape of this interview

available to him.

\noindent Kohn, W. 2013b. October 8 2013 interview conducted

by the author and subsequent correspondence.

\noindent Kohn, W.  and N. Bloembergen. 1950. Remarks on  the

nuclear resonance shift in metallic lithium. {\it Physical

Review} 80: 913.

\noindent Kohn, W. and M. Luming. 1963. Orbital susceptibility

of dilute alloys. {\it Journal of the Physics and

Chemistry of Solids} 24: 851-862.

\noindent Kohn, W. and J.M. Luttinger. 1955a. Hyperfine splitting

 of donor states in silicon. {\it Physical Review} 97:
883-888.

\noindent Kohn, W. and J.M. Luttinger. 1955b. Theory of donor

states in silicon. {\it Physical Review} 98: 915-922.

\noindent Kohn, W. and J.M. Luttinger. 1957. Quantum theory of

electrical transport phenomena. {\it Physical Review} 108:

590-611.

\noindent Kohn, W. and J.M. Luttinger 1960. Ground state of a

many-fermion system. {\it Physical Review} 118: 41-45.

\noindent Kohn, W. and S.J. Nettel. 1960. Giant fluctuations in a

degenerate fermi gas. {\it Physical Review Letters} 5:
8-9.

\noindent Kohn,  W. and N. Rostoker.  1954. Solution of the

Schr\"{o}dinger equation in periodic lattices with an

application to metallic lithium. {\it Physical Review}

94: 1111-1120.

\noindent Kohn, W., D. Ruelle, and A. Wightman. 1992. Res Jost.

{\it Physics Today} 45: 120-121.

\noindent Kohn, W. and D. Schechter. 1955. Theory of acceptor

levels in germanium. {\it Physical Review} 99: 1903-1904.

\noindent Kohn, W. and L.J. Sham. 1965a. Quantum density

oscillations in an inhomogeneous electron gas.

{\it Physical Review} 137: A1697-A1705.

\noindent Kohn, W. and L.J. Sham. 1965b. Self-consistent

equations including exchange and correlation effects.

{\it Physical Review} 140: A1133-A1138.

\noindent Kohn, W. and S.H. Vosko. 1960. Theory of nuclear

resonance intensity in dilute alloys. {\it Physical

Review} 119: 912-918.

\noindent Korringa, J. 1947. On the calculation of the energy of a

Bloch wave in a metal. {\it Physica} 13:392–400.

\noindent Korringa, J. 1958. Dispersion theory for electrons

in a random lattice with applications to the electronic

structure of alloys. {\it Journal of the Physics and

Chemistry of Solids} 7: 252-258.

\noindent Korringa, J. 1994. Early history of multiple scattering

theory for ordered systems. {\it Physics Reports}. 238:

341-360.

\noindent Kryachko, E.S. and C.V. Lude$\tilde{\rm n}$a. 1990. {\it Energy Density

Functional Theory of Many-Electron Systems}

Dordrecht: Kluwer Academic Publishers.

\noindent Kubo, R. 1957. Statistical mechanical theory of irrever-

sible processes. I. {\it Journal of the Physical Society

of Japan}. 12: 570-586.

\noindent Lambek, J. 1980. Reminiscences of Fritz Rothberger.

Published in the September 2000 issue of the

newsletter {\it CMS Notes de la SMC} 32: (5) 29.

\noindent Landau, L.D. 1956. The theory of a Fermi liquid.

{\it Journal of Experimental and Theoretical Physics}

30: 1058-1064.

\noindent Langer, J.S. 2003. Reminiscences on the occasion

of Walter Kohn's $80^{\rm th}$ birthday. pp. 124-126 in:

M. Scheffler  and P. Weinberger (eds.) {\it Walter Kohn--

Personal Stories and Anecdotes}. Berlin: Springer.

\noindent Langer, J.S. and S.H. Vosko. 1959. The shielding of a

fixed charge in a high-density electron gas. {\it Journal

of Physics and Chemistry of Solids} 12: 196-205.

\noindent Langreth, D.C. and M.J. Mehl. 1983. Beyond the

local density approximation in calculations of

ground state electronic properties. {\it Physical Review B}

28: 1809-1834.

\noindent Lee, C., Yang, W., and R.G. Parr. 1988. Develop-

ment of the Colle-Salvetti correlation energy into a

density functional of the electron density. {\it Physical

Review A} 37: 785-789.

\noindent LJPS 1985. Proceedings of the La Jolla Physics

Symposium. September 6-8, 1985. pp. 142-145.

http://www.physics.ucsd.edu/dept/department\textunderscore\

\!history.pdf

\noindent L\"{o}wdin, P.-O. 1959. Correlation problem in many-

electron quantum mechanics. {\it Advances in Chemical

Physics} 2: 207-322.

\noindent Luttinger, J.M. 1960. Fermi surface and some

simple equilibrium properties of a system of interaction

fermions. {\it Physical Review} 119: 1153-1163.

\noindent Luttinger, J.M. 1961. Theory of the de Haas-van Alphen

effect for a system of interacting electrons. {\it Physical

Review} 121: 1251-1258.

\noindent Luttinger, J.M. and W. Kohn. 1955. Motion of electrons

and holes in perturbed periodic fields. {\it Physical

Review} 97: 869-883.

\noindent Luttinger, J.M. and W. Kohn. 1958. Quantum theory of

electrical phenomena. II. {\it Physical Review} 109:

1892-1909.

\noindent Luttinger, J.M. and P. Nozi\`{e}res. 1962. Derivation of the

Landau theory of Fermi liquids.  II. Equilibrium

Properties and Transport Equation.  {\it Physical

Review} 127: 1431-1440.

\noindent March, N.H. 1957. The Thomas-Fermi approximation in

quantum mechanics. {\it Advances in Physics} 6:1-101.

\noindent March, N.H. 1975. {\it Self-Consistent Fields in Atoms}.

 Oxford: Pergamon Press.

\noindent March, N.H. and A.M. Murray. 1961. Self-consistent

perturbation treatment of impurities and imperfec-

tions in metals. {\it Proceedings of the Royal Society A}

261: 119-133.

\noindent Markoff, J. 2011. Jacob E. Goldman, founder of Xerox

lab dies at 90. {\it New York Times} December 23 2011:

B17.

\noindent Martin, P.C. and S.L. Glashow. 1995. Julian Schwinger:

prodigy, problem solver, pioneering physicist. {\it Physics

Today} 48 (10): 40-46.

\noindent McWeeny, R. 1960. Some recent advances in density

matrix theory. {\it Reviews of Modern Physics} 32:

335-369.

\noindent Mehra, J. and K.A. Milton. 2000. {\it Climbing the Moun-

tain}. Oxford: University Press.

\noindent Mermin, N.D. 1965. Thermal properties of the inhomo-

geneous electron gas. {\it Physical Review} 137:

A1441-A14443.

\noindent Mermin, N.D. 2003. Memorable moments with Walter

Kohn. pp. 155-159 in: M. Scheffler and P. Weinberger

(eds.) {\it Walter Kohn--Personal Stories and

Anecdotes}. Berlin: Springer.

\noindent Mermin, N.D. 2013. February 12 2013 correspondence

with the author.

\noindent Meyenn, K. von. 1989. Physics in the Making in Pauli's

Z\"{u}rich. pp. 93-130 in: A. Sarlemijn and M.J. Sparnaay

(editors): {\it Physics in the Making}. Amsterdam:

North-Holland.

\noindent Miedema, A.R., F.R. de Boer, and P.F. de Chatel. 1973.

Empirical description of the role of electronegativity in

alloy formation. {\it Journal of Physics F: Metal Physics}

3: 1558-1576.

\noindent Millman, S. 1983. {\it A History of Engineering and Science

in the Bell System: Physical Science (1935-1980)}.

Murray Hill, New Jersey, Bell Telephone

Laboratories.

\noindent Morse, P.M. 1982. John Clarke Slater. {\it Biographical

Memoir} Washington, D.C.: National Academy of

Sciences Press.

\noindent Mott, N.F. 1936. Resistivity of dilute solid solutions.

{\it Proceedings of the Cambridge Philosophical Society} 32:

281-290.

\noindent Mott, N.F. 1937. The energy of the superlattice in $\beta$

brass. {\it Proceedings of the Physical Society} 49:

258-263.

\noindent Mott, N.F. 1949. The basis of the electron theory of

metals, with special reference to the transition metals.

{\it Proceedings of the Physical Society (London)} 62:

416-422.

\noindent Mott, N.F. and H. Jones. 1936. {\it The Theory of the

Properties of Metals and Alloys}. Oxford: University

Press.

\noindent Musher, J.I. 1966. Comment on some theorems of

quantum chemistry. {\it American Journal of Physics}

34: 267-268.

\noindent Muto, T. 1938. On the electronic structure of alloys.

{\it Scientific  Papers of the Institute of Physical and

Chemical  Research (Tokyo)} 34: 377-390.

\noindent NBA 1951a. Niels Bohr Archive. Guest Book of the

Institute for Theoretical Physics. Copenhagen.

Denmark

\noindent NBA 1951b. Niels Bohr Archive. Program and

List of Participants for the Conference on Problems

of Quantum Physics. July 6-10 1951. Institute for

Theoretical Physics. Copenhagen. Denmark.

\noindent Nesbet, R.K. 2003. {\it Variational Principles and Methods

in Theoretical Physics and Chemistry}. Cambridge:

University Press.

\noindent Neuhaus, H. 2003. A class with class. pp. 173-176  in:

 M. Scheffler and P. Weinberger (eds.) {\it Walter

Kohn--Personal Stories and Anecdotes}. Berlin:

Springer.

\noindent Nobel Media. 2013. The Nobel Prize in Chemistry 1998.

Online at http://www.nobelprize.org/nobel\textunderscore\ prizes/

chemistry/laureates/1998/

\noindent Nordheim, L.H. 1931.  Zur electronentheorie der metalle

II. {\it Annalen der Physik} 9: 641-678.

\noindent Nozi\`{e}res, P. 1963. {\it Le Probl\`{e}me a N Corps} Paris: Dunod.

\noindent Nozi\`{e}res, P. 2012a. Sixty years of condensed matter

physics: an everlasting adventure. {\it Annual Reviews

of Condensed Matter Physics} 3: 1-7.

\noindent Nozi\`{e}res, P. 2012b. July 6 2012 correspondence with the

author.

\noindent Nozi\`{e}res, P. and J.M. Luttinger. 1962. Derivation of the

Landau theory of Fermi liquids I. formal prelimi-

naries. {\it Physical Review} 127: 1423-1431.

\noindent Old, B.S. 1961. The evolution of the Office of Naval

Research. {\it Physics Today} 14(8): 30-35.

\noindent Pais, A. 1982. {\it Subtle is the Lord: the Science and Life

of Albert Einstein}. Oxford: Clarendon Press.

\noindent Pais, A. 1996. Res Jost. pp. 1-9 in : K. Hepp (editor)

{\it Das M\"{a}rchen vom Elfenbeinernen Turm: Reden

und Aufs\"{a}tze}. Berlin: Springer.

\noindent Park, B.S. 2009. Between accuracy and manageability:

computational imperatives in quantum chemistry.

{\it Historical Studies in the Natural Sciences}

39: 32-62.

\noindent Parr, R.G. and W. Yang. 1989. {\it Density-Functional

Theory of Atoms and Molecules} Oxford: Clarendon

Press.

\noindent Pauling, L. 1949. A resonating-valence-bond theory of

metals and intermetallic compounds. {\it Proceedings of

the Royal Society of London A} 196: 343-262.

\noindent Peierls, R. 1933. Zur theorie des diamagnetismus von

leitungselektronen.  {\it Zeitschrift f\"{u}r Physik}.

80: 763-791.

\noindent Percus, J.K. 1963 {\it The Many-Body Problem}. New York:

Interscience.

\noindent Perdew, J.P. 1986. Density functional approximation for

the correlation energy of the inhomogeneous electron

gas. {\it Physical Review B} 33: 8822-8824.

\noindent Perutz, M. 1985. Enemy Alien. {\it The New Yorker}.

August 12 1985: 35-54.

\noindent Pincherle, L. 1960. Band structure calculations in

solids. {\it Reports on Progress in Physics} 23: 355-394.

\noindent Pines, D. 1961. {\it The Many-Body Problem}. Reading, MA:

Benjamin.

\noindent Pines, D. 1963. {\it Elementary Excitations in Solids}. New

York: Benjamin.

\noindent Pippard, A.B. 1957. An experimental determination of

the Fermi surface in copper. {\it Philosophical

Transactions  of the Royal Society A}. 250: 325-357.

\noindent Pople, J.A. 1965. Two-dimensional chart of quantum

chemistry. {\it Journal of Chemical Physics} 43: S229-S230.

\noindent Pople, J.A. 1991. The computation of molecular energies.

Video recording of an invited talk to the VII$^{\rm th}$

International Congress of Quantum Chemistry.

Courtesy of Prof. Axel Becke.

\noindent Powell, C.F. and G.P.S. Occhialini. 1947. {\it Nuclear

Physics in Photographs}. Oxford: Clarendon Press.

\noindent PT 1961. APS-AAPT Annual Joint Meeting in New

York. {\it Physics Today} 14(3): 52-56.

\noindent Pugh, E.M. and N. Rostoker. 1953. Hall effect in

ferromagnetic materials. {\it Reviews of Modern Physics}

25: 151-157.

\noindent Raimes, S.  1963. The rigid-band model. pp. X-1-9 in:

J. Friedel and A. Guinier (eds.)  {\it Metallic Solid

 Solutions}. New York: Benjamin.

\noindent Ren, M.L. 2013. October 20 2013 telephone interview

with Max Luming Ren.

\noindent Robinson, G. de B. 1979. {\it The Mathematics Department

in the University of Toronto 1827-1978} Toronto:

Department of Mathematics, University of Toronto.

\noindent Rostoker, N. 2003. I am happy that the R stands for

Rostoker. pp. 206-207 in: M. Scheffler  and P.

Weinberger (eds.) {\it Walter Kohn--Personal

Stories and Anecdotes}. Berlin: Springer.

\noindent Rostoker, N. 2013. July 23 2013 telephone interview with

the author.

\noindent Rowland, T.J. 1960. Nuclear magnetic resonance in

copper alloys: electron distribution  around solute

atoms. {\it Physical Review} 119: 900-912.

 \noindent Rozental, S. 1967. The forties and the fifties.

 pp. 149-190 in: S. Rozental (editor).  {\it Niels Bohr}

Amsterdam: North Holland.

\noindent R\"{u}rup R. 2008. {\it Schicksale und Karrieren} Gottingen:

Wallstein.

\noindent Rudnick, J. 2003. It started with image charges. pp.

208-210  in:  M. Scheffler and P. Weinberger (eds.)

{\it Walter Kohn--Personal Stories and Anecdotes}.

Berlin: Springer.

\noindent Sapolsky, H.M. 1990. {\it Science and the Navy} Princeton:

University Press.

\noindent Schwinger, J. 1947. A variational principle for scattering

problems. {\it Physical Review} 72: 742.

\noindent Schweber, S.S. 1994. {\it QED and the Men who Made it:

Dyson, Feynman, Schwinger, and Tomonaga}.

Princeton: University Press.

\noindent SDUT. 2010. Lois M. Kohn. Obituary. San Diego Union

Tribune, January 22 2010.

\noindent Sears E. 1990. The Life and Work of William S.

Heckscher. {\it Zeitschrift f\"{u}r Kunstgeschichte}

53: 107-133.

\noindent Seitz, F. 1940. {\it The Modern Theory of Solids}. New York:

McGraw-Hill.

\noindent Seitz, F. 1994. {\it On the Frontier: My Life in Science}. New

York: American Institute of Physics Press.

\noindent Seitz, F. and D. Turnbull. 1955.  {\it Solid State Physics}.

Volume 1. New York: Academic.

\noindent Senft, G. 2003. Economic development and economic

policies in the St\"{a}ndestaat era. pp. 32-55 in: G.

Bischof, A. Pelinka and A. Lassner (eds.): {\it The

Dollfuss-Schuschnigg Era in Austria: A Reassess-

ment}. New Brunswick, NJ: Transaction Publishers.

\noindent Sham, L.J. 1965. A calculation of the phonon frequencies

in sodium. {\it Proceedings of the Royal Society

of London A} 283: 33-49.

\noindent Sham, L.J. 2014. January 10 2014 correspondence with

the author.

\noindent Sham, L.J. and W. Kohn. 1966. One-particle properties

of an inhomogeneous interacting electron gas. {\it Physical

Review} 145: 561-567.

\noindent Sherwood, A.I. 2013. April 30 2013 telephone interview

with the author. Arnold Sherwood was a PhD student

of Maria Goeppert Mayer at UCSD.

\noindent Shockley, W. 1950. {\it Electrons and Holes in Semi-

conductors}. Princeton: Van Nostrand Co.

\noindent Silverman, R.A. 1951. {\it The Fermi Energy of Metallic

Lithium}. PhD thesis. Harvard University.

\noindent Silverman, R. and W. Kohn. 1950. On the cohesive

energy of metallic lithium. {\it Physical Review} 80:

912-913.

\noindent Siochi, C. 2013. March 20 2013 correspondence between

the author and Carlos Siochi, University of Toronto

Alumni Relations Officer for the Faculty of Arts and

Science.

\noindent Slater, J.C. 1939. {\it Introduction to Chemical Physics}. New

York: McGraw Hill.

\noindent Slater, J.C. 1951. A simplification of the Hartree-Fock

method. {\it Physical Review} 81: 385-390.

\noindent Slater, J.C. 1953. An augmented plane wave method for

the periodic potential problem. {\it Physical Review} 92:

603-608.

\noindent Slater, J.C. 1956. Band theory of bonding in metals.

pp. 1-12 in: {\it Theory of Alloy Phases}. Cleveland, OH,

American Society for Metals.

\noindent Slater, J.C. 1963. The electronic structure of atoms--

the Hartree-Fock method and correlation. {\it Reviews of

Modern Physics} 35: 484-487.

\noindent Slichter, C.P. 2010. Frederick Seitz.  {\it Biographical

Memoir} Washington, D.C.: National Academy of

Sciences Press.

\noindent Sonnert, G. and G. Holton. 2006. {\it What Happened

to the Children Who Fled Nazi Persecution}. New

York: Palgrave Macmillan.

\noindent Stevenson, A.F. and M.F. Crawford. 1938. A lower limi

for the theoertical energy of the normal state of helium.

 {\it Physical Review} 54: 375-379.

\noindent Swaminathan, S. 2000. Obituary of Fritz Rothberger.

{\it CMS Notes de la SMC} 32: (5) 29.

\noindent Taylor, P.L. 1963. Theory of Kohn anomalies in the

phonon spectra of metals. {\it Physical Review} 131:

1995-1999.

\noindent Taylor, P.L. 2013. March 10, 2013 interview with

the author.

\noindent Titchmarsh, E.C. 1949. Godfrey Harold Hardy. {\it Obituary

Notices of Fellows of the Royal Society} 6: 446-461.

\noindent Tong, B.Y. and L.J. Sham. 1966. Application of a self-

consistent scheme including exchange and correlation

effects in atoms. {\it Physical Review} 144: 1-4.

\noindent UCSDA 1960. University of California San Diego

Archives. University Communications News Releases.

RSS 6020. September 15 1960. Mandeville Special

Collections Library.

\noindent Varley, J.H.O. 1954. The calculation of heats of forma-

tion of binary alloys. {\it Philosophical Magazine} 45:

887-916.

\noindent Wannier, G.H. 1937. The structure of electronic

excitation levels in insulating crystals. {\it Physical

Review} 52: 191-197.

\noindent Weinstein, A. 1941. Les vibrations et le calcul des

variations. {\it Portugaliae Mathematica} 2: 36-55.

\noindent Weinstein, A. 1942. The spherical pendulum and

complex integration. {\it American Mathematical

Monthly}. 49: 521-523.

\noindent Wilson, A.H. 1936 {\it The Theory of Metals} Cambridge:

University Press.

\noindent Wilson, E.B. 1962. Four-dimensional electron density

function. {\it Journal of Chemical Physics} 36: 2232-2233.

\noindent WKP 1953a. Walter Kohn Papers. January 12 1953

letter to Walter Kohn from R.A. Deller, Bell Telephone

Laboratory; February 13 letter to Walter Kohn from

H. Tate, McGill University; February 24 letter to

Walter Kohn from Edward Creutz, Carnegie Institute

of Technology. Department of Special Collections,

Davidson Library, University of  California, Santa

Barbara.

\noindent WKP 1953b. Walter Kohn Papers. Final program of the

July 1953 Gordon Research Conference on the Physics

and Chemistry of Metals. Department of Special

Collections, Davidson Library, University of

California, Santa Barbara.

\noindent WKP 1955. Walter Kohn Papers. April 20 1955

letter from Walter Kohn to Harry Jones.

Department of Special Collections, Davidson Library,

University of California, Santa Barbara.

\noindent WKP 1957. Walter Kohn Papers. October 7 1957

letter from Frederick Reif to Walter Kohn.

Department of Special Collections, Davidson

Library, University of California, Santa Barbara.

\noindent WKP 1958. Walter Kohn Papers. January 29 1958

letter from Carnegie Tech President J.C. Warner

to Walter Kohn. Department of Special

Collections, Davidson Library, University of

California, Santa Barbara.

\noindent WKP 1959a. Walter Kohn Papers. October 21 1959

letter from Conyers Herring to Walter Kohn.

Department of Special Collections, Davidson

Library, University of California, Santa Barbara.

\noindent WKP 1959b. Walter Kohn Papers. October 26 1959

letter from Walter Kohn to Keith Brueckner.

Department of Special Collections, Davidson

Library, University of California, Santa Barbara.

\noindent WKP 1959c. Walter Kohn Papers. October 9 1959

letter from Keith Brueckner to Ben Mottelson.

Department of Special Collections, Davidson

Library, University of California, Santa Barbara.

\noindent WKP 1963a. Walter Kohn Papers. June 10 1963

letter from Lu Jeu Sham to Walter Kohn. Department

of Special Collections, Davidson Library, University of

California, Santa Barbara.

\noindent WKP 1963b. Walter Kohn Papers. August 6 1963

letter from Alan R. Liss (Vice President, Academic

Press, Inc.) to Walter Kohn. Department of Special

Collections, Davidson Library, University of

California, Santa Barbara.

\noindent WKP 1964. Walter Kohn Papers. June 15 1964 letter

from Walter Kohn to Pierre Hohenberg. Department

of Special Collections, Davidson Library,

University of California, Santa Barbara.

\noindent Woll, Jr., E.J. and W. Kohn. 1962. Images of the Fermi

surface in phonon spectra of metals. {\it Physical Review}

126: 1693-1697.

\noindent Young, A. 2013. May 2 2013 correspondence between the

author and Alice Carroll Young.

\noindent Zabloudil, J., R. Hammerling, P. Weinberger, and L.

Szunyogh. 2005. {\it Electron Scattering in Solid Matter}.

Berlin: Springer.

\noindent Zangwill, A. 2013. Hartree and Thomas: the forefathers

of density functional theory. Archive for History of

Exact Sciences 67: 331-348.

\noindent Ziman, J. 1960. {\it Electrons and Phonons}. Oxford:

Clarendon Press.

\noindent Ziman, J. 1964. {\it Principles of the Theory of Solids}.

Cambridge: University Press.


\end{document}